%% file: radvalley.tex
\documentclass[twocolumn]{emulateapj}
\usepackage[backref,breaklinks,bookmarks,bookmarksopen,colorlinks,linkcolor={blue},
  citecolor={blue},urlcolor={blue}]{hyperref}
\usepackage[all]{hypcap}

\usepackage{amsmath}
\usepackage{natbib}
\usepackage{color}
\usepackage[x11names]{xcolor}
\usepackage{graphics}
\usepackage{enumerate}
\usepackage{setspace}
\newcommand{\gaia}[1]{\emph{Gaia}#1}
\newcommand{\kepler}[1]{\emph{Kepler}#1}
\newcommand{\ktwo}[1]{\emph{K2}#1}
\newcommand{\tess}[1]{\emph{TESS}#1}
\newcommand{\plato}[1]{\emph{PLATO}#1}
\newcommand{\teff}[1]{$T_{\text{eff}}$#1}
\newcommand{\logg}[1]{$\log{g}$#1}

\defcitealias{fulton17}{F17}

\newcommand{\cdbox}[1]{%
  \colorlet{currentcolor}{.}%
  {\color{Blue1}%
    \dbox{\color{currentcolor}#1}}%
}
\usepackage{dashbox,framed,color,ocg-p}
\newcommand{\ToggleLayer}[2]{%
  \leavevmode
  \pdfstartlink user {
    /Subtype /Link
    /Border [0 0 0]%
    /A <<
      /S/JavaScript
      /JS (
         var aOCGs = this.getOCGs(), Layer;
         var Layers = "#1".split(","), Active = -1, i, l;
         for (l=0; l<Layers.length; l++) {
           Layer = Layers[l];
           for (i=0; aOCGs && i<aOCGs.length; i++) {
             if (aOCGs[i].state && aOCGs[i].name == Layer) {
               Active = l;
               aOCGs[i].state = false;
             }
           }
           if (Active >= 0) break;
         }
         if (Active == -1) {
           for (l=0; l<Layers.length; l++) {
             if (Layers[l] == "") Active = l;
           }
         }
         Active = Active + 1;
         if (Active == Layers.length) Active = 0;
         Layer = Layers[Active];
         for (i=0; aOCGs && i<aOCGs.length; i++) {
           if (aOCGs[i].name == Layer) aOCGs[i].state = true;
         }
      )
    >>
  }#2%
  \pdfendlink
}

\shortauthors{Cloutier \& Menou}
\shorttitle{Evolution of the radius valley with stellar mass}

\begin{document}
\title{Evolution of the radius valley around low mass stars from \emph{KEPLER} and \emph{K2}}
\author{Ryan Cloutier\altaffilmark{1,2}}
\author{Kristen Menou\altaffilmark{3,2,4}}

\email{ryan.cloutier@cfa.harvard.edu}

\altaffiltext{1}{Center for Astrophysics $|$ Harvard \& Smithsonian,
  60 Garden Street, Cambridge, MA, 02138, USA}
\altaffiltext{2}{Dept. of Astronomy \& Astrophysics, University
of Toronto, 50 St. George Street, Toronto, ON, M5S 3H4, Canada}
\altaffiltext{3}{Physics \& Astrophysics Group, Dept. of Physical \& Environmental Sciences,
  University of Toronto Scarborough, 1265 Military Trail, Toronto, ON, M1C 1A4, Canada}
\altaffiltext{4}{Dept. of Physics,
  University of Toronto, 60 St. George Street, Toronto, ON, M5S 1A7, Canada}

\begin{abstract}
  We present calculations of the occurrence rate of small close-in planets around low mass dwarf stars
  using the known planet populations from the \kepler{} and \ktwo{} missions.
  Applying completeness corrections 
  clearly reveals the radius valley in the maximum a-posteriori occurrence rates as a
  function of orbital separation and planet radius. We measure the slope of the valley
  to be $r_{p,\text{valley}} \propto F^{-0.060\pm 0.025}$ which
  bears the opposite sign from that measured around Sun-like stars thus suggesting that
  thermally driven atmospheric mass loss may not dominate the evolution of planets in the low stellar mass
  regime or that we are witnessing the emergence of a separate channel of planet formation.
  The latter notion is supported by the relative occurrence of rocky to non-rocky planets
  increasing from $0.5\pm 0.1$ around mid-K dwarfs to $8.5\pm 4.6$ around mid-M dwarfs.
  Furthermore, the center of the radius valley at $1.54\pm 0.16$ R$_{\oplus}$ is
  shown to shift to smaller sizes with
  decreasing stellar mass in agreement with physical models of photoevaporation,
  core-powered mass loss, and gas-poor formation. Although current measurements are
  insufficient to robustly identify the dominant formation pathway of the radius valley,
  such inferences may be obtained by \tess{} with $\mathcal{O}(85,000)$
  mid-to-late M dwarfs observed with 2-minute cadence.
  The measurements presented herein also precisely designate the subset of planetary orbital periods
  and radii that should be targeted in radial velocity surveys to resolve the rocky to non-rocky
  transition around low mass stars. 
\end{abstract}

\section{Introduction}
NASA's \kepler{} space telescope has discovered thousands of exoplanets over its lifetime and
consequently enabled robust investigations of the occurrence rate of planets within our galaxy.
One striking outcome of such studies was that the so-called super-Earths and sub-Neptunes---whose
radii span sizes intermediate between those of the Earth and Neptune---represent the most common
type of planet around Sun-like stars and M dwarfs alike
\citep[e.g.][]{youdin11,howard12,dressing13,fressin13,petigura13b,morton14,dressing15a,mulders15a,gaidos16,fulton17,hardegree19}.
Furthermore, mass measurements of many of these transiting planets via transit-timing variations
or precision radial velocity measurements revealed that the majority of planets
smaller than $\sim 1.6$ R$_{\oplus}$ are consistent with having bulk rocky compositions
\citep[e.g.][]{weiss14,dressing15b,rogers15}.

Early studies of the \kepler{} planet population
hinted that planets at small orbital separations exhibited a
bimodal radius distribution \citep[e.g.][]{owen13}---commonly referred to as the radius valley---that
is thought to be representative of a population of small, predominantly rocky planets plus a population
of inflated non-rocky planets that have retained significant H-He envelopes.
Consequently, numerous studies of planet formation and evolution sought to explain the
apparent bimodality. One such proposed mechanism is
that of photoevaporation wherein the gaseous envelopes of small close-in planets may be stripped by
X-ray and extreme ultraviolet (XUV) radiation from their host stars during the first $\sim 100$ Myrs
of the planet's lifetime
\citep{jackson12,owen13,jin14,lopez14,chen16,owen17,jin18,lopez18}.
The radius valley may also be explained by core-powered mass loss wherein the
luminosity from a planetary core's primordial energy reservoir from formation drives atmospheric escape
over Gyr timescales \citep{ginzburg18,gupta19a,gupta19b}.
Impact erosion by planetesimals may also drive the emergence of the radius valley either by atmospheric
stripping or by the growth of volatile-rich secondary atmospheres \citep{shuvalov09,schlichting15,wyatt19}. 
An alternative explanation to the processing of primordial atmospheres is the formation of distinct rocky and
non-rocky planet populations with the former invoking gas-poor formation wherein gas accretion is delayed
by dynamical friction whilst the
planetary core is still embedded within the protoplanetary disk until a point at which the gaseous disk
has almost completely dissipated after just a few Myrs \citep{lee14,lee16,lopez18}.

Observational tests of the aforementioned theoretical frameworks have become feasible in recent years due to 
the precise refinement of measured planet radii following improved stellar host characterization via  
spectroscopy, asteroseismology, and \gaia{} parallaxes
\citep[e.g.][]{fulton17,berger18,fulton18,vaneylen18,martinez19}. Each of these independent studies clearly
resolved the radius valley among small close-in planets orbiting Sun-like stars.
A variety of trends were also observed in either
the raw or in the completeness-corrected (i.e. the occurrence rate) distributions. Firstly,
the location of the radius valley around FGK stars is period-dependent with slope
$\mathrm{d}\log{r_p} / \mathrm{d}\log{P} \sim -0.1$ \citep{vaneylen18,martinez19}. This result is consistent
with both photoevaporation and core-powered mass loss models but is largely inconsistent with the late formation
of terrestrial planets in a gas-poor environment. Secondly, the feature locations (i.e. the weighted
average radius of the peaks and valley) appear to exist at smaller planet radii with decreasing stellar
mass \citep{fulton18,wu19}.

\begin{figure}
  \centering
  \includegraphics[width=0.98\hsize]{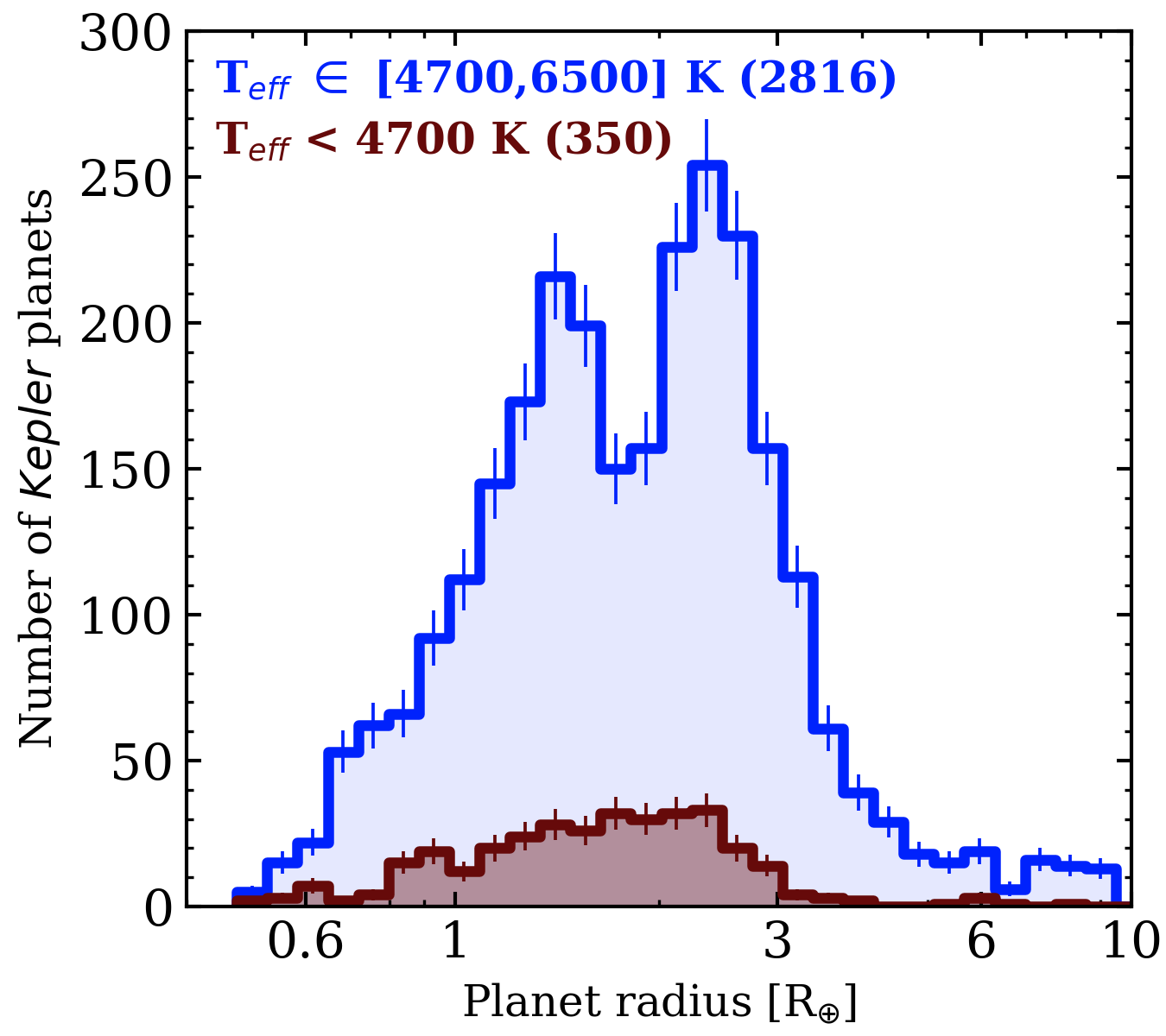}
  \caption{Empirical distributions of \kepler{} planet radii. Histograms of \kepler{} planet radii
    from \cite{berger18} for planets with host stellar effective temperatures \teff{} $\in [4700,6500]$ K
    (\emph{blue}) and \teff{} $<4700$ K (\emph{red}). The former subset of 2816 planets corresponds to the
    effective temperature range considered in the California Kepler Survey \citep{fulton17}
    wherein the radius valley is clearly
    resolved in the empirical distribution even without completeness corrections. A similar bimodal
    structure is not resolved in the empirical distribution of the latter subset around low mass stars
    due in-part to the relatively poor counting statistics with just 350 planets.}
  \label{fig:berger}
\end{figure}

In this study, we extend the investigation of the occurrence rate of small close-in planets to the low mass
stellar regime by considering planetary systems hosted by low mass dwarf stars later than mid-K dwarfs.
The empirical population of known planets in this stellar mass regime features nearly an order of magnitude
fewer planets than around Sun-like stars thus making the clear detection of the radius valley more
difficult and at a lower signal-to-noise. This fact is clearly evidenced in the empirical \kepler{} planet population
for which the radius valley around Sun-like stars (\teff{} $\in [4700,6500]$ K) is clearly exhibited whereas a similar
feature around low mass stars (\teff{} $< 4700$ K) is not easily discernible by eye (\autoref{fig:berger} based on
the data from \citealt{berger18}). Herein we leverage the precise stellar parallaxes from the \gaia{} DR2
for low mass stars observed by \kepler{} and \ktwo{} to refine the stellar parameters and compute precise
occurrence rates of close-in planets with the goal of resolving the radius valley and accurately measuring the
locations of its features and their uncertainties. Although it is unlikely that a single physical
mechanism is solely responsible for sculpting the radius valley, investigation the evolution of the valley features
with stellar mass can allude to which process---if any---dominates the evolution of close-in planets.

In Sects.~\ref{sect:stars} and ~\ref{sect:planets} we define our stellar sample from \kepler{} and \ktwo{}
and compile our sample of confirmed planets from each mission.
In Sect.~\ref{sect:completeness} we derive the transiting planet detection completeness and use those results
to calculate the occurrence rate of small close-in planets in which the structure of the radius valley around
low mass stars is resolved (Sect.~\ref{sect:occurrence}). In Sect.~\ref{sect:models} we compare our results to
model predictions and to results from planet population studies around Sun-like stars. 
Sect.~\ref{sect:discussion} presents a discussion of our results
and its implications followed by a summary of our main findings in Sect.~\ref{sect:conclusions}.

\section{Low mass dwarf stellar sample} \label{sect:stars}
The goal of this study is to extend measurements of the occurrence rate of close-in planets to planetary systems hosted
by low mass dwarf stars with effective temperatures \teff{} $<4700$ K: the lower limit of \teff{}
considered in the California Kepler Survey \citep[CKS;][]{fulton17}.
This adopted temperature threshold approximately corresponds to spectral types later than
K3.5V \citep{pecaut13}. In the following subsections we define our stellar samples from each of \kepler{} or \ktwo{.}

\subsection{Kepler stellar sample} \label{sect:kep}
Following the release of \gaia{} DR2 \citep{lindegren18}, \cite{berger18} cross-matched \kepler{} target stars
with DR2 and compiled a catalog of stellar
parallaxes $\varpi$, 2MASS $K_s$-band magnitudes, and spectroscopic measurements of \teff{,} \logg{,} and [Fe/H]
for $\sim 178,000$ stars observed as part of the primary \kepler{} mission. Spectroscopic measurements were obtained from
either the Data Release 25 (DR25)
Kepler Stellar Properties Catalog \citep[KSPC;][]{mathur17}, the California
Kepler Survey \citep[CKS;][]{petigura17} where available, and \teff{} values for stars with \teff{} $<4000$ K were compiled from
\cite{gaidos16}. The full set of available stellar parameters were used as input within the spectral classification code
\texttt{isoclassify} \citep{huber17} to calculate stellar luminosities. The resulting luminosity values were consequently combined
with \teff{} measurements to refine the stellar radii using the Stefan-Boltzmann law for the majority of \kepler{} FGK stars.
However, bolometric corrections for \kepler{} M dwarfs with \teff{} $<4100$ K
and absolute $K_s$-band magnitudes $M_{K_s}>3$ are known to suffer significant inaccuracies owing to incomplete
molecular line lists. For these stars, \cite{berger18} instead adopted the empirically-derived M dwarf radius-luminosity
relation from \cite{mann15} to refine the M dwarf stellar radii. \cite{berger18} also combined \teff{} and the 
luminosity measurements to derive stellar evolutionary flags aimed at classifying stars as either a dwarf, a subgiant, or a
red giant.

Stellar masses $M_s$ are not reported by \cite{berger18}. In order to study the \kepler{} planet population as a function of $M_s$
we derive $M_s$ values given the measured stellar radii $R_s$ and using the mass-radius relation from \cite{boyajian12} 
applicable to both K and M dwarfs.
\cite{boyajian12} acquired interferometric measurements with the \emph{CHARA} array of 21 nearby K and M dwarfs
to measure the angular size of each stellar disk at the level of $\lesssim 5$\%. Their stellar sample was supplemented by 12
literature measurements of $R_s$ from interferometry. Mass measurements were then derived using the $K_s$-band mass-luminosity
relation from \cite{henry93} which was valid for their full stellar sample spanning 0.13-0.90 R$_{\odot}$. \cite{boyajian12}
parameterized the stellar mass-radius relationship as a quadratic in $M_s$ and reported values and uncertainties for each polynomial
coefficient. Here, we assume independent Gaussian probability density functions (PDF) for each coefficient and sample their values
along with each star's $R_s$ from their respective measurement uncertainties to derive the $M_s$ PDF for all of the low mass dwarfs
in our preliminary \kepler{} sample.

We define our final \kepler{} stellar sample by focusing on stars that satisfy the following criteria:

\begin{enumerate}
\item \kepler{} magnitude $K_p < 16$,
\item $T_{\text{eff}} - \sigma_{T_{\text{eff}}} \leq 4700$ K,
\item $R_s - \sigma_{R_s} \leq 0.8$ R$_{\odot}$,
\item $M_s - \sigma_{M_s} \leq 0.8$ M$_{\odot}$, and
\item and an evolutionary flag corresponding to a dwarf star. 
\end{enumerate}

\begin{figure}
  \centering
  \includegraphics[width=0.98\hsize]{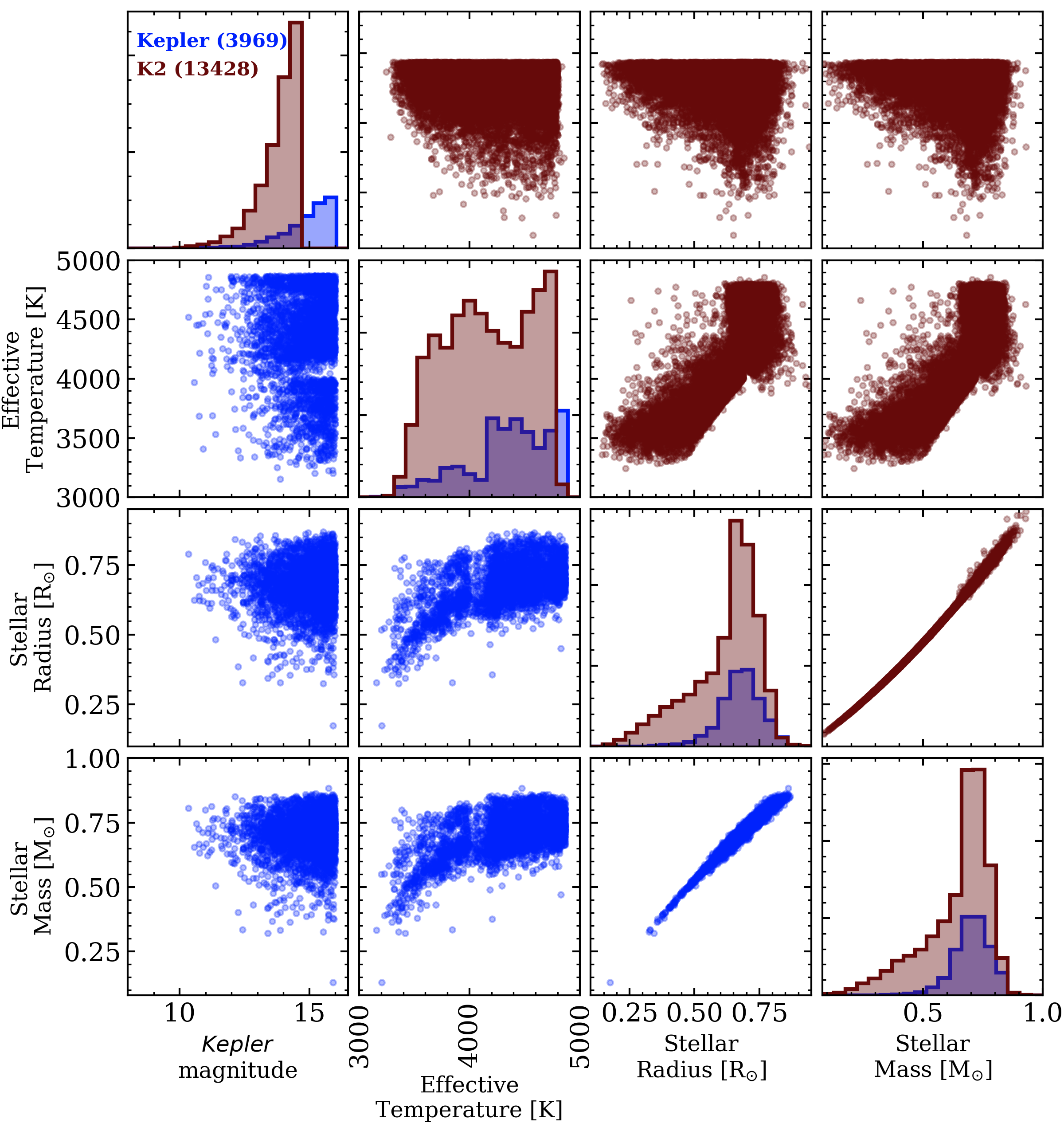}
  \caption{Low mass dwarf stellar samples from \kepler{} and \ktwo{.} Distributions of \kepler{} magnitudes,
    effective temperatures, stellar radii, and stellar masses for stars in our final stellar sample from either
    \kepler{} (\emph{blue histogram and markers}) or \ktwo{} (\emph{red histogram and markers}).}
  \label{fig:stars}
\end{figure}

\noindent We also only consider \kepler{} stars for which reliable completeness products from DR25
are available (see Sect.~\ref{sect:kepsens}). Based on these criteria, we retrieve 3965 low mass \kepler{}
stars whose stellar parameters are depicted in \autoref{fig:stars}.
In our \kepler{} sample, the \kepler{} magnitudes span $K_p \in [10.35, 16.00]$ with a median value of 15.16,
effective temperatures span \teff{} $\in [3154, 4870]$ K with a median value of 4394 K,
stellar radii span $R_s \in [0.17, 0.87]$ R$_{\odot}$ with a median value of 0.68 R$_{\odot}$, and
stellar masses span $M_s \in [0.13, 0.88]$ M$_{\odot}$ with a median value of 0.70 M$_{\odot}$.
Our final \kepler{} sample boasts a median fractional $R_s$ uncertainty of $\sim 6.7$\% which is $\sim 4-5$
times smaller than the typical $R_s$ uncertainty reported in the KSPC. The median fractional uncertainty on
$M_s$ is $\sim 5.5$\%.

\subsection{K2 stellar sample}
We first retrieved the list of probable low mass dwarf stars observed in any \ktwo{} campaign by querying
MAST\footnote{Mikulski Archive for Space Telescopes, \url{https://archive.stsci.edu/k2/}.}. Our initial
search was restricted to \ktwo{} stars with \teff{} $<4900$ K, \logg{} $>4$, and $R_s<1$ R$_{\odot}$. Note that these
criteria are not intended to represent the parameter ranges for low mass dwarf stars but are intended as
conservative conditions to encapsulate all such stars prior to their refinement using the \gaia{} DR2
data. From MAST we retrieve each star's Ecliptic Plane Input Catalog
(EPIC) numerical identifier, stellar photometry in the \kepler{} bandpass $K_p$ and 2MASS bands $JHK_s$, along
with measured values of \teff{,} \logg{,} [Fe/H], and $R_s$.

We proceed with refining the stellar parameters by cross-matching our initial \ktwo{} sample with \gaia{}  
DR2 using the \gaia{-}\ktwo{} data products from Megan Bedell\footnote{\url{https://gaia-kepler.fun/}}. Where
available, we retrieve reach star's celestial coordinates, stellar parallaxes $\varpi$, and \gaia{} photometry.
Measurements of $R_s$ then follow from the methodology of \cite{berger18} and outlined as follows.
The formalism of \cite{bailerjones18} is used to transform the assumed
Gaussian-distributed $\varpi$ PDFs into stellar distance PDFs which need not remain Gaussian.
Using the measured distances $d$ and celestial coordinates, we interpolate over the $E_{B-V}$ extinction maps using the
\texttt{mwdust} software \citep{bovy16} to derive both the $V$ and $K_s$-band extinction coefficients $A_V$ and
$A_{K_s}$. We then calculate each star's absolute $K_s$-band magnitude $M_{K_s} = K_s - \mu - A_{K_s}$ where
the distance modulus is $\mu = 5\log_{10}(d/10\text{ pc})$.

For the earliest stars in our sample ($M_{K_s}\leq 4.6$) , for which the bolometric corrections
are still reliable, we interpolate the MIST bolometric correction grids \citep{choi16} over \teff{,} \logg{,} [Fe/H],
and $A_V$ to derive the $K_s$-band bolometric corrections $BC_{K_s}$. We then compute the
absolute bolometric magnitudes $M_{\text{bol}}=M_{K_s} + BC_{K_s}$ and consequently the bolometric stellar
luminosities as 

\begin{equation}
  L_{\text{bol}} = L_0 \cdot 10^{-0.4 M_{\text{bol}}},
\end{equation}

\noindent where $L_0 = 3.0128 \times 10^{28}$ W \citep{mamajek15}. The refined $R_s$ values
are then calculated using the Stefan-Boltzmann law given $L_{\text{bol}}$ and \teff{} with measurement uncertainties
propagated throughout.

For the remaining late type stars with $M_{K_s}>4.6$, we revert to the empirically-derived radius-luminosity relation
from \cite{mann15} to calculate the M dwarf stellar radii. \cite{mann15} fit a second-order polynomial to $R_s$ as
a function of $M_{K_s}$ which has a characteristic dispersion in the fractional radius uncertainty of 2.89\%. To quantify
the final $R_s$ uncertainty we sample $M_{K_s}$ from its posterior PDF and transform each $M_{K_s}$ draw to an $R_s$ value
using the aforementioned radius-luminosity relation. To each star's derived $R_s$ PDF, we add an additional
dispersion term, in quadrature, whose fractional uncertainty is 2.89\%. Stellar masses within our \ktwo{} sample are derived
identically to the method applied to the \kepler{} sample using the \cite{boyajian12} stellar
mass-radius relation (see Sect.~\ref{sect:kep}). 

We define our final \ktwo{} stellar sample of low mass dwarf stars similarly to our definition of the \kepler{} sample.
Explicitly, we focus on stars that obey the following criteria:

\begin{enumerate}
\item $K_p < 14.7$,
\item $T_{\text{eff}} - \sigma_{T_{\text{eff}}} \leq 4700$ K,
\item $R_s - \sigma_{R_s} \leq 0.8$ R$_{\odot}$,
\item $M_s - \sigma_{M_s} \leq 0.8$ M$_{\odot}$, and
\item $R_s < R_{s,\text{max}}$.
\end{enumerate}

\noindent Because our \ktwo{} sample lacks any evolutionary flags, we adopt the following ad hoc upper limit on $R_s$
from \cite{fulton17} that aims to reject evolved stars:

\begin{equation}
  R_{s,\text{max}} = \text{R}_{\odot} \cdot 10^{0.00025(T_{\text{eff}}/\text{K}-5500)+0.2}.
\end{equation}

\noindent Based on these criteria, we retrieve 13428 low mass \ktwo{} stars whose
stellar parameters are also depicted in \autoref{fig:stars}.
In our \ktwo{} sample, the \kepler{} magnitudes span $K_p \in [8.47, 14.68]$ with a median value of 14.04,
effective temperatures span \teff{} $\in [3246, 4856]$ K with a median value of 4017 K,
stellar radii span $R_s \in [0.14, 0.94]$ R$_{\odot}$ with a median value of 0.70 R$_{\odot}$, and
stellar masses span $M_s \in [0.09, 0.93]$ M$_{\odot}$ with a median value of 0.69 M$_{\odot}$.
The stars in this sample exhibit a median fractional $R_s$ uncertainty of $\sim 3.5$\% which is $\sim 2$
times smaller than the typical $R_s$ uncertainty obtained for stars in our \kepler{} sample.
The median fractional uncertainty on $M_s$ is $\sim 3.9$\%.

Our full stellar sample therefore contains 17393 stars.
Each of the \kepler{} and \ktwo{} stellar samples are dominated by mid-to-late K dwarfs
with temperatures and radii $\gtrsim 3800$ K and $\gtrsim 0.6$ R$_{\odot}$ respectively. This
fact will have important implications on our ability to precisely measure the planet occurrence
rate around the lowest mass stars in our sample.

\section{Population of small close-in planets around low mass dwarf stars} \label{sect:planets}
Here we define the population of small close-in planets orbiting stars contained in our stellar sample.
Our initial sample of transiting planets from either \kepler{} or \ktwo{} were retrieved from the
NASA Exoplanet Archive \citep{akeson13} on June 15, 2019. Only confirmed
planets---based on their Exoplanet Archive dispositions---with orbital periods
$P\in [0.5,100]$ days are included. By considering confirmed
planets only, we naturally focus on the true empirical population of small close-in planets
without being contaminated by astrophysical false positives that may plague the planet
candidate sample that is excluded from our initial sample.

The refined stellar radii derived in Sect.~\ref{sect:stars} enable us to derive more precise
planetary radii.
We refine the planetary radii $r_p$ by retrieving point estimates of each planet's scaled planetary radius
$r_p/R_s$, which often includes a median value 
accompanied by the 16$^{\text{th}}$ and 84$^{\text{th}}$ percentiles of the $r_p/R_s$ posterior PDF.
In cases for which the $r_p/R_s$ uncertainties
are symmetric, we assume that the $r_p/R_s$ posterior PDF is Gaussian. For planets with asymmetric reported
uncertainties, we fit the $r_p/R_s$ percentiles with a skew-normal distribution using the
\texttt{scipy.skewnorm python} class. We fit for the location, scale, and shape parameters of the
distribution such that its resulting percentiles are consistent with
the $r_p/R_s$ point estimates reported for each planet. The refined planetary radii are then derived by
sampling the product of the $r_p/R_s$ and $R_s$ distributions. Our planet sample is then updated by
only considering planets whose radii are consistent with $r_p = 0.5-4$ R$_{\oplus}$.

From the distributions of $R_s$, \teff{,} $M_s$, and $P$ or each planet and host star, we derive the planets'
semimajor axes $a$ and insolations $F$ via

\begin{equation}
  \frac{F}{F_{\oplus}} = \left( \frac{R_s}{R_{\odot}} \right)^2  \left( \frac{T_{\text{eff}}}{5777 \text{ K}} \right)^4 \left( \frac{a}{1 \text{ AU}} \right)^{-2}.
\end{equation}

\begin{figure*}
  \centering
  \includegraphics[width=\hsize]{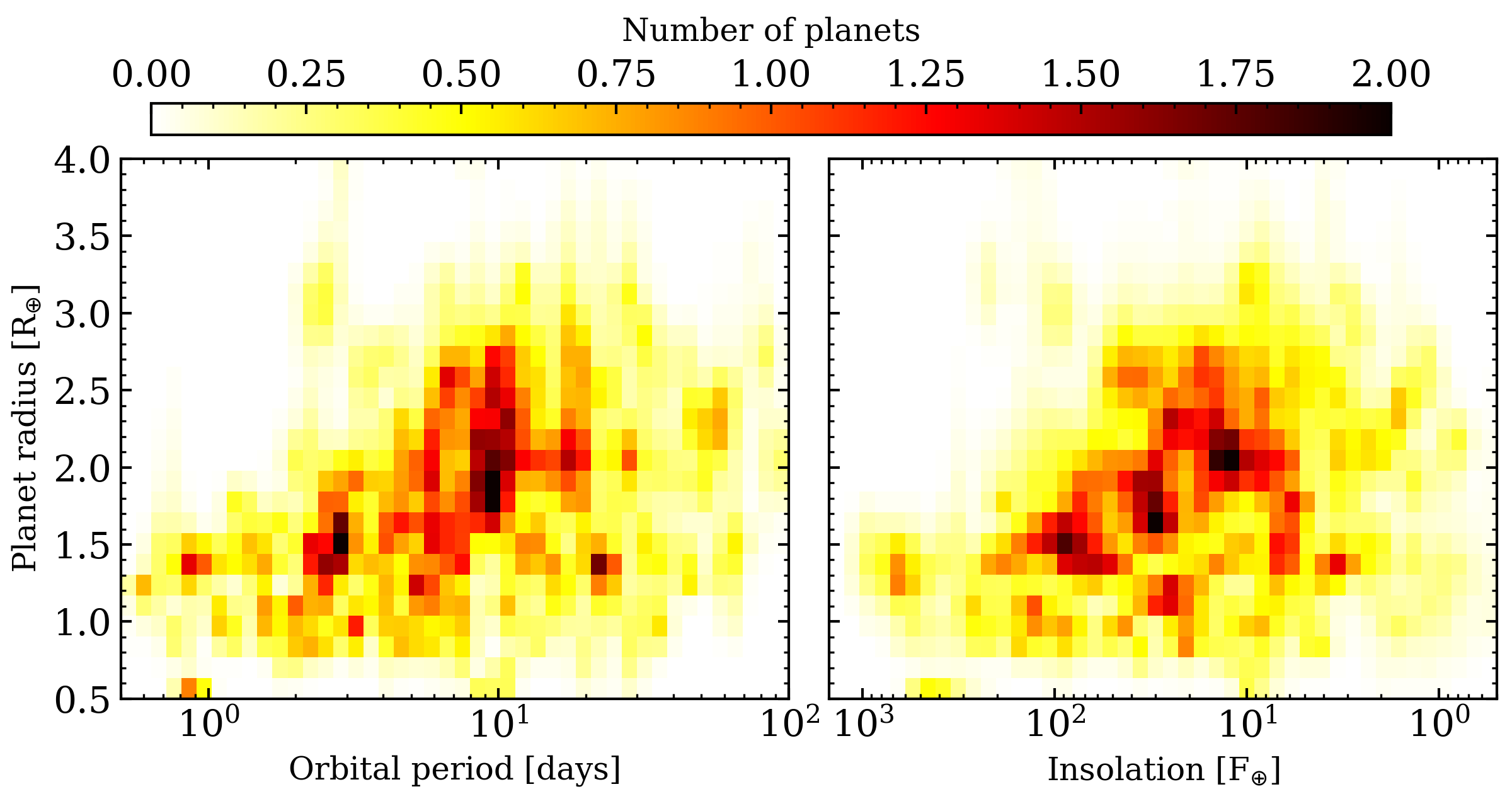}%
  \hspace{-\hsize}%
  \begin{ocg}{fig:offNd}{fig:offNd}{0}%
  \end{ocg}%
  \begin{ocg}{fig:onNd}{fig:onNd}{1}%
  \includegraphics[width=\hsize]{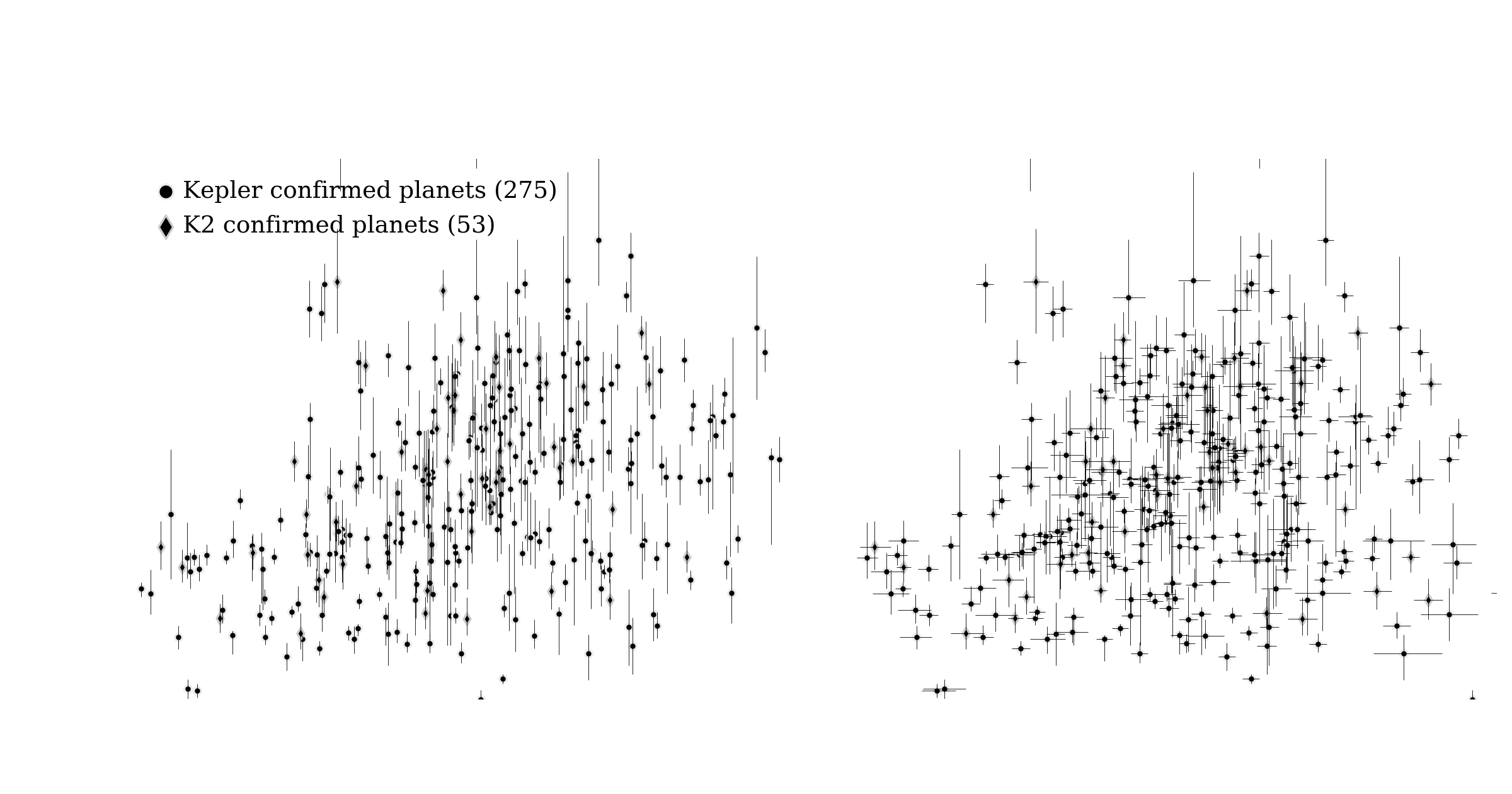}%
  \end{ocg}
  \hspace{-\hsize}%
  \caption{(\emph{Interactive figure}) Empirical population of confirmed close-in planets around low mass stars.
    The distribution of 275 and 53 confirmed planets from \kepler{}
    and \ktwo{}
    \ToggleLayer{fig:onNd,fig:offNd}{\protect\cdbox{(\emph{circles and diamonds} respectively)}}
    as a function of orbital period, insolation, and planet radius. The two-dimensional maps are
    Monte-Carlo sampled from the measurement uncertainties on the planetary radii and insolations while 
    the fractional uncertainties on the orbital periods are inflated to
    20\%.} 
  \label{fig:Ndet}
\end{figure*}

Our final sample of confirmed small close-in planets 
contains 275 \kepler{} and 53 \ktwo{} planets respectively. Their respective median fractional
radius uncertainties are 7.1\% and 9.0\%. Properties of the 328 confirmed planets in our sample
are reported in \autoref{table:planetsKep} and \autoref{table:planetsK2}. Our planet sample is
depicted in \autoref{fig:Ndet} as two-dimensional maps of the number of planet detections in
the $P-r_p$ and $F-r_p$ spaces. The two-dimensional histogram maps are computed by
Monte-Carlo sampling planets from their $F$ and $r_p$
measurement uncertainties and with a fractional precision on $P$ inflated to 20\%.

\input{Keplerplanettable}
\input{K2planettable}

The empirical planet population in \autoref{fig:Ndet} exhibits many recognizable features in the distribution of
planets orbiting low mass stars \citep[e.g.][]{morton14,dressing15a,gaidos16}. Namely, the dearth of planets with
$r_p\gtrsim 2$ R$_{\oplus}$ at short orbital periods known as the Neptunian desert \citep{lundkvist16,mazeh16},
the prominence of super-Earth and sub-Neptune-sized planets with orbital periods of a few to tens of days,
and the lack of small planets at long orbital periods ($P\gtrsim 40$ days) due to the poor transit detection
completeness in this region. Any features resembling the radius valley are not prominent in the empirical planet
distribution. Assuming that the radius valley around Sun-like stars persists in some form around low
mass stars, the fact that a distinct valley is not visible in the empirical planet population
highlights the importance of measuring valley
features from the completeness-corrected planet distribution. Alternatively, the valley---close to the expected rocky to non-rocky
transition of $\sim 1.5-1.8$ R$_{\oplus}$ \citep{weiss14}---may not be entirely void of planets. Indeed there exists a
significant subset of confirmed planets between 1.5-1.8 R$_{\oplus}$ with periods out to $\sim 12$ days indicating that
the mechanism responsible for producing the radius valley might not be as efficient as it is when operating on planetary
systems around Sun-like stars.

\section{Transiting planet detection completeness}  \label{sect:completeness}
Derivation of the planet occurrence rate requires the empirical distribution of planet detections to be corrected
for imperfect survey completeness. The completeness correction is treated separately for each subset
of planets from \kepler{} or \ktwo{} in the following subsections. Each set of corrections is designed to account
for detection biases arising from the imperfect transit detection sensitivity and for the 
geometric probability of a planetary transit to occur.

\subsection{Kepler sensitivity} \label{sect:kepsens}
The derivation of the \kepler{} planet detection sensitivity follows from the methodology outlined in
\cite{christiansen16} and used by \cite{fulton17} to resolve the radius valley around FGK stars. Per-target
\kepler{} completeness products for DR25 and the SOC 9.3 version of the \kepler{} pipeline
\citep{jenkins10} are available
for all of the stars in our \kepler{} sample \citep{burke15,burke17}. Detection sensitivities
(or efficiencies) were calculated via transiting planetary signal injections at the pixel level
which are subsequently processed by the \kepler{} pipeline Transiting Planet Search (TPS) module from
which the detection sensitivity is computed as the fraction of
injected signals that are successfully recovered by the pipeline as a function of the Multi-event statistic
\citep[MES;][]{christiansen15,christiansen17}.

The MES represents the level of significance of a repeating transit signal at a specified transit duration ranging
from 1.5-15 hours. Following \cite{petigura18}, we adopt an alternative diagnostic for the transit signal significance
in the form of the transit signal-to-noise ratio

\begin{equation}
  \text{S/N} = \frac{Z}{\text{CDPP}_D} \sqrt{n_{\text{transits}}(\mathbf{t},P,T_0)}  \label{eq:snr}
\end{equation}

\noindent where $Z=(r_p/R_s)^2$ is the transit depth assuming a non-grazing transit (i.e. $b\lesssim 0.9$),
CDPP$_D$ is the Combined Differential Photometric
Precision on the timescale of the transit duration $D$ \citep{koch10}, and $n_{\text{transits}}$ is the number of
observed transits given the target's data span and duty cycle
of the observations $\mathbf{t}$, the planet's orbital period $P$, and its time of mid-transit $T_0$.

\begin{figure}
  \centering
  \includegraphics[width=0.98\hsize]{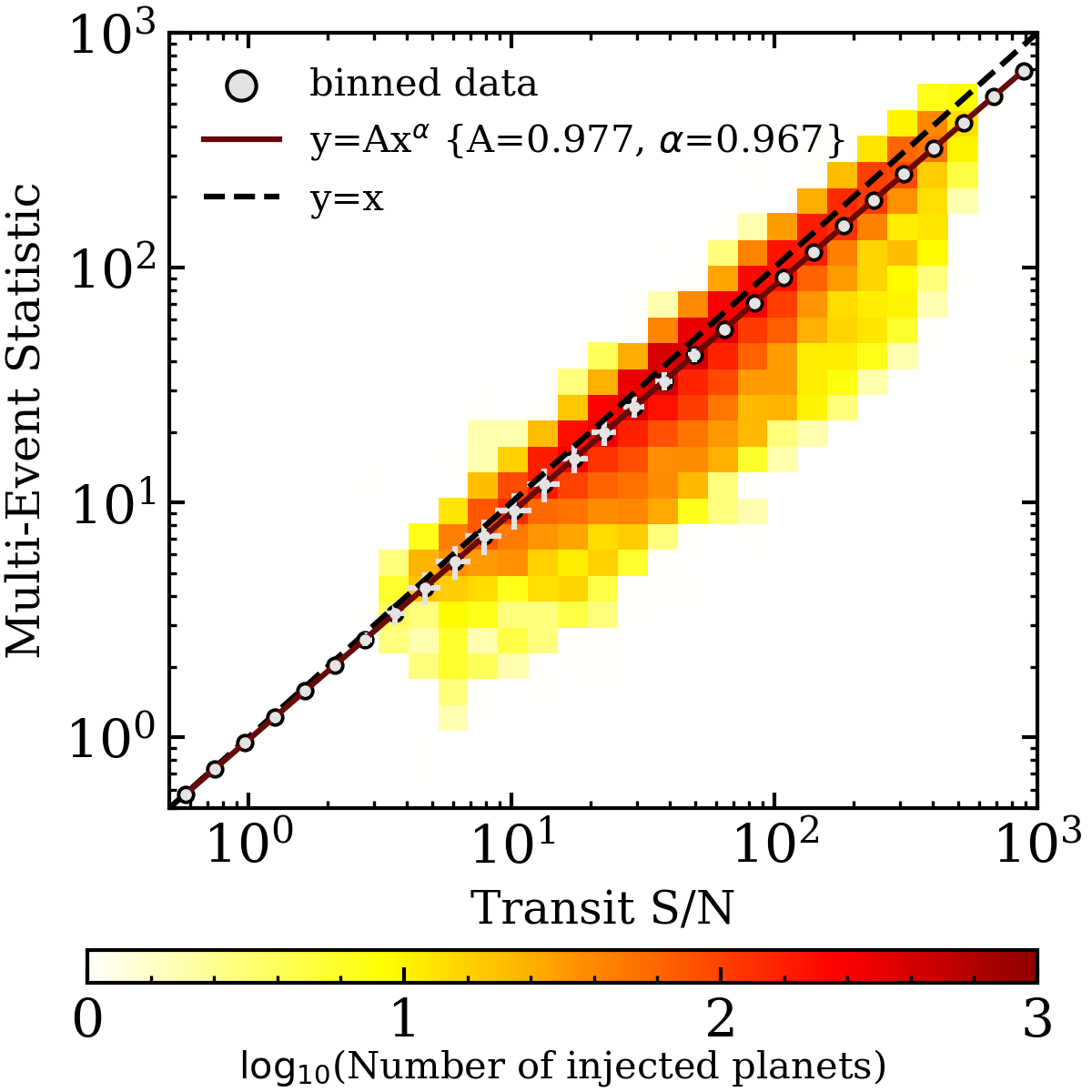}
  \caption{Correlation between the \kepler{} multi-event statistic and transit S/N. The mapping between the MES and
    S/N based on the synthetic planetary signals injected into the \kepler{} pipeline \citep{christiansen15}.
    The number-weighted powerlaw fit (\emph{solid line}) to the correlation differs slightly from a one-to-one relation
    (\emph{dashed line}) with a marginally lower amplitude and shallower slope.}  
  \label{fig:messnr}
\end{figure}

\begin{figure}
  \centering
  \includegraphics[width=0.98\hsize]{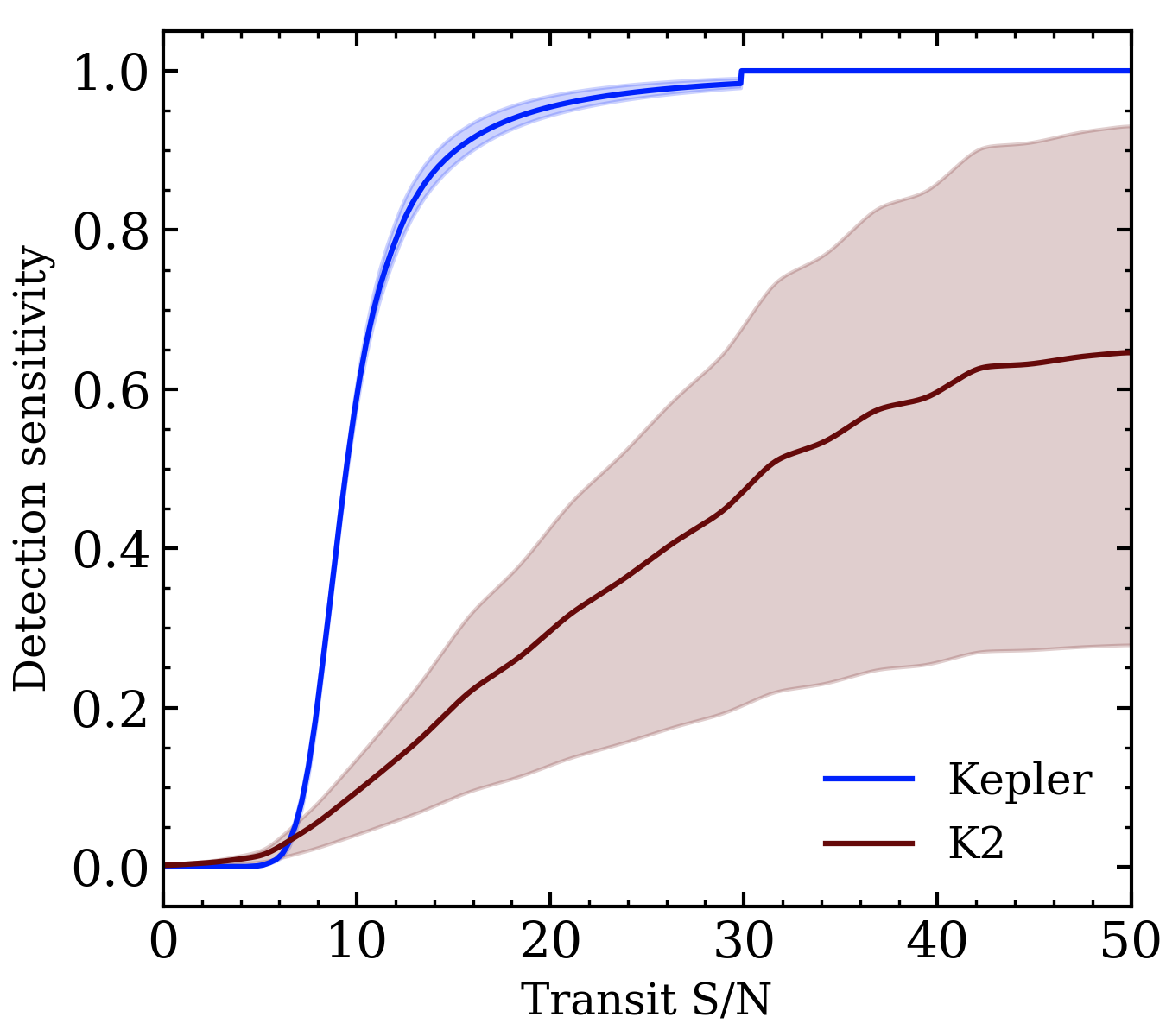}
  \caption{Average detection sensitivity for \kepler{} and \ktwo{.} The \emph{solid curves} represent the
    average transiting planet detection sensitivity for the \kepler{} and \ktwo{} stars in our sample as
    a function of the transit S/N (Eq.~\ref{eq:snr}). The shaded regions mark the 16$^{th}$ and 84$^{th}$
    percentiles of the measured detection sensitivities.} 
  \label{fig:senscurves}
\end{figure}

To compute the \kepler{} detection sensitivity as a function of S/N, we first
derive the mapping between the MES and the transit S/N using the data from \cite{christiansen15} who
derived the detection sensitivity of the \kepler{} pipeline from one year of data. 
The parameters of the injected planets are provided along with their corresponding MES and CDPP at each value of
$D$ considered. For each injected planet we interpolate its MES and CDPP values to $D$ and calculate
the transit S/N using Eq.~\ref{eq:snr}. The mapping between MES and S/N is shown in \autoref{fig:messnr}
for the full set of injected planets whose transit S/N values span 2.7-4843. Given the large number of injected planetary
signals ($>10^4$), we fit the number-weighted S/N to MES mapping using the \texttt{scipy.curve\_fit} non-linear least
squares algorithm with a powerlaw function of the form $\text{MES} = A\cdot \text{S/N}^{\alpha}$. We find a best-fit
amplitude and powerlaw index of $A=0.977$ and $\alpha=0.967$ respectively with negligible uncertainties.
This relation is used to map the transit S/N to MES
which is then mapped to the detection sensitivity. The average \kepler{} detection sensitivity curve as a function of
transit S/N, along with the $16^{\text{th}}$ and $84^{\text{th}}$ percentiles for the stars in our \kepler{} sample
are shown in \autoref{fig:senscurves}.

\subsection{K2 sensitivity} \label{sect:k2sens}
Unlike the primary \kepler{} mission, the \ktwo{} data products do not feature detailed completeness and reliability
products. To derive the detection sensitivity among the \ktwo{} stars in our sample we employ the transit detection
pipeline \texttt{ORION} \citep{cloutier19b}.

The failure of the second reaction wheel on board the \kepler{} spacecraft in 2013 prevented the observatory from
maintaining the fine pointing accuracy required to continue to obtain ultra precise photometry. The re-purposed \ktwo{}
mission exploited the solar wind pressure by enabling the observatory to continue pointing along the ecliptic
plane with realignments via thruster firings \citep{howell14}.
\texttt{ORION} does not feature a specialized module to correct for the temporally correlated pointing corrections.
This requires that pointing-corrected light curves be used as input. We adopt
the \texttt{EVEREST}-reduced \ktwo{} light curves which use a pixel level decorrelation to remove systematics from
the spacecraft's variable pointing \citep{luger16,luger18}. We favor the \texttt{EVEREST} \ktwo{} light curves over
light curves produced by analogous pipelines (e.g. \texttt{K2SFF}; \citealt{vanderburg14}, \texttt{K2SC};
\citealt{aigrain15,aigrain16}) due to its demonstrated performance
in obtaining improved photometric precision by a factor of $\sim 20-50$\% \citep{luger16}.

We quantify the \ktwo{} detection sensitivity using \texttt{ORION} by first retrieving the \texttt{EVEREST} light
curve from MAST for each star in our sample. We only consider light curves from individual campaigns. As
\texttt{ORION} input we supply the time sampling $\mathbf{t}$ in BJD, the corrected flux, and flux uncertainties
in e$^-$/second, from the \texttt{EVEREST} keywords \texttt{TIME}, \texttt{FCOR}, and \texttt{FRAW\_ERR}.
The duty cycle is derived by restricting to light curve measurements for which the \texttt{QUALITY} flag is zero.
In light curves with known signals from planets or planet candidates, those signals are modeled and removed from the
light curve based on their reported transit parameters and using the \texttt{batman} \citep{kreidberg15} implementation
of the \cite{mandel02} transit model. We then inject transiting planetary signals directly into the light curve
by sampling planets from the linear transit S/N grid $\mathcal{U}(0,50)$. The per-system
multiplicity is drawn
from the cumulative occurrence rate of small planets out to 200 days around
mid-K to early M dwarfs from \kepler{} \citep[$2.5 \pm 0.2$;][]{dressing15a}.
Each planet's time of mid-transit $T_0$ is
drawn from $\mathcal{U}(\text{min}(\mathbf{t}),\text{max}(\mathbf{t}))$.
In a given light curve, with fixed $\mathbf{t}$ and CDPP$_D$, for a star
whose $R_s$ and $M_s$ values are fixed to their maximum likelihood values, we
draw each planet's logarithmic orbital period from $\mathcal{U}(\log_{10}(0.5\text{ days}),\log_{10}(200\text{ days}))$
which allows us to compute the number of
transits that occur within $\mathbf{t}$. Note that some injected planets
will exhibit $n_{\text{transits}}=0$ due to the limited \ktwo{} baselines of typically $\sim 80$ days.
The drawn orbital period also uniquely determines the planet's radius corresponding to its drawn value of the S/N.
To ensure dynamical stability in multi-planet
systems, we compute the maximum likelihood planet mass from the probabilistic mass-radius relation \texttt{forecaster}
\citep{chen17} and analytically assess the Lagrange stability of each neighboring planet pair assuming circular
orbits \citep{barnes06}. Each planet's scaled semimajor axis $a/R_s$ and scaled radius $r_p/R_s$ follow from
their sampled radius $r_p$ and the stellar parameters $R_s$ and $M_s$. We sample impact parameters $b$ from
$\mathcal{U}(0,0.9)$ to compute the orbital inclinations. Furthermore, we adopt fixed quadratic limb darkening
coefficients by interpolating the \kepler{} bandpass coefficient grid along \teff{,} \logg{,} and [Fe/H], assuming
solar metallicity when [Fe/H] measurements are absent \citep{claret12}.
These parameters are used to compute transit models in the absence of any transit timing variations.
Transit signals are then injected into the cleaned \ktwo{} light curves and fed to
\texttt{ORION} to conduct a blind search for transiting signals.

The detection sensitivity as a function of S/N for each \ktwo{} star is computed by considering
$10^2$ injected planetary systems per star and computing the recovery fraction of injected planets with
$P \leq 100$ days. The average \ktwo{} detection sensitivity curve, along with the
$16^{\text{th}}$ and $84^{\text{th}}$ percentiles, are also included in \autoref{fig:senscurves}. The quality
of the pointing corrections within the \texttt{EVEREST} light curves can vary widely within our sample such
that there is considerably more variance in the \ktwo{} detection sensitivity relative to \kepler{.} Furthermore,
the average detection sensitivity is significantly reduced compared to \kepler{.} 
The reduced sensitivity is due in-part to the imperfect corrections of the reduced pointing accuracy and
to the limited time baseline of $\sim 80$ days in a typical \ktwo{} light curve compared to \kepler{.} 
Furthermore, \texttt{ORION} was originally developed for use on the 2-minute cadence data
from the \tess{} mission. Here
we have not attempted to optimize the performance of \texttt{ORION} on \ktwo{} light curves 
beyond slight modifications to the algorithm's performance hyperparameters that were made to ensure the
detection of 52/53 confirmed \ktwo{} planets. The planet K2-21c (EPIC 206011691.02, $P=15.5$ days)
remains undetected by \texttt{ORION} because of the algorithm's requirement to discard putative signals
that are commensurate with other high S/N signals in the light curve. The presence of K2-21b at $P=9.32$ days
is within 1\% of a 5:3 period ratio with K2-21c and thus prohibits the identification of the 15.5 day signal
as being independent and planetary.

\subsection{Two-dimensional sensitivity maps}
The sensitivity curves depicted in \autoref{fig:senscurves} enable us to extend the visualization of the
detection sensitivity to two dimensions. Explicitly, we consider the detection sensitivity
$s_{nij}$ for each star (indexed by $n$) and as a function of $P$ and $r_p$ which are indexed by $i$ and
$j$ respectively. Consideration of the sensitivity in $P-r_p$ space is needed to evaluate the
occurrence rates in that parameter space and ultimately for understanding the structure of
the radius valley around low mass stars due to the dependence of the efficiency of atmospheric loss on
both planet size and separation, regardless of the physical mechanism involved. 

\begin{figure*}
  \centering
  \includegraphics[width=0.98\hsize]{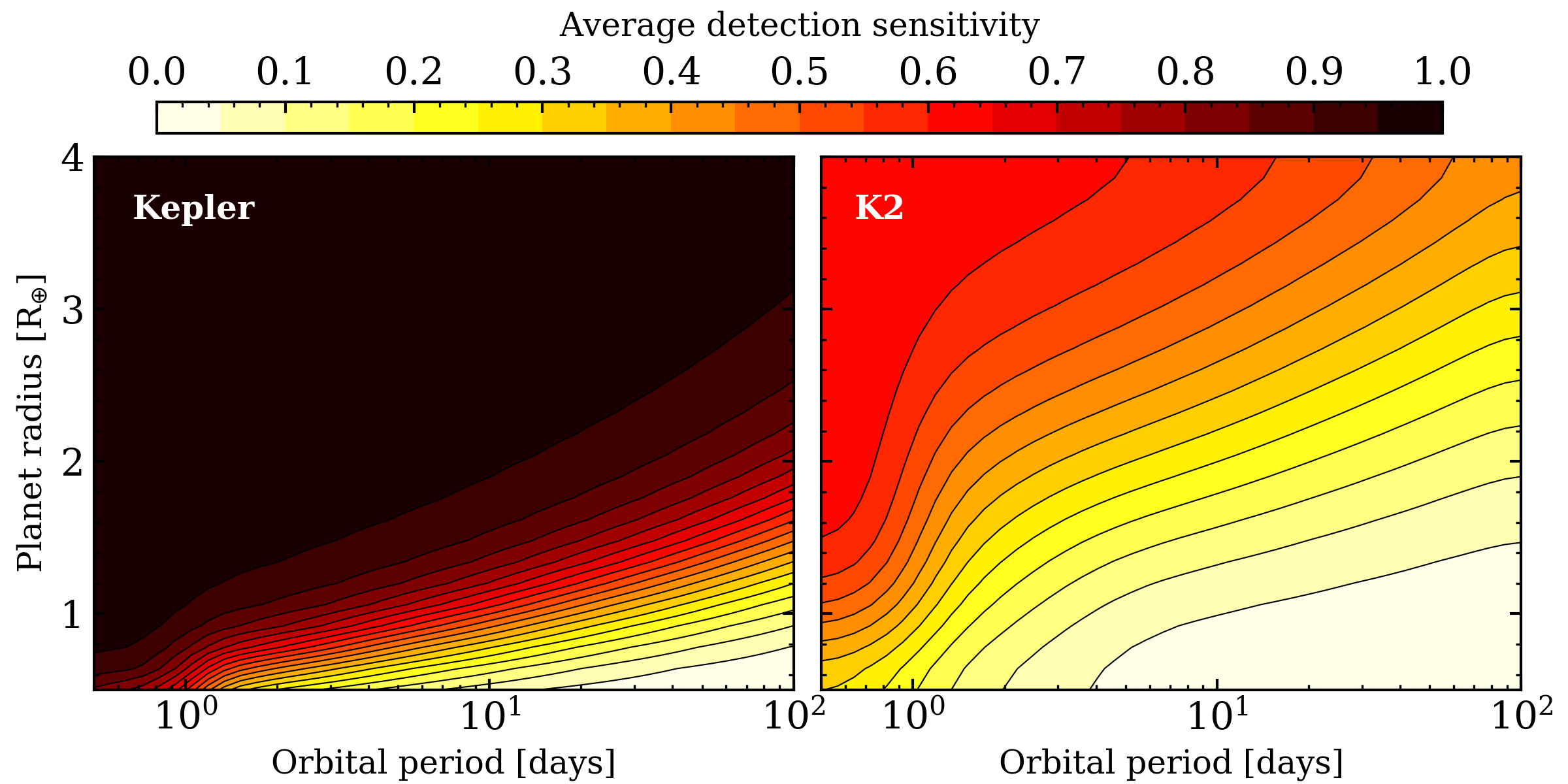}
  \caption{Average detection sensitivity versus orbital period and planetary radius.
    The detection sensitivity maps averaged over \kepler{} stars (\emph{left panel)} and over \ktwo{} stars
    (\emph{right panel}) from our sample of low mass dwarf stars.} 
  \label{fig:sensmap}
\end{figure*}

We consider orbital periods $P \in [0.5,100]$ days and planet radii $r_p \in [0.5,4]$ R$_{\oplus}$. At
each grid cell $nij$ we compute the average S/N within the cell and map that value to the detection sensitivity
using the data in \autoref{fig:senscurves}. The detection sensitivity maps for \kepler{} and \ktwo{,}
averaged over the index $n$, are shown in \autoref{fig:sensmap}.

\subsection{Survey completeness} \label{sect:comp}
Only transiting planets are detectable in transit surveys. To correct for the non-detection of otherwise
detectable but non-transiting planets we compute
the geometric transit probability for each star $n$ and at each grid cell $ij$ in the $P-r_p$ space to be

\begin{equation}
  p_{t,nij} = \frac{R_{s,n} + r_{p,j}}{a_{ni}}. \label{eq:ptransit}
\end{equation}

\noindent Note that we are only interested in the relative planet occurrence rate and therefore do not consider
constant scalar modifications to $p_{t,nij}$ from effects such as grazing transits or non-zero eccentricities 
\citep{barnes07b}.

\begin{figure*}
  \centering
  \includegraphics[width=0.98\hsize]{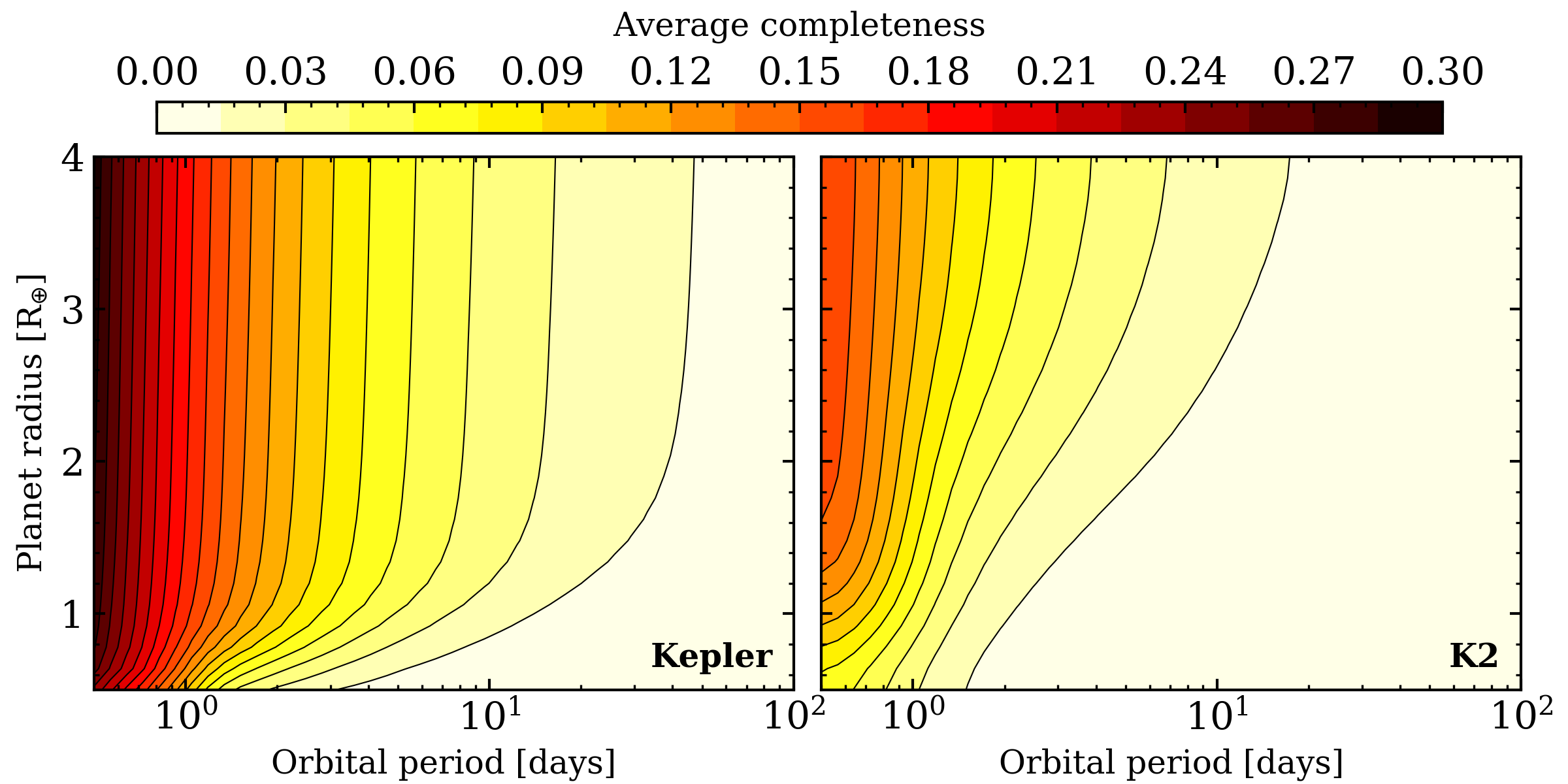}
  \caption{Average completeness versus orbital period and planetary radius.
    Maps of the product of the detection sensitivity and geometric transit probability averaged over \kepler{} stars
    (\emph{left panel)} and over \ktwo{} stars (\emph{right panel}) from our sample of low mass dwarf stars.} 
  \label{fig:compmap}
\end{figure*}

The product of each star's detection sensitivity with its geometric transit probability yield completeness
maps as a function of $P$ and $r_p$. The average completeness maps for our \kepler{} and \ktwo{} stars are
shown in \autoref{fig:compmap}.

\section{The occurrence rate of small close-in planets around low mass dwarf stars} \label{sect:occurrence}
\subsection{Occurrence rates versus orbital period and planet radius} \label{sect:fmap}
The detection and validation of planets from the \kepler{} and \ktwo{} missions enables the measurement of the
occurrence rate of planets given the completeness corrections derived in Sect.~\ref{sect:completeness}.
For the index $i$ representing a planet's orbital period and $j$ representing the planetary
radius, the probability of detecting an integer number of planets within that grid cell ($k_{ij}$) around
$N_s$ stars is given by the binomial likelihood function

\begin{equation}
  \mathcal{L}_{nij}(k_{ij}|N_s,P_{nij}) = \binom{N_s}{k_{ij}} \prod_{n=1}^{N_s} P_{nij}^{k_{ij}} (1-P_{nij})^{N_s-k_{ij}}
  \label{eq:lnL}
\end{equation}

\noindent where

\begin{equation}
  P_{nij} = s_{nij} \cdot p_{t,nij} \cdot f_{ij},
  \label{eq:prob}
\end{equation}

\noindent is the probability of detecting a planet in the $ij$ grid cell around the $n^{\text{th}}$ star.
This quantity is dependent on the detection sensitivity $s_{nij}$, the transit probability $p_{t,nij}$, and the 
intrinsic occurrence rate of planets in the grid cell $ij$ $f_{ij}$ which is assumed to be common to all
of the $N_s$ stars in the sample.
Recall that the number of planet detections $k_{ij}$ was depicted in \autoref{fig:Ndet} and
calculations of $s_{nij}$ and $p_{t,nij}$ produced the completeness maps shown in \autoref{fig:compmap}. 
Taken together, and noting from Bayes theorem that the posterior probability of $f_{ij}$ is 

\begin{equation}
  p(f_{ij}|N_s,s_{nij},p_{t,nij},k_{ij}) \propto \mathcal{L}_{nij}(k_{ij}|N_s,s_{nij},p_{t,nij},f_{ij}),
  \label{eq:feq}
\end{equation}
  
\noindent modulo the coefficient of proportionality which we set to unity, we are able to compute
the maximum a-posteriori (MAP) occurrence rate and uncertainty maps according to Eq.~\ref{eq:feq}.

Before proceeding, first recall that our planet sample contains $\sim 5$ times more confirmed planets
from \kepler{} than from \ktwo{}
(see \autoref{fig:Ndet}) despite our stellar sample containing $\sim 4.4$ times more stars observed by
\ktwo{} than by \kepler{} 
(see \autoref{fig:stars}). These factors compound to produce a lower planet occurrence rate measured
from \ktwo{} confirmed planets than with \kepler{} as the reduced \ktwo{} detection completeness
(see \autoref{fig:compmap}) is insufficient to account for the lower measured
planet occurrence rates. Explicitly, we measure cumulative occurrence rates of
$2.48\pm 0.32$ and $0.75\pm 0.11$ confirmed planets per star with $P\in [0.5,100]$ days
and $r_p\in [0.5,4]$ R$_{\oplus}$ from \kepler{} and \ktwo{} respectively.
The discrepancy arises from the disparate resources that have
been dedicated to the validation of planet candidates from \kepler{} and \ktwo{.} The result being that
the number of real planets within the full set of planet candidates from \ktwo{} is underestimated
by the number of planet candidates that have been reported as validated to date. We address this discrepancy
by scaling the cumulative occurrence rate measured by \ktwo{} to that of \kepler{.}
In this way, we inherently assume that the planet populations studied by each mission are inherently
equivalent despite existing around distinct stellar populations within the galaxy, albeit with similar
physical properties. In Sect.~\ref{sect:comparef} we will revisit the comparison of the planet occurrences
rates from \kepler{} and \ktwo{} following the inclusion of \ktwo{} planet candidates.

\begin{figure*}
  \centering
  \includegraphics[width=.9\hsize]{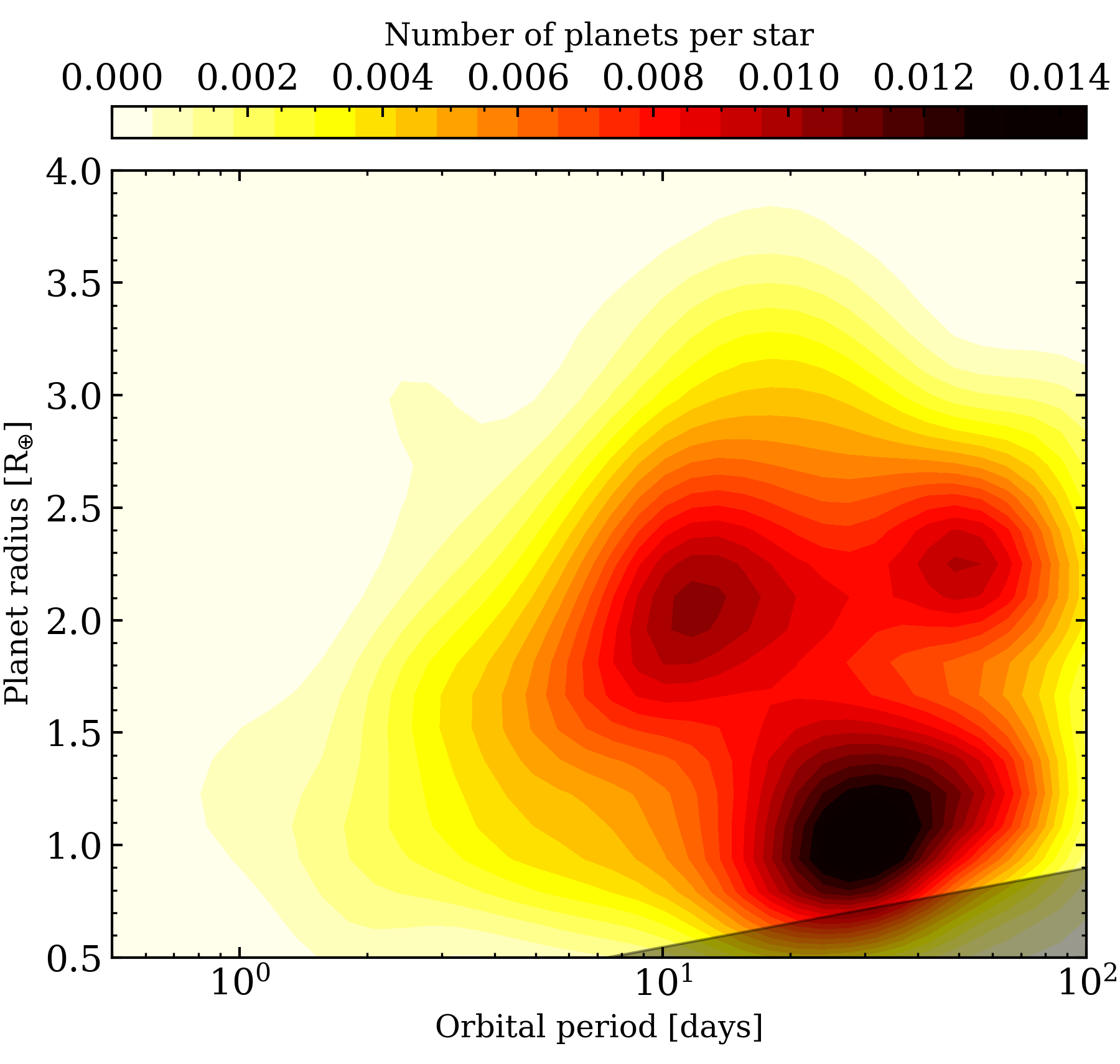}%
  \hspace{-.9\hsize}%
  \begin{ocg}{fig:off3}{fig:off3}{0}%
  \end{ocg}%
  \begin{ocg}{fig:on3}{fig:on3}{1}%
  \includegraphics[width=.9\hsize]{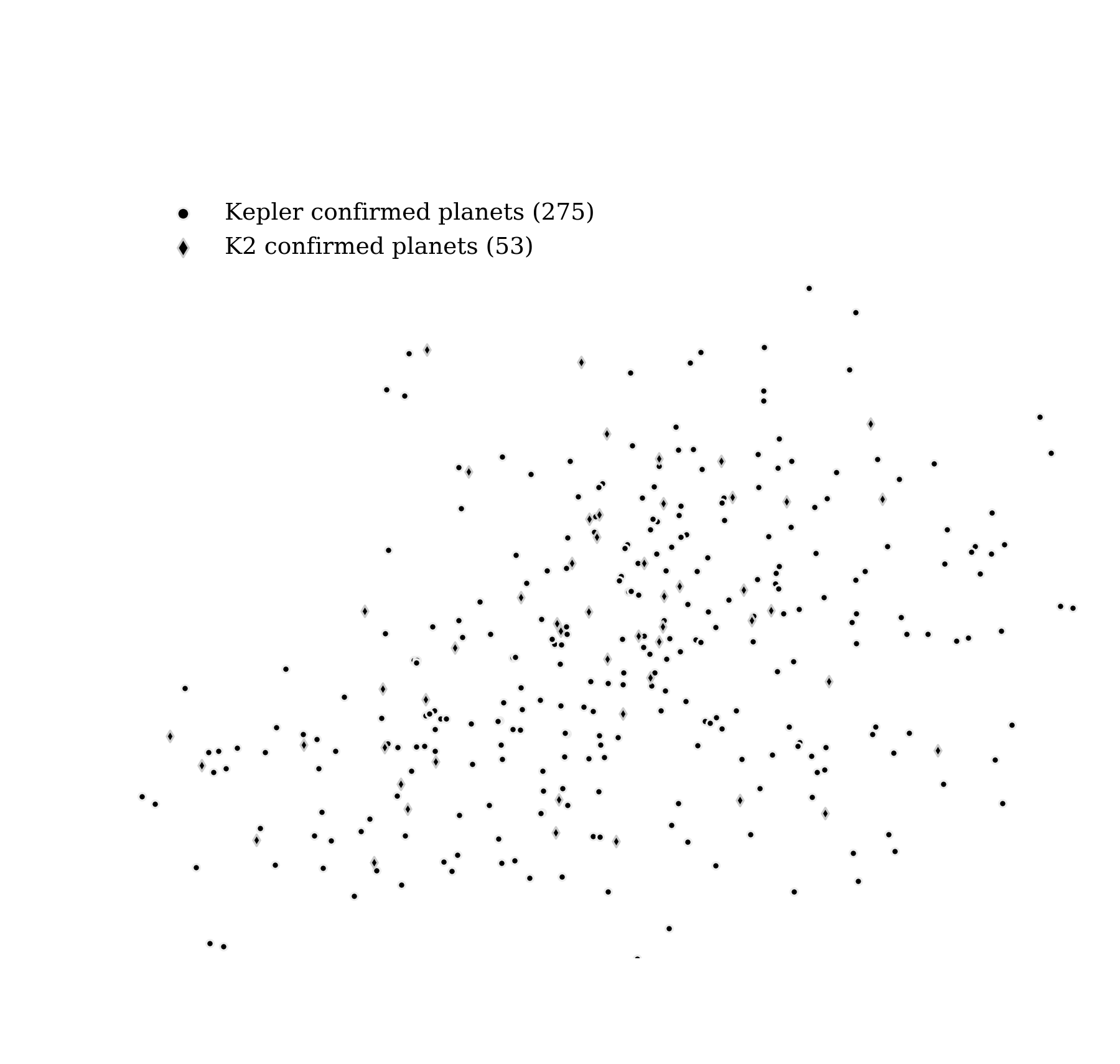}%
  \end{ocg}
  \hspace{-.9\hsize}%
  \begin{ocg}{fig:off1}{fig:off1}{0}%
  \end{ocg}%
  \begin{ocg}{fig:on1}{fig:on1}{1}%
  \includegraphics[width=.9\hsize]{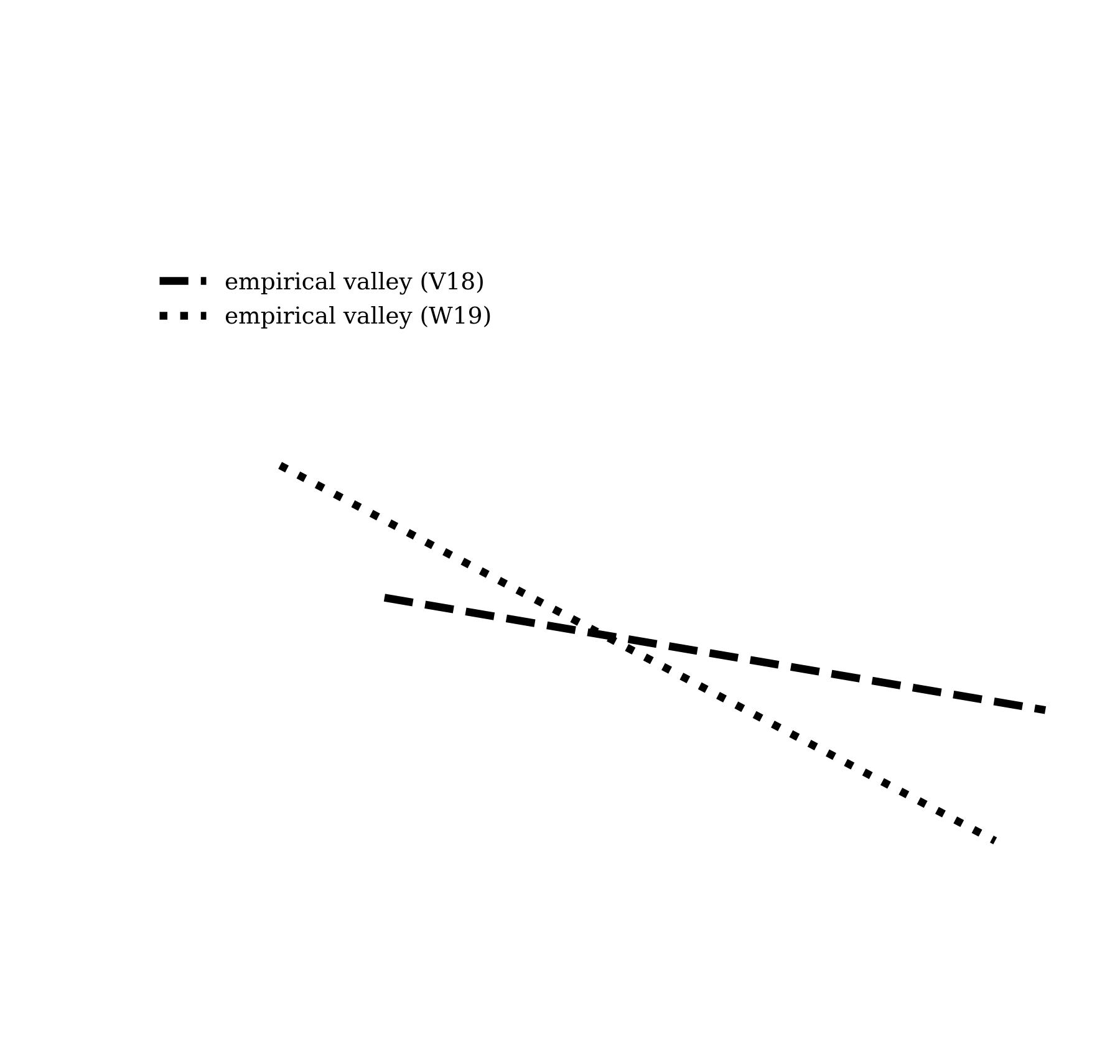}%
  \end{ocg}
  \hspace{-.9\hsize}%
  \begin{ocg}{fig:off4}{fig:off4}{0}%
  \end{ocg}%
  \begin{ocg}{fig:on4}{fig:on4}{1}%
  \includegraphics[width=.9\hsize]{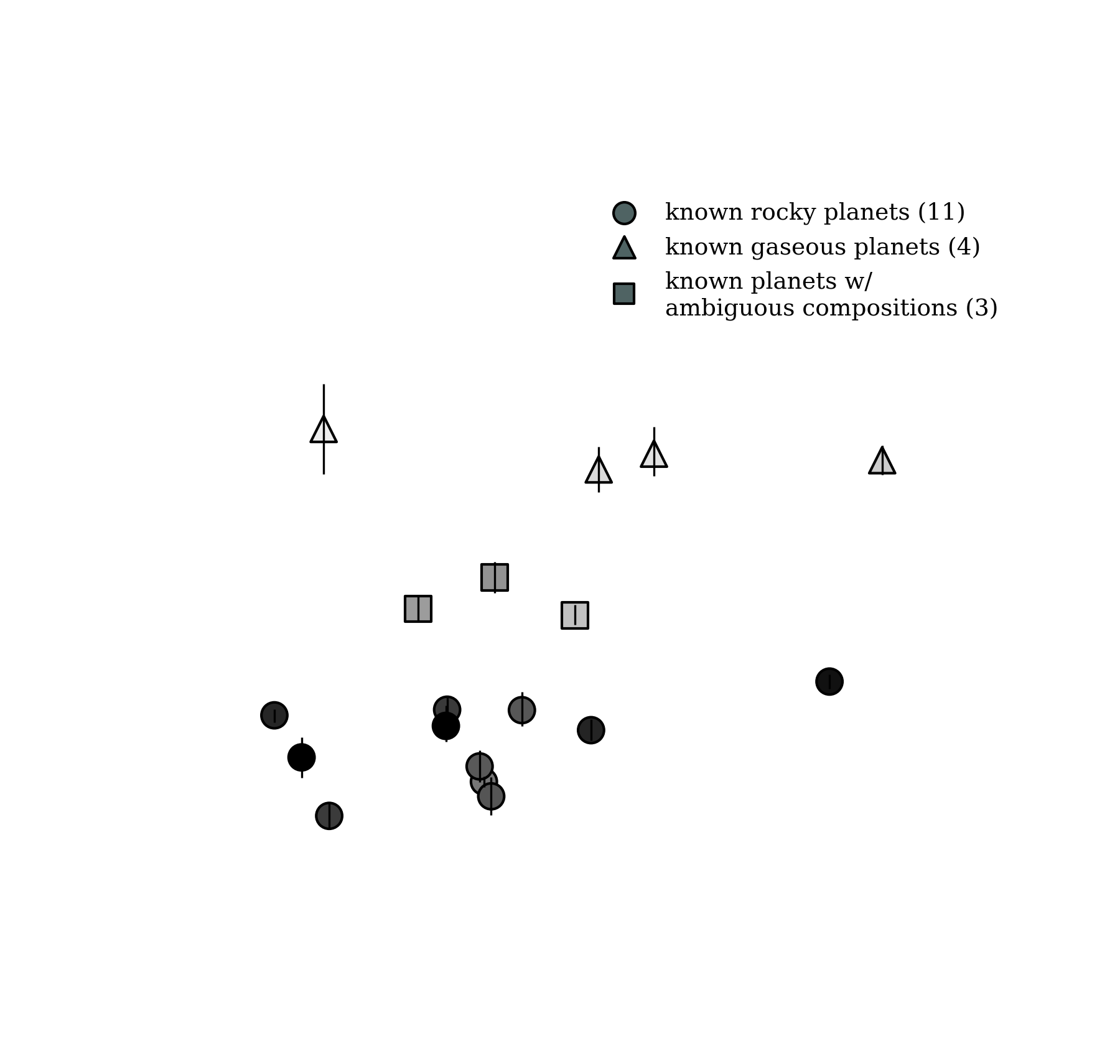}%
  \end{ocg}
  \hspace{-.9\hsize}%
  \caption{(\emph{Interactive figure}) Planet occurrence rate versus orbital period and planetary radius. The maximum a-posteriori occurrence
    rate map calculated from the population of confirmed planets from \kepler{} and
    \ktwo{} around low mass dwarf stars \ToggleLayer{fig:on3,fig:off3}{\protect\cdbox{(\emph{small circles and diamonds})}}.
    For comparison, overplotted are the \ToggleLayer{fig:on1,fig:off1}{\protect\cdbox{empirical locations}}
    of the radius valley around FGK stars 
    characterized via asteroseismology \citep[\emph{dashed line},][]{vaneylen18} and the approximate radius
    valley around mid-K to early-M dwarfs \citep[\emph{dotted line},][]{wu19}. Also overplotted are
    \ToggleLayer{fig:on4,fig:off4}{\protect\cdbox{planets}}
    with $\geq 3\sigma$ bulk density measurements from the literature that are classified as having either
    a rocky (\emph{circles}), a gaseous (\emph{triangles}), or an ambiguous (\emph{squares}) bulk
    composition. Marker colors are indicative of the MAP planet bulk densities. 
    The shaded region indicates where the \kepler{} sensitivity falls below 10\%.} 
  \label{fig:fmap}
\end{figure*}

The MAP $f_{ij}$ map is depicted in \autoref{fig:fmap}.
Here the existence of the radius valley around low mass stars is clearly
evident. Distinct peaks in the planet frequency are separated along the planetary radius axis and span
$\sim 0.9-1.4$ R$_{\oplus}$ and $\sim 1.9-2.3$ R$_{\oplus}$ respectively.
Note however that the lower limit on the former peak 
approaches the region in which the \kepler{} sensitivity falls below 10\% and the $f$ values become
unreliable.
The occurrence rates also highlight the relative dearth of planets larger than $\sim 3$ R$_{\oplus}$
including the Neptunian desert at short orbital periods \citep{lundkvist16,mazeh16}. The large scale
structure of the measured occurrence rates are also broadly consistent with previous investigations
of the planet population around low mass \kepler{} stars \citep{morton14,dressing15a,gaidos16} such
as the prominence of planets $\lesssim 2$ R$_{\oplus}$ with $P \sim 10-60$ days and the measured
cumulative occurrence rate of $2.48\pm 0.32$ planets per star with $P\in [0.5,100]$ days and
$r_p\in [0.5,4]$ R$_{\oplus}$.

The location and slope of the radius valley (i.e. $\text{d}r_p/\text{d}\log{P})$ appear broadly consistent
with the valley structure measured from the empirical planet population of FGK stars characterized via
asteroseismology \citep{vaneylen18}. \cite{wu19} also provided a visual approximation to the location of the
radius valley around stars with $M_s \in [0.5,0.76]$ M$_{\odot}$ in their \gaia{-}\kepler{} sample. However
we find the location of the terrestrial-sized planet peak to exist at longer $P \sim 30$ days compared its
location at $\sim 5$ days from \cite{wu19} (c.f. Figure 2). The discrepancy likely originates from differences
in the method of correcting for survey incompleteness.
Recall that in this study the detection sensitivity for \kepler{} stars is computed on a per star basis
given the unique completeness products from the \kepler{} pipeline whereas \cite{wu19} adopt the piecewise
completeness levels of 10, 50, or 90\% complete as a function of $P$ and $r_p$ from \cite{zhu18b}.

Also included in \autoref{fig:fmap} are planets
with $\geq 3\sigma$ bulk density measurements ($\rho_p$) from either precision radial velocities or
transit timing variations.
Planet parameters are retrieved from the NASA Exoplanet Archive for planets orbiting stars with
\teff{} $\in[2800,4700]$ K, whose orbital periods and radii span the domain considered in \autoref{fig:fmap},
and whose masses are inconsistent with zero (i.e. no mass upper limits).
The properties of the resulting 18 planets are listed in \autoref{table:rhop}. Based on the
planetary masses, radii, and compositional mass-radius relations from \cite{zeng13},
we define the following composition dispositions: \emph{rocky}
planets have bulk densities that are greater than or consistent with a purely rocky composition
(i.e. 100\% MgSiO$_3$) given the planet's size, \emph{gaseous} planets have bulk densities that are less
than and inconsistent with that of a pure water world (i.e. 100\% H$_2$O) given the planet's size, 
and all intermediate planets are flagged as having an \emph{ambiguous} bulk composition (i.e. not clearly
terrestrial-like or likely to be hosting a significant gaseous envelope).

\input{densitytable}

The retrieved planets in \autoref{fig:fmap} demonstrate clear compositional clustering with planet radius.
All rocky planets appear to be smaller than 1.8 R$_{\oplus}$ independently of orbital period.
Similarly, all four gaseous planets are larger than 2.6 R$_{\oplus}$ while the three remaining planets
with intermediate radii correspond to those with ambiguous bulk compositions. 
Thus we are justified in classifying the occurrence rate peak spanning $\sim 0.9-1.4$ R$_{\oplus}$
as representing rocky planets. The second peak between $\sim 1.9-2.3$ R$_{\oplus}$ hosts
planets with ambiguous bulk compositions but are clearly inconsistent with being rocky
such that we will refer to this feature as the non-rocky peak in what follows.

Lastly, we note that the radius valley as a function of $P$ is not completely void of planets.
This may allude to the efficiency of any gap clearing mechanism around low mass stars and is
discussed further in Sect.~\ref{sect:void}.

\vspace{1cm}

\subsection{Occurrence rates versus planet radius}
\begin{figure*}
  \centering
  \includegraphics[scale=.8]{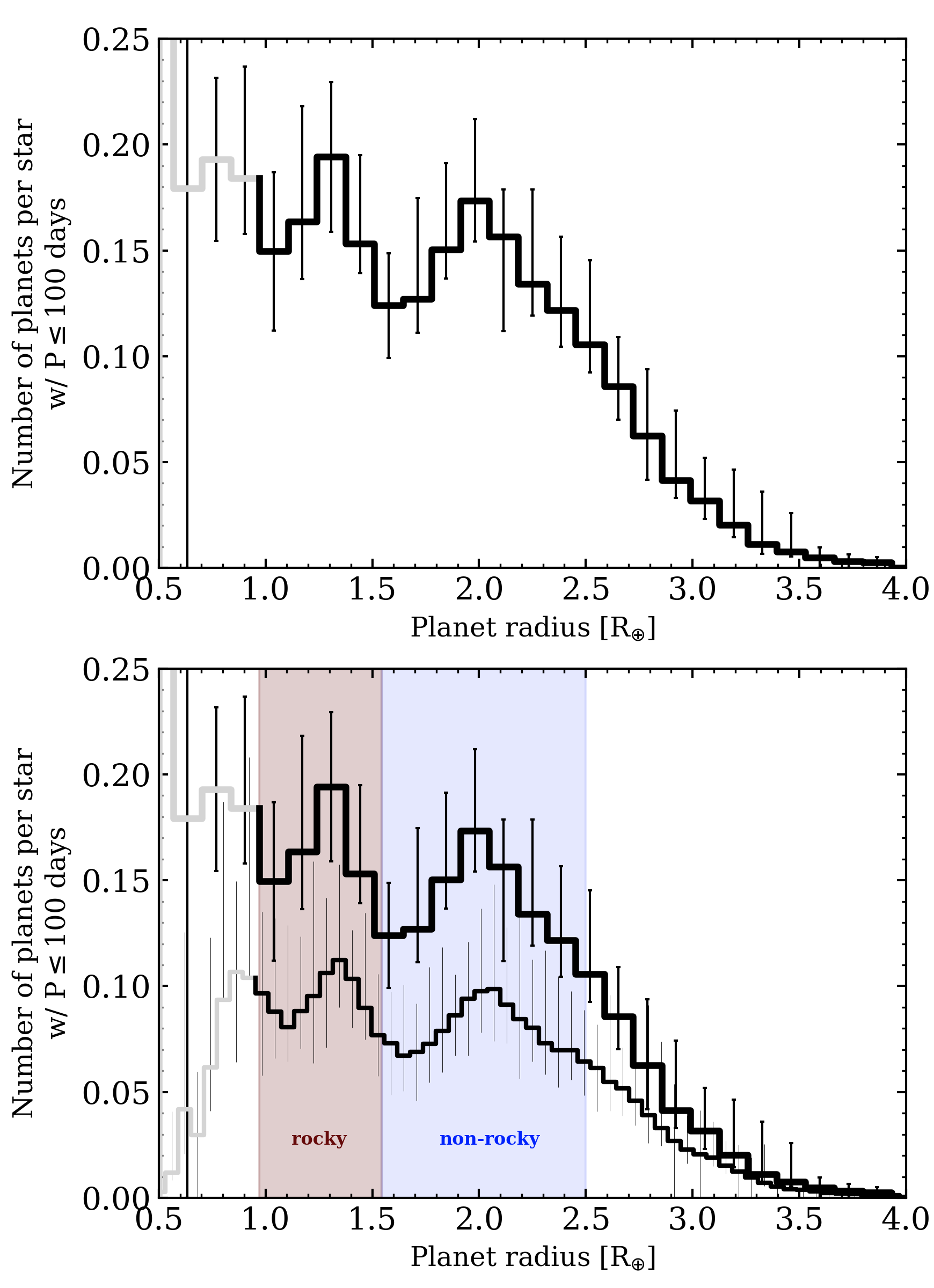}
  \caption{Occurrence rate of planets as a function of size. \emph{Upper panel}:
    histogram depicting the relative occurrence
    rate of close-in planets with orbital periods $<100$ days derived from the population of confirmed
    \kepler{} and \ktwo{} planets around low mass stars. The bimodal distribution of planet radii peaking
    at occurrence rate-weighted radii of 1.12 and 2.07 R$_{\oplus}$ indicates the presence of the radius valley
    around low mass stars centered at 1.54 R$_{\oplus}$. Uncertainties in the planet occurrences follow from binomial
    statistics and are limited by
    the relatively small number of confirmed planets around low mass stars from \kepler{} and \ktwo{.} 
    The measured occurrence rates below $\sim 1$ R$_{\oplus}$ (shown in \emph{grey}) should be ignored due to
    poor detection sensitivity. \emph{Lower panel}: identical occurrence rates as in the upper panel 
    accompanied by the same occurrence rates but with finer radius bins.
    The corresponding occurrence rate uncertainties per bin are inflated but the bimodal structure continues
    to be exhibited in the MAP occurrence rates. The shaded regions highlight our approximate
    definitions of rocky planets ($r_p \in [0.97,1.54]$ R$_{\oplus}$), down to reasonable sensitivity limits,
    and non-rocky planets ($r_p > 1.54$ R$_{\oplus}$) around low mass stars. Note the 2.5 R$_{\oplus}$ outer limit of
    the shaded region is chosen arbitrarily.}
  \label{fig:rphist}
\end{figure*}

Next, we marginalize over $P$ and compute the one-dimensional occurrence rate of small close-in planets
as a function of $r_p$. The resulting occurrence rates are shown in \autoref{fig:rphist} in which the bimodal
distribution of planet sizes is again clearly discernible in the MAP occurrence rates.
The uncertainties on each $f_j$ bin are computed from the $16^{\text{th}}$ and $84^{\text{th}}$ percentiles
of the $f_j$ posterior. In \autoref{fig:rphist} we ignore the measured occurrence rate in bins with
$r_p\lesssim 1$ R$_{\oplus}$ where the detection sensitivity is poor.

From the bimodal distribution we highlight the approximate radii likely corresponding to planets with rocky
bulk compositions ($r_p\lesssim 1.54$ R$_{\oplus}$) versus planets with significant size fractions in a volatile-rich
or extended H-He gaseous
envelope ($r_p\gtrsim 1.54$ R$_{\oplus}$) around
low mass stars. Also depicted in \autoref{fig:rphist} is $f_j$ with a bin width half that of the principal
$f_j$ depiction (i.e. 0.06 R$_{\oplus}$ compared to 0.13 R$_{\oplus}$).
With finer binning the fractional uncertainties on $f_j$ are sufficiently
large to eliminate the significance of the distinct bimodal peaks. Despite this, the bimodality in the MAP
occurrence rate continues to persist with the location of the valley features only being marginally affected.
We interpret this as further evidence for the existence of the radius valley in the close-in planet population
around low mass stars.

\vspace{1cm}

\subsection{Inclusion of supplemental K2 planet candidates}
In an attempt to improve the counting statistics in the occurrence rate calculations,
we will consider an enlarged planet sample. This sample is the union of our existing sample of confirmed planets
with a set of additional planet candidates (PCs) from the \ktwo{}
mission. Specifically, we consider the set of PCs reported by \cite{kruse19}
from \ktwo{} campaigns 0-8 that includes 126 PCs not already included in our sample of confirmed planets
and orbiting stars contained within our stellar sample.

By definition we cannot identify which PCs are true planets of interest for this study and
which PCs are instead produced by an astrophysical false positive. 
The inclusion of \ktwo{} PCs therefore requires that we account for sample contamination by false positives
probabilistically. We do so by considering
a number of studies from the literature that perform a transiting planet search in \ktwo{---}from any subset of its
campaigns---and attempt to validate their uncovered PCs statistically based on follow-up observations
\citep{montet15,crossfield16b,dressing17,hirano18,livingston18a,mayo18}. Each of these studies utilized some combination of
ground-based photometry to validate planet ephemerides, reconnaissance
spectroscopy to identify spectroscopic binaries, and speckle or AO-assisted imaging to search for nearby stellar companions.
Each of the aforementioned studies used their respective set of follow-up observations together with the
statistical validation tool \texttt{vespa} \citep{morton12,morton15} to statistically classify their PCs as either a validated
planet (VP)\footnote{Validated and confirmed planets are equivalent dispositions.},
a false positive (FP), or some other inconclusive disposition (e.g. remains a PC). The FP rate around cool
stars (\teff{} $< 4700$ K) from each study is estimated by calculating the ratio of the number of reported FPs to
the total number of FPs plus VPs. Notably, \cite{crossfield16b} showed that the FP rate in their \ktwo{} sample is dependent
on the measured planet radius as giant PCs have a larger likelihood of
being a FP. Hence, we only focus on PCs with $r_p<4$ R$_{\oplus}$ when deriving FP rates.

The resulting FP rates are reported in \autoref{tab:FP}. Half of the studies do not find any probable
FP signals among the small PCs orbiting cool stars in their samples. In such cases, only upper limits on the FP rate
can be derived which all agree that the FP rate is $\lesssim 20$\% at 95\%. The remaining studies
each detect at least one FP such that a non-zero maximum likelihood FP rate is measured. Their average FP rate is 5.7\%
which is also in agreement with the derived upper limits from the aforementioned studies.
We proceed by constructing $10^3$ realizations of the planet population
that includes all confirmed planets from both \kepler{} and \ktwo{} plus a subset of the 126 \ktwo{} PCs from
\cite{kruse19}. The subset of included PCs are randomly sampled from the full set of PCs
according to the adopted FP rate such that each realization contains $(1-0.057)\cdot 126 \approx 119$ PCs.

\input{FPtable}

\begin{figure*}
  \centering
  \includegraphics[scale=.8]{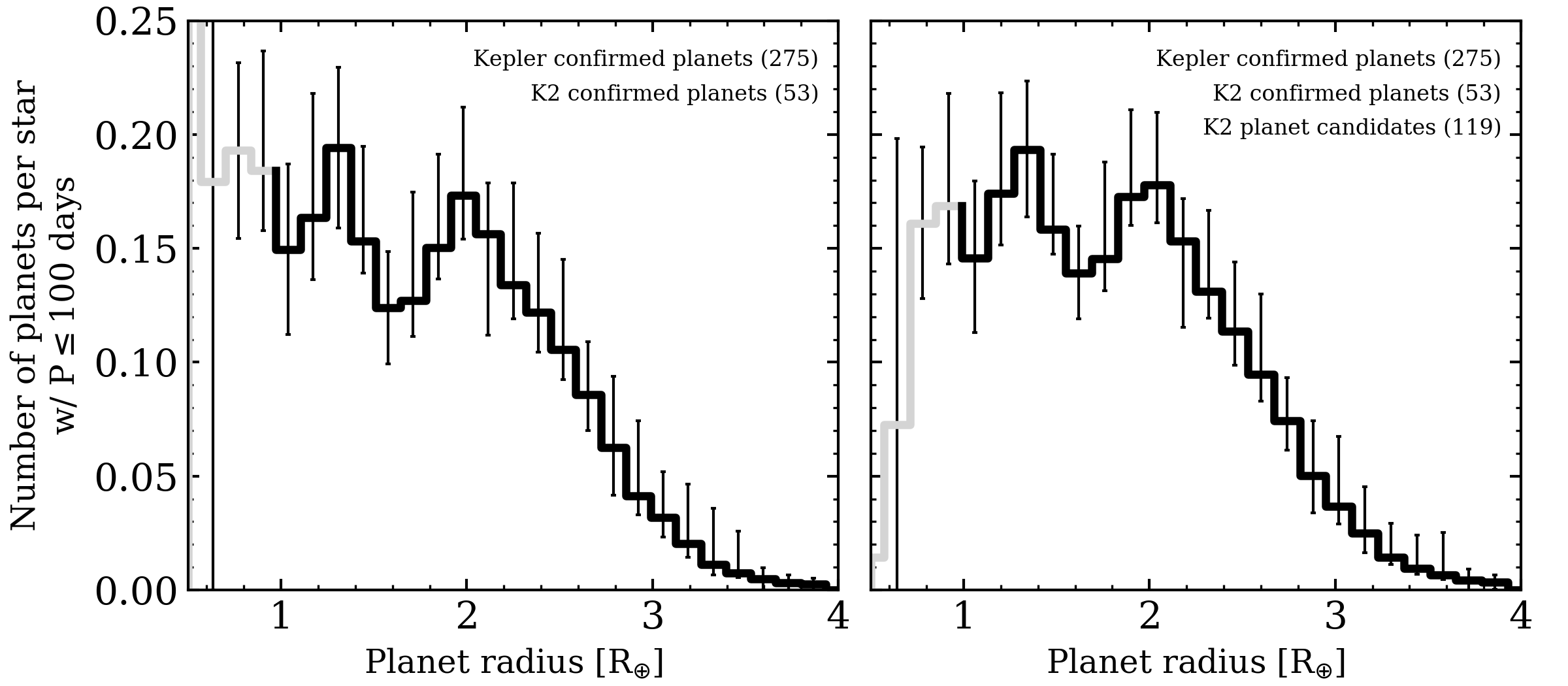}
  \caption{Comparison of occurrence rates with and without planet candidates included. \emph{Left panel}:
    same as \autoref{fig:rphist}. \emph{Right panel}: histogram depicting the relative occurrence
    rate of close-in planets with orbital periods $<100$ days derived from the population of confirmed
    planets from \kepler{} and \ktwo{} and supplemented by 119 randomly selected PCs
    around low mass stars from \cite{kruse19}. The radius valley continues to be resolved with
    the inclusion of PCs improving the median uncertainty on the occurrence rate bins although the
    bimodality becomes less prominent with numerous PCs partially filling the valley.}
  \label{fig:rphistPCs}
\end{figure*}

The effect of including PCs on the derived occurrence rates is assessed by comparing the $f_j$ distributions
measured with and without the inclusion of PCs (\autoref{fig:rphistPCs}). The radius valley continues to be resolved in
the MAP occurrence rates. Furthermore, the addition of PCs reduces the median $f_j$ uncertainty among planets
with $r_p>1$ R$_{\oplus}$ from
0.0216 to 0.0186 planets per star (i.e. $\sim 15$\% improvement). However, the partial filling of the gap is
further substantiated as the contrast between the maximum $f_j$ of the rocky planet peak ($r_p \sim 1.3$ R$_{\oplus}$)
and the minimum $f_j$ of the valley ($r_p \sim 1.6$ R$_{\oplus}$) decreases from 0.070
to 0.054 R$_{\oplus}$ (i.e. $3.2\sigma \to 2.9\sigma$). Contamination by true FPs in the planet sample containing
PCs may contribute to the reduced significance of the valley so we revert to considering confirmed planets only
for the remainder of this study.

\subsection{Comparison of the recovered planet occurrence rates from Kepler and K2} \label{sect:comparef}
Here we present a broad comparison of the occurrence rates of small close-in planets around
low mass dwarf stars from \kepler{} and \ktwo{.} Recall from Sect.~\ref{sect:fmap} that
we measure cumulative occurrence rates from confirmed \kepler{} and \ktwo{} planets of 
$2.48\pm 0.32$ and $0.75\pm 0.11$ planets per star respectively. Note that the former is consistent with
previous measurements from \kepler{} \citep{morton14,dressing15a,gaidos16}. However, noting that 
the FP rate of small \ktwo{} PCs is small ($\sim 5.7$\%, \autoref{tab:FP}), many \ktwo{} PCs should contribute
to the calculation of the occurrence rates from \ktwo{.}

We consider the 126 PCs from the transiting planet search in \ktwo{} campaigns 0-8 \citep{kruse19}.
Statistically correcting
for FPs results in 119 PCs plus 52 confirmed planets around 7227 \ktwo{} stars within our stellar sample that
were observed in any of the \ktwo{} campaigns 0-8. Given the period and radius of each of these 171 planets,
we apply the completeness corrections computed in Sect.~\ref{sect:k2sens}, divide out the number of low mass stars
observed in campaigns 0-8, and measure a cumulative occurrence rate of $2.26\pm 0.38$ planets per star. The inclusion
of PCs from \cite{kruse19} boosts the cumulative occurrence rate measurement from \ktwo{} to a value that is
consistent with that from \kepler{} (i.e. $2.48\pm 0.32$). This indicates that the \kepler{} and
\ktwo{} planet populations are consistent and that the reduced detection sensitivity of \ktwo{} compared to
\kepler{} is genuine (c.f. \autoref{fig:senscurves}). Note that the lesser \ktwo{} completeness continues to
have a negative
effect on the precision of the occurrence rate measurement despite \ktwo{} campaigns 0-8 having $\sim 1.8$ times
as many low mass dwarf stars as \kepler{.} We also note that in this comparison we have not corrected for any stellar
mass dependence of the occurrences rates which may produce a true discrepancy between the \kepler{} and \ktwo{}
values as their respective stellar samples within this study are not equivalent (c.f. \autoref{fig:stars}).  

\section{Evolution of the radius valley around low mass stars} \label{sect:models}
\subsection{Slope of the radius valley}
\begin{figure*}
  \centering
  \includegraphics[width=0.85\hsize]{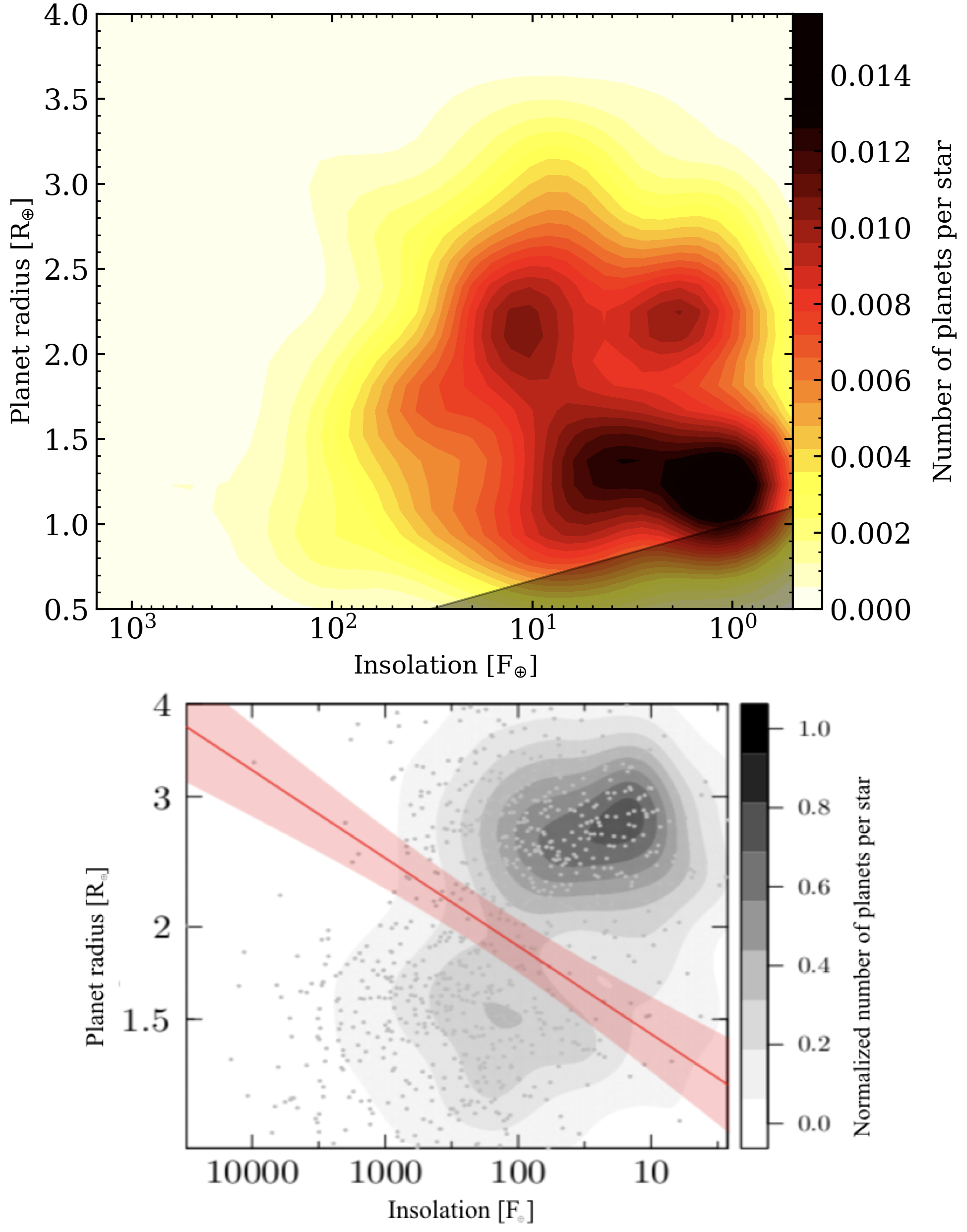}%
  \hspace{-0.85\hsize}%
  \begin{ocg}{fig:off1F}{fig:off1F}{0}%
  \end{ocg}%
  \begin{ocg}{fig:on1F}{fig:on1F}{1}%
  \includegraphics[width=0.85\hsize]{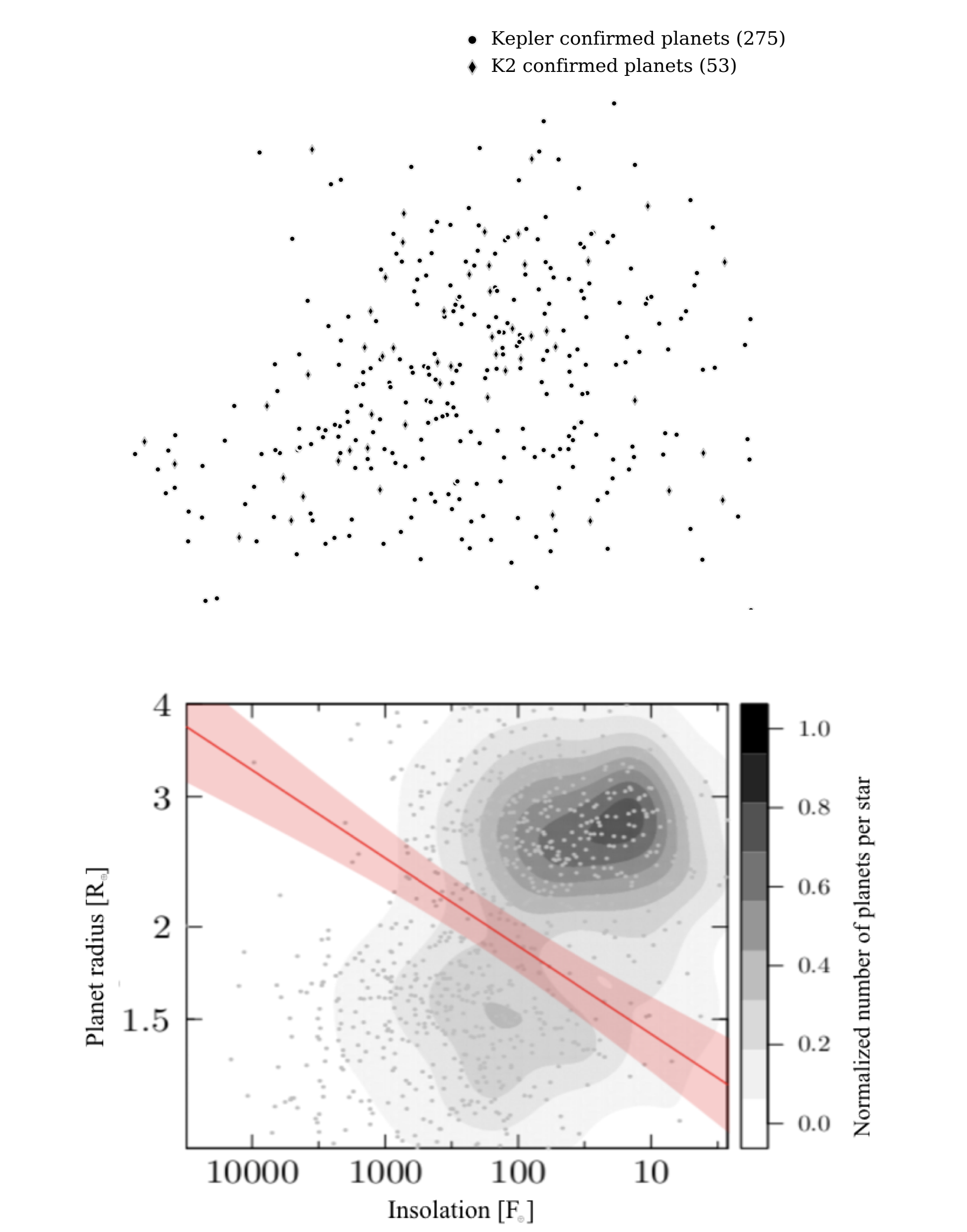}%
  \end{ocg}
  \hspace{-0.85\hsize}%
  \begin{ocg}{fig:off2F}{fig:off2F}{0}%
  \end{ocg}%
  \begin{ocg}{fig:on2F}{fig:on2F}{1}%
  \includegraphics[width=0.85\hsize]{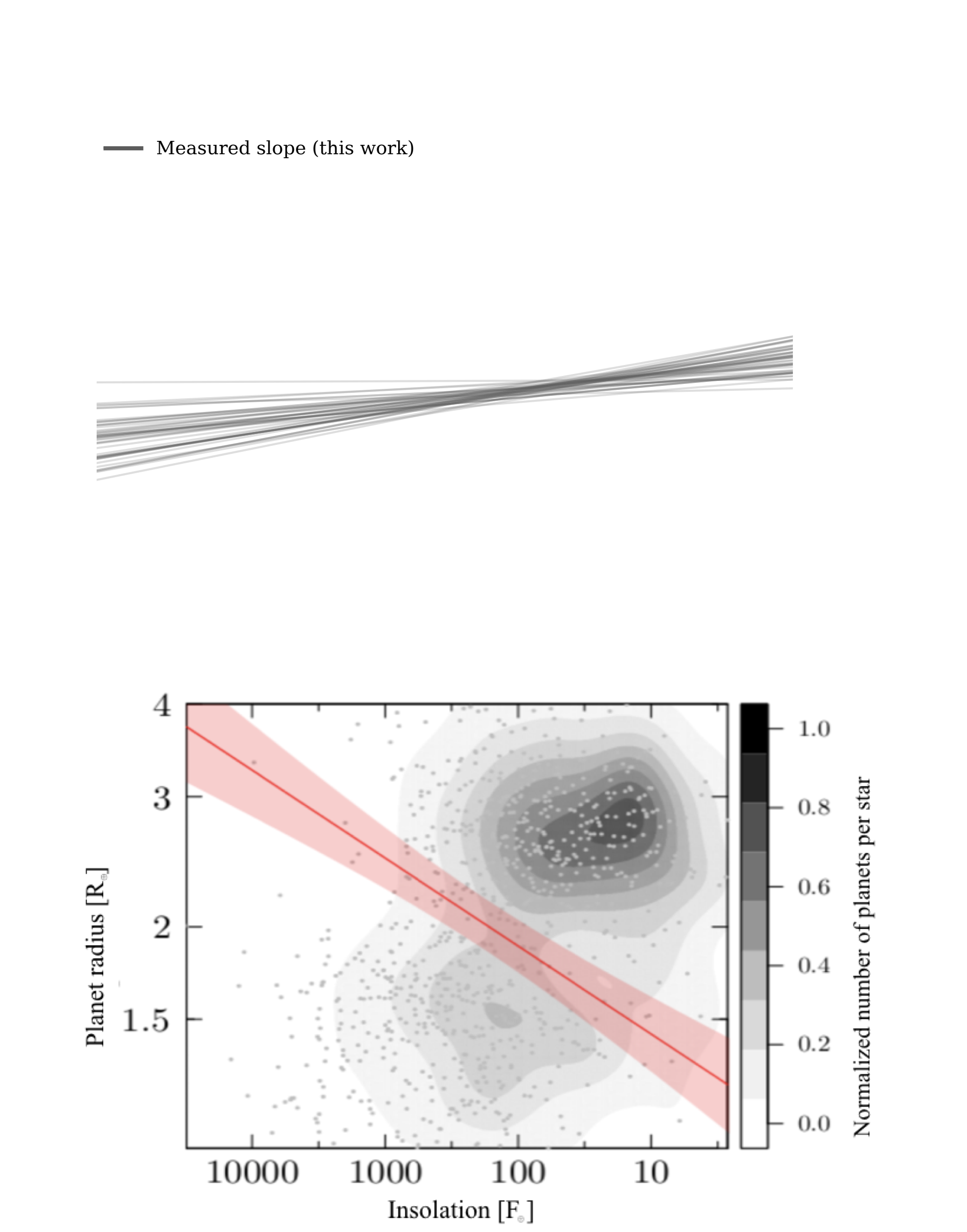}%
  \end{ocg}
  \hspace{-0.85\hsize}%
  \begin{ocg}{fig:off3F}{fig:off3F}{0}%
  \end{ocg}%
  \begin{ocg}{fig:on3F}{fig:on3F}{1}%
  \includegraphics[width=0.85\hsize]{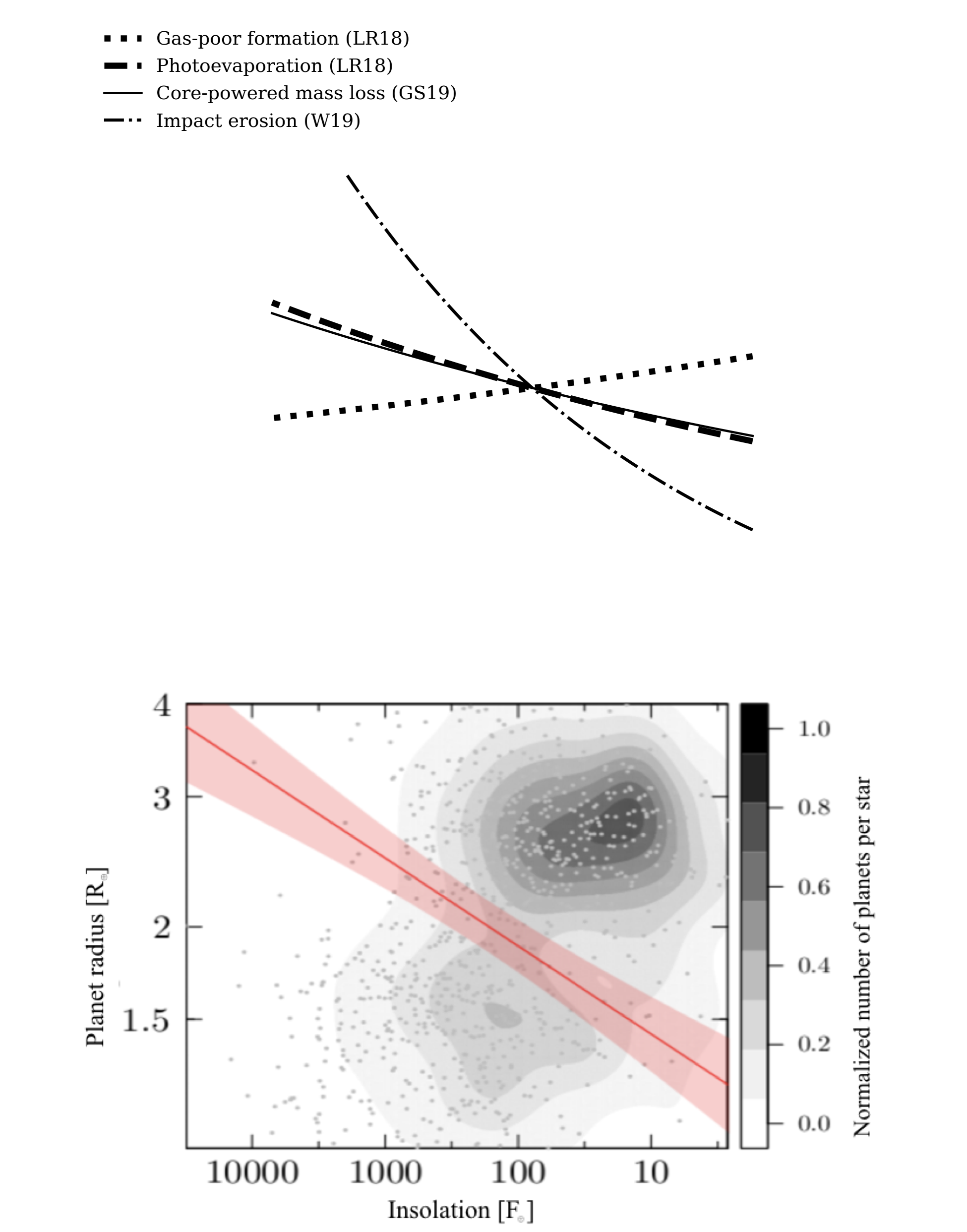}%
  \end{ocg}
  \hspace{-0.85\hsize}%
  \begin{ocg}{fig:off4F}{fig:off4F}{0}%
  \end{ocg}%
  \begin{ocg}{fig:on4F}{fig:on4F}{1}%
  \includegraphics[width=0.85\hsize]{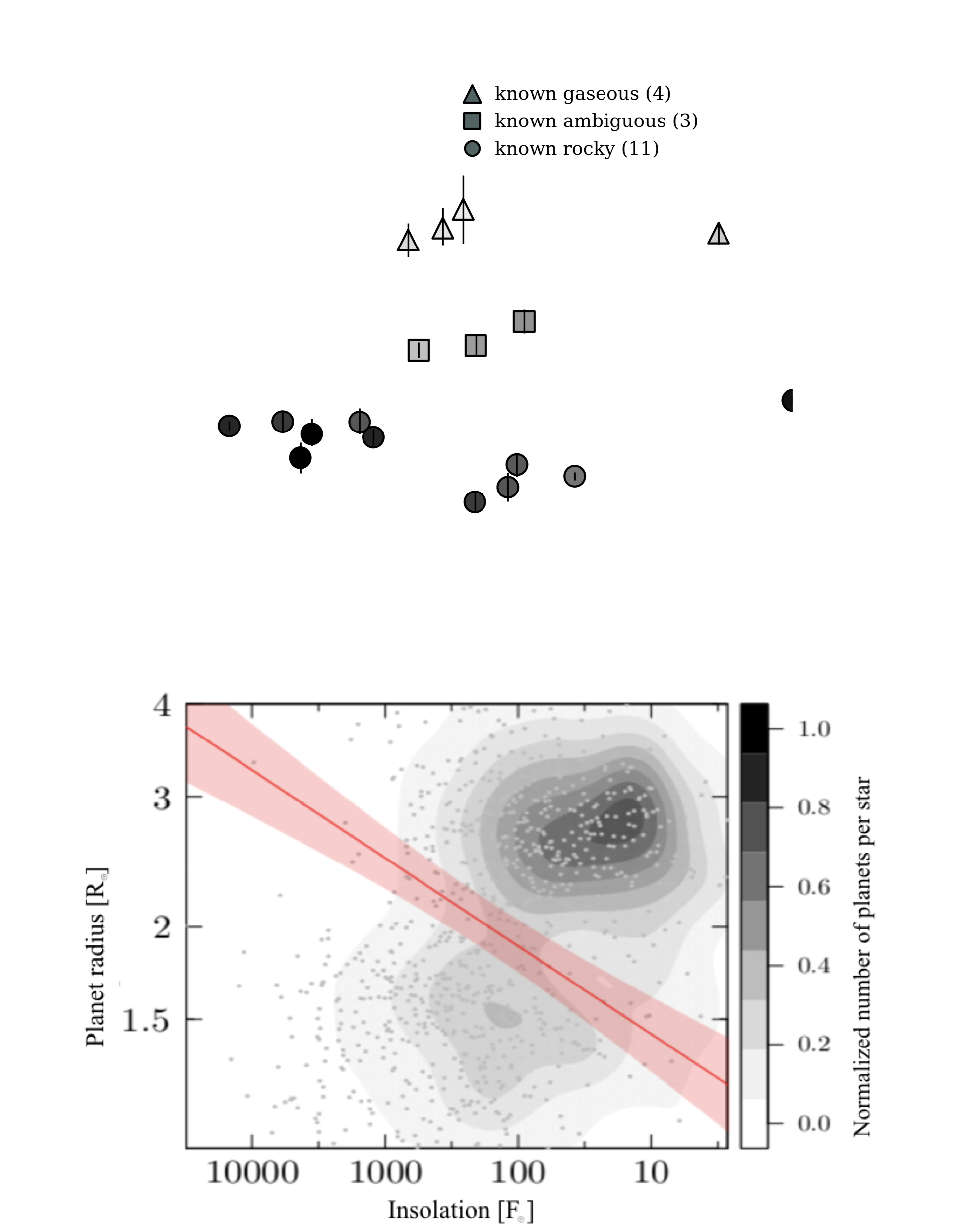}%
  \end{ocg}
  \hspace{-0.85\hsize}%
  \caption{(\emph{Interactive figure})
    Planet occurrence rates versus insolation and planet radius around low mass and Sun-like stars.
    \emph{Upper panel}: the maximum a-posteriori occurrence
    rate map calculated from the population of confirmed planets from \kepler{} and
    \ktwo{} around low mass dwarf stars
    \ToggleLayer{fig:on1F,fig:off1F}{\protect\cdbox{(\emph{small circles and diamonds})}}.
    Overplotted in black are
    \ToggleLayer{fig:on3F,fig:off3F}{\protect\cdbox{model predictions}} of the transition from rocky to non-rocky
    planets in the following scenarios: core-powered mass loss \citep{gupta19b}, photoevaporation
    \citep{lopez18}, and gas-poor formation \cite{lopez18}. We measure the
    \ToggleLayer{fig:on2F,fig:off2F}{\protect\cdbox{slope of the radius valley}}
    to be $r_{p,\text{valley}} \propto F^{-0.060\pm 0.025}$ which is consistent with predictions
    from gas-poor formation of terrestrial planets. Also overplotted are
    \ToggleLayer{fig:on4F,fig:off4F}{\protect\cdbox{planets}}
    with $\geq 3\sigma$ bulk density measurements from the literature that are classified as having either
    a rocky (\emph{circles}), a gaseous (\emph{triangles}), or an ambiguous (\emph{squares}) bulk
    composition. Marker colors are indicative of the MAP planet bulk densities.
    \emph{Lower panel}: the occurrence rate map of close-in
    \kepler{} planets around Sun-like stars \citep[$r_{p,\text{valley}} \propto F^{0.12\pm 0.02}$;][]{martinez19}.
    Note the unique $F$ and $r_p$ scales depicted in each panel.}
  \label{fig:fmapF}
\end{figure*}

\autoref{fig:fmapF} shows the two-dimensional planetary occurrence rates in the $F-r_p$ space for
our planet sample as well as for the close-in \kepler{} planets around Sun-like stars
from \cite{martinez19}.
In this parameter space we calculate the slope of the radius valley with $F$ and compare the measured
value to model predictions of the transition from rocky to non-rocky planets
versus insolation. We measure the slope using a similar methodology to \cite{martinez19} wherein
the two-dimensional occurrence rates are first resampled from its MAP value and uncertainties in
$10^3$ realizations. In each realization the
one-dimensional occurrence rate distribution $f_j(r_p)$ is computed in logarithmically spaced $F$
bins. From each distribution the $f_j$-weighted radii of the rocky and non-rocky peaks are then calculated.
The central radius of the valley is computed as the average radius between the two peaks
assuming a uniform weighting as we are interested in measuring the location of the valley and not the relative
strength of the two peaks as a function of $F$. The resulting radius valley locations in the $\log{F}-\log{r_p}$
space are then fit with a linear model as depicted in \autoref{fig:fmapF}. 
Over the $10^3$ realizations of resampled planet
populations, we measure an average slope and standard deviation of
$r_{p,\text{valley}} \propto F^{-0.060\pm 0.025}$. Similarly, repeating this exercise in the $\log{P}-\log{r_p}$
space yields $r_{p,\text{valley}} \propto P^{-0.058\pm 0.022}$. For ease of reference, the slopes measured
in this work and around FGK stars from the CKS \citep{martinez19} are reported in \autoref{table:slopes}.

\input{slopetable}

The negative slope of $r_{p,\text{valley}} \propto F^{-0.06}$ indicates that the location of the radius valley
drops to smaller planet radii with increasing insolation (i.e. towards smaller orbital separations).
This behavior is broadly consistent with models of the formation small rocky planets in a gas-poor environment
\citep{lee14,lee16,lopez18}. This formation scenario
leads to the transition from rocky to non-rocky planets emerging due to the
superposition of rocky and non-rocky planet populations whose formation timescales differ. In this scenario, the
transition radius as a function of orbital separation is set by the maximum mass of a bare rocky core which
itself is set by the amount of available solid material for the proto-planet to form out of via collisional
growth. According to the
minimum-mass extrasolar nebula \citep{chiang13}, the solid surface density radial profile is
$\sigma_{\text{solid}} \propto a^{-1.6}$ where $a$ is the semimajor axis. The amount of solid material 
accreted by a proto-planet
is proportional to its Hill radius $r_H = a(m_p/3M_s)^{1/3}$ such that integrating over the feeding zone within
the disk surface results
in the maximum mass of a bare rocky core $m_{p,\text{max}} \propto a^{0.6} M_s^{-0.5}$, or
$r_{p,\text{valley}} \propto a^{0.16} M_s^{-0.14}$ after applying the rocky planet mass-radius relation
\citep{zeng16}. Hence, the transition radius is predicted to occur at larger planet radii with increasing
separation for a given host spectral type \citep{lopez18}. The corresponding theoretical scaling of the transition
radius with insolation for a given spectral type is $r_{p,\text{valley}} \propto F^{-0.08}$
which is consistent with our measured scaling of $r_{p,\text{valley}}\propto F^{-0.060\pm 0.025}$.

In addition to being consistent with predictions from gas-poor terrestrial planet formation models,
our slope measurements are inconsistent with models of thermally driven atmospheric
escape from photoevaporation or core-powered mass loss that predict
an increasing transition radius with increasing insolation ($r_{p,\text{valley}} \propto F^{0.11}$; \citealt{lopez18},  
$r_{p,valley} \propto F^{0.10}$; \citealt{gupta19b} respectively).
The negative slope of the radius valley around low mass stars differs in sign from the trend seen around
Sun-like stars \citep[$r_{p,\text{valley}}\propto F^{0.12\pm 0.02}, r_{p,\text{valley}}\propto P^{-0.11\pm 0.03}$;][]{martinez19}.
These differing observational findings in each stellar mass regime may either be interpreted as
a signature of distinct planet formation processes wherein gas-poor planet formation is more prominent
around low mass stars, or that the efficiency of atmospheric post-processing is weakened around lower mass stars.

The inclusion of planets with $\geq 3\sigma$ bulk density measurements (see \autoref{table:rhop}) in
\autoref{fig:fmapF} reveals that all planets that are inconsistent with having bulk rocky
compositions lie above the transition radius predictions from all physical models considered
(i.e. gas-poor formation, photoevaporation, core-powered mass loss, and impact erosion). However, the
temperate rocky planet LHS 1140b ($F=0.5$ F$_{\oplus}$, $r_p=1.73$ R$_{\oplus}$) sits in the $F-r_p$ parameter
space below the predicted transition radius from gas-poor formation but above the predicted transition from
photoevaporation, core-powered mass loss, and impact erosion. Although LHS 1140b is the only instance of a planet
existing between the radius valley predictions from gas-poor formation and thermally driven or impact driven
mass loss in \autoref{fig:fmapF}, the location and rocky composition of LHS 1140b
provide supporting evidence for the applicability of models of 
gas-poor terrestrial planet formation to the emergence of the radius valley around low mass stars.

\subsection{Planet populations versus stellar mass} \label{sect:Msbin}
\begin{figure*}
  \centering
  \includegraphics[width=.98\hsize]{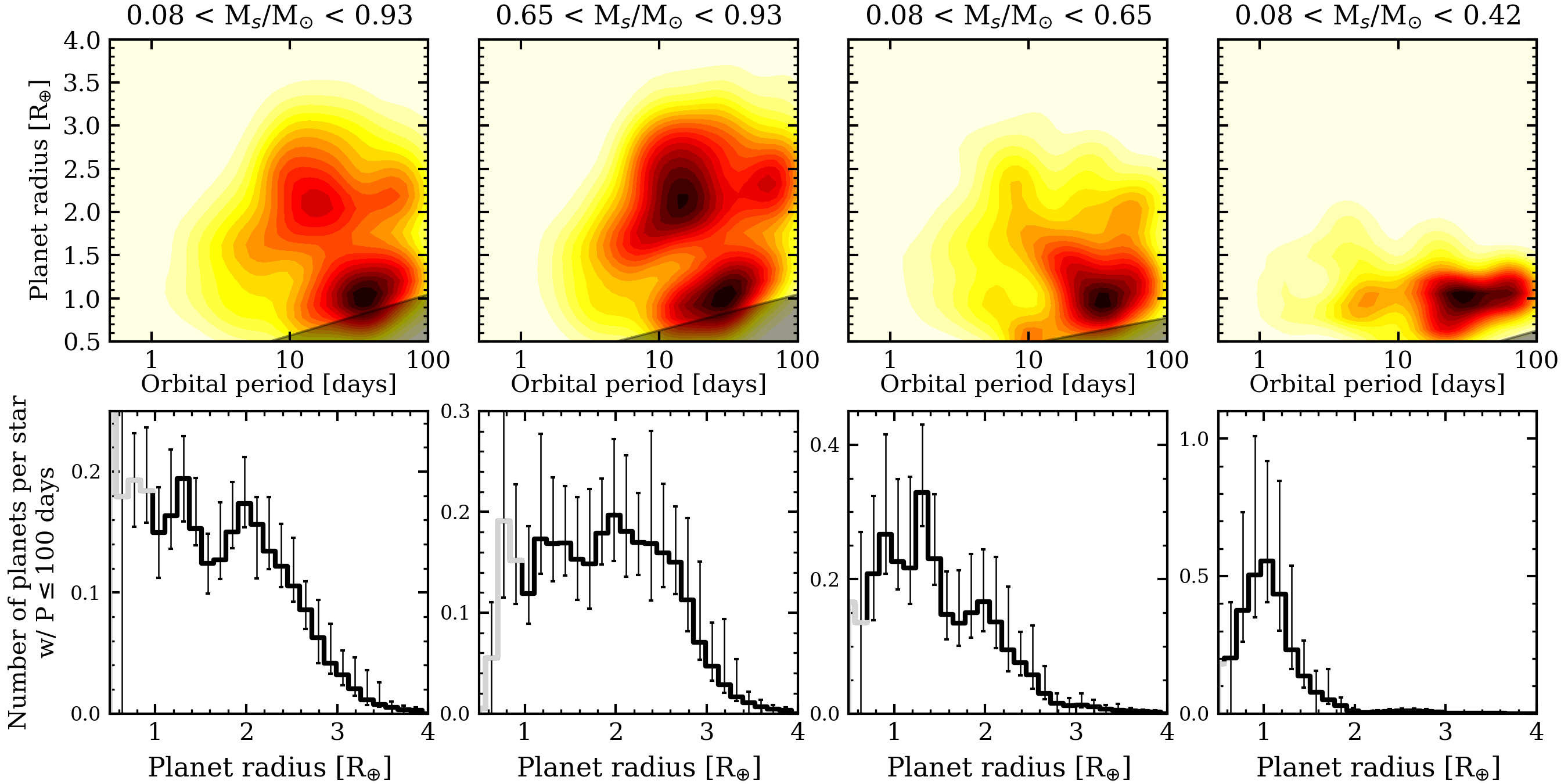}%
  \hspace{-.98\hsize}%
  \begin{ocg}{fig:off4r}{fig:off4r}{0}%
  \end{ocg}%
  \begin{ocg}{fig:on4r}{fig:on4r}{1}%
  \includegraphics[width=.98\hsize]{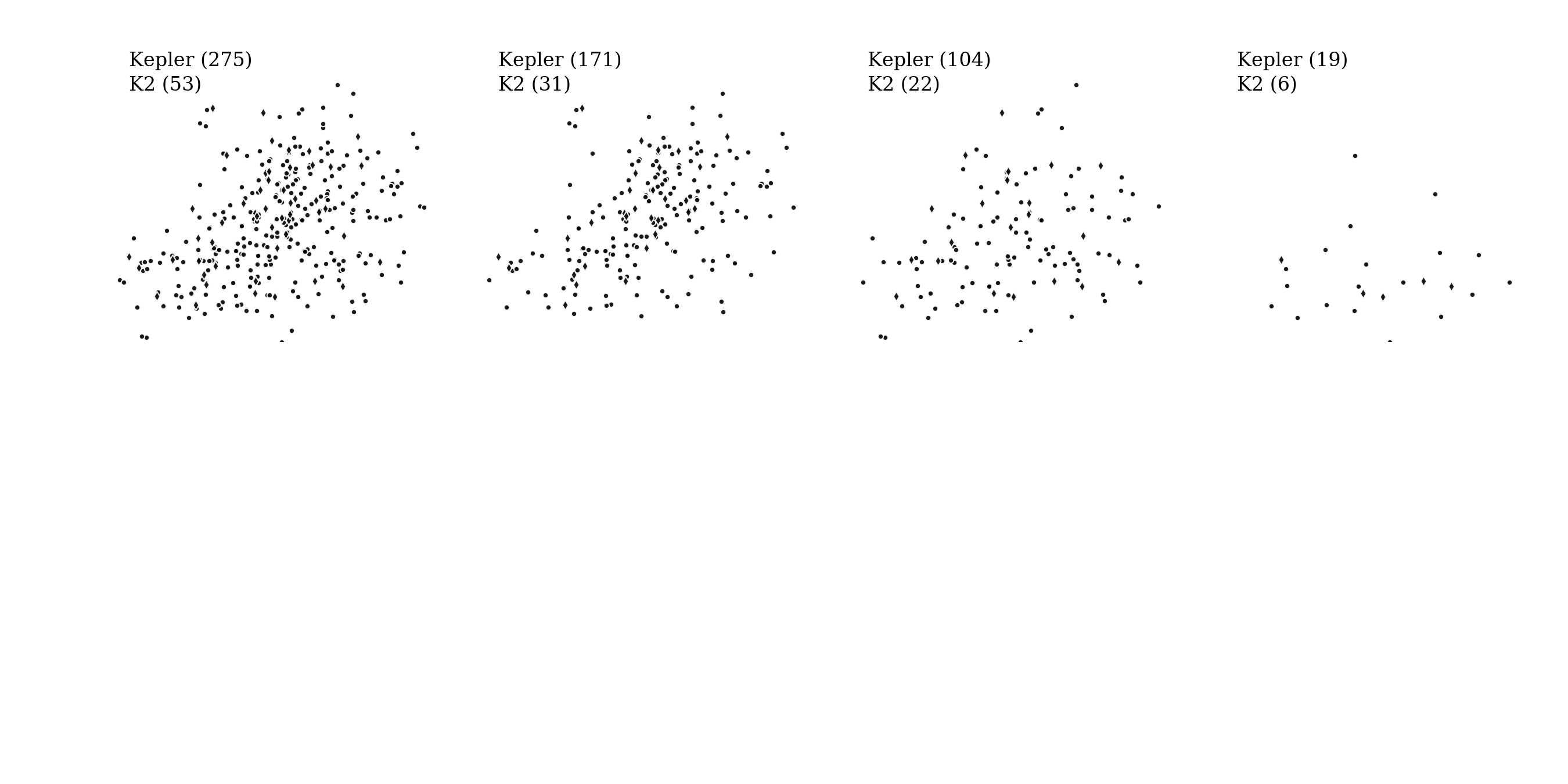}%
  \end{ocg}
  \hspace{-.98\hsize}%
  \caption{(\emph{Interactive figure}) 2D and 1D planet occurrence rates in various stellar mass bins. \emph{Top panels}: planet occurrence
    rate maps as a function of orbital period and planet radius. Overplotted are the relevant subsets of the
    population of confirmed planets from \kepler{} and
    \ktwo{} \ToggleLayer{fig:on4r,fig:off4r}{\protect\cdbox{(\emph{small circles and diamonds})}}.
    \emph{Bottom panels}: distributions of the relative
    planet occurrence rate as a function of planet radius. Note the differing occurrence rate scales.
    Each column corresponds to a unique cut in stellar masses
    which represent the full stellar sample ($M_s \in [0.08,0.93]$ M$_{\odot}$), the massive half of the stellar sample
    ($M_s \in [0.65,0.93]$ M$_{\odot}$), the low mass half of the stellar sample ($M_s \in [0.08,0.65]$ M$_{\odot}$),
    and the lowest mass bin ($M_s \in [0.08,0.42]$ M$_{\odot}$) depicting a subset of the low mass half of the stellar
    sample. Rocky planets appear to increase in prominence around lower mass stars.}
  \label{fig:rphistcomp}
\end{figure*}

In addition to calculating the occurrence rates $f_{ij}$ among our full stellar sample, here we consider
the evolution of the planet population in unique host stellar mass bins. \autoref{fig:rphistcomp} shows
the MAP $f_{ij}$ maps in $P-r_p$ space and the marginalized $f_j$ distributions in four stellar mass bins
representing our full stellar sample
($M_s \in [0.08,0.93]$ M$_{\odot}$), the massive half of the sample ($M_s>0.65$ M$_{\odot}$),  
the low mass half of the sample ($M_s<0.65$ M$_{\odot}$), and a subset of the latter focusing on
increasingly lower mass stars ($M_s<0.42$ M$_{\odot}$). The statistically significant resolution of the
radius valley in the $f_j$ occurrence rates is only accomplished in the full stellar sample. The reduction
of the sample size in the three remaining $M_s$ bins inflates the $f_j$ uncertainties such that the valley
is observed at $<1\sigma$ and hence not significant. However, the characteristic bimodality continues to be
exhibited in the
MAP $f_{ij}$ for the more massive half of our stellar sample. Furthermore, the $f_{ij}$ structures from
the full and massive samples are similar
as the majority of our full planet sample orbit stars more massive than the median stellar mass of 0.65
M$_{\odot}$ (i.e. $\sim 62$\% of our confirmed planet sample).

In considering stars less massive than 0.65 M$_{\odot}$, the non-rocky planet peak begins to diminish relative
to the terrestrial-sized planets. As evidenced in the MAP $f_j$ distribution around stars with
$M_s \in [0.08,0.65]$ M$_{\odot}$, the radius valley might persist around 1.6 R$_{\oplus}$ but the non-rocky planet
peak does not appear distinct from the rocky planet peak in the MAP $f_{ij}$ map. That is that the relative
frequency of rocky to non-rocky planets appears to increase significantly around M dwarfs compared to the more
massive K dwarfs. This feature is further accentuated around the lowest mass stars in our sample ($<0.42$
M$_{\odot}$) for which terrestrial-sized planets clearly dominate the distribution of close-in planets. The relative
frequency of rocky to non-rocky planets in each stellar mass bin are reported in \autoref{tab:rel}
for fixed definitions of $r_p \in [1,1.6]$ R$_{\oplus}$ and $r_p \in [1.6,2.5]$ R$_{\oplus}$ respectively. The inner
limit of 1 R$_{\oplus}$ restricts this analysis to where the detection sensitivity is still informative. The outer
limit of 2.5 R$_{\oplus}$ is chosen such that the full width at half maximum of the non-rocky planet peak in the
$f_j$ distribution from the full stellar sample is approximately identical for each peak (\autoref{fig:rphist})
but is still somewhat arbitrary.

\input{reltable}

The values in \autoref{tab:rel} indicate the significant increase in the relative occurrence of rocky
planets with decreasing stellar mass that is illustrated in \autoref{fig:rphistcomp}. Our measurements show
that non-rocky planets are nearly twice as common as rocky planets
around mid to late K dwarfs ($M_s \in [0.65,0.93]$ M$_{\odot}$) while the relative frequency approaches unity
around the full suite of M dwarfs ($M_s \in [0.08,0.65]$ M$_{\odot}$). Focusing on mid-to-late M dwarfs only in
the lowest stellar mass bin considered, rocky planets become much more prominent as they outnumber non-rocky
planets by a factor of $\sim 8.5\pm 4.6$. This result is broadly
consistent with the calculations from \cite{hardegree19} who find that terrestrial-sized 
planets ($r_p \in[0.5,1.5]$ R$_{\oplus}$) are about $4-5$ times as common as non-rocky planets ($r_p \in [1.5,2.5]$
R$_{\oplus}$) around M3-5.5 dwarfs ($M_s \in [0.12,0.38]$ M$_{\odot}$). 
Our calculations provide supporting evidence for an increase in the frequency of close-in rocky planets
around increasingly lower mass stars even with the small number of confirmed transiting planets in that mass
regime. More robust statements regarding the absolute occurrence rate of rocky planets around mid-to-late
M dwarfs will require a larger stellar sample in transit surveys with sensitivity to wider separations out to
hundreds of days where giant planets begin to emerge around these stars \citep{bonfils13,morales19}.

Although our data show a significant increase in the relative occurrence of rocky to non-rocky planets
around increasingly lower mass stars, we are unable to firmly identify the cause of this trend with these data.
That is that there are two outstanding hypotheses that cannot be ruled out by our data. The first scenario
requires the preferential formation of rocky planets around low mass stars such that atmospheric processing
by photoevaporation, core-powered mass loss, or impact erosion, have little to no effect. The alternative scenario
is that planet formation processes around low mass stars continues to produce non-rocky planets with significant
primordial atmospheres that are subsequently stripped by any of the aforementioned processes. The slope of the
valley measured from \autoref{fig:fmapF} suggests that the former scenario is important but our calculations of
the occurrence rates in various stellar mass bins are unable to solely confirm or reject any physical mechanism
at this time.

\subsection{Dependence of radius valley features on stellar mass}
Here we measure the locations and uncertainties of features in the radius valley in each of the stellar
mass bins considered in Sect.~\ref{sect:Msbin}. For each stellar mass bin
we measure the occurrence rate-weighted radius of the rocky planet peak,
the non-rocky planet peak (where applicable), and the radius valley. The uncertainties in the feature locations
are largely determined by uncertainties in the measured occurrence rates but are also directly affected by
the following hyperparameters: the $f_{ij}$ smoothing parameter, the minimum detection sensitivity still
considered reliable,
the $P$ bin width, the $r_p$ bin width, and the imposed upper and
lower $P$ and $r_p$ limits on each peak. The upper and lower $r_p$ limits are defined based on the visual
inspection of the $f_{ij}$ maps in \autoref{fig:rphistcomp} and are used to demarcate the
boundaries of each peak---and by extension---the valley separating the peaks.
As an example, if the prescribed boundaries on the rocky peak are set to 1-50 days and 0.8-1.4 R$_{\oplus}$,
then only the occurrence rates over that subset of the $P-r_p$ parameter space are used to calculate the
$f_{ij}$-weighted rocky peak radius. The range of boundary values for each peak are listed in
\autoref{tab:bounds}. In practice, we derive $10^3$ realizations of each $f_{ij}$ map with each realization
having a unique set of the aforementioned hyperparameters.
The resulting $f_{ij}$ maps are marginalized over $P$ and the $f_j$-weighted radius of
each peak is computed over the domain bounded by the relevant hyperparameters.
The same is done for the radius valley using the inverse occurrence rates.

\input{boundstable}

\begin{figure*}
  \centering
  \includegraphics[width=.97\hsize]{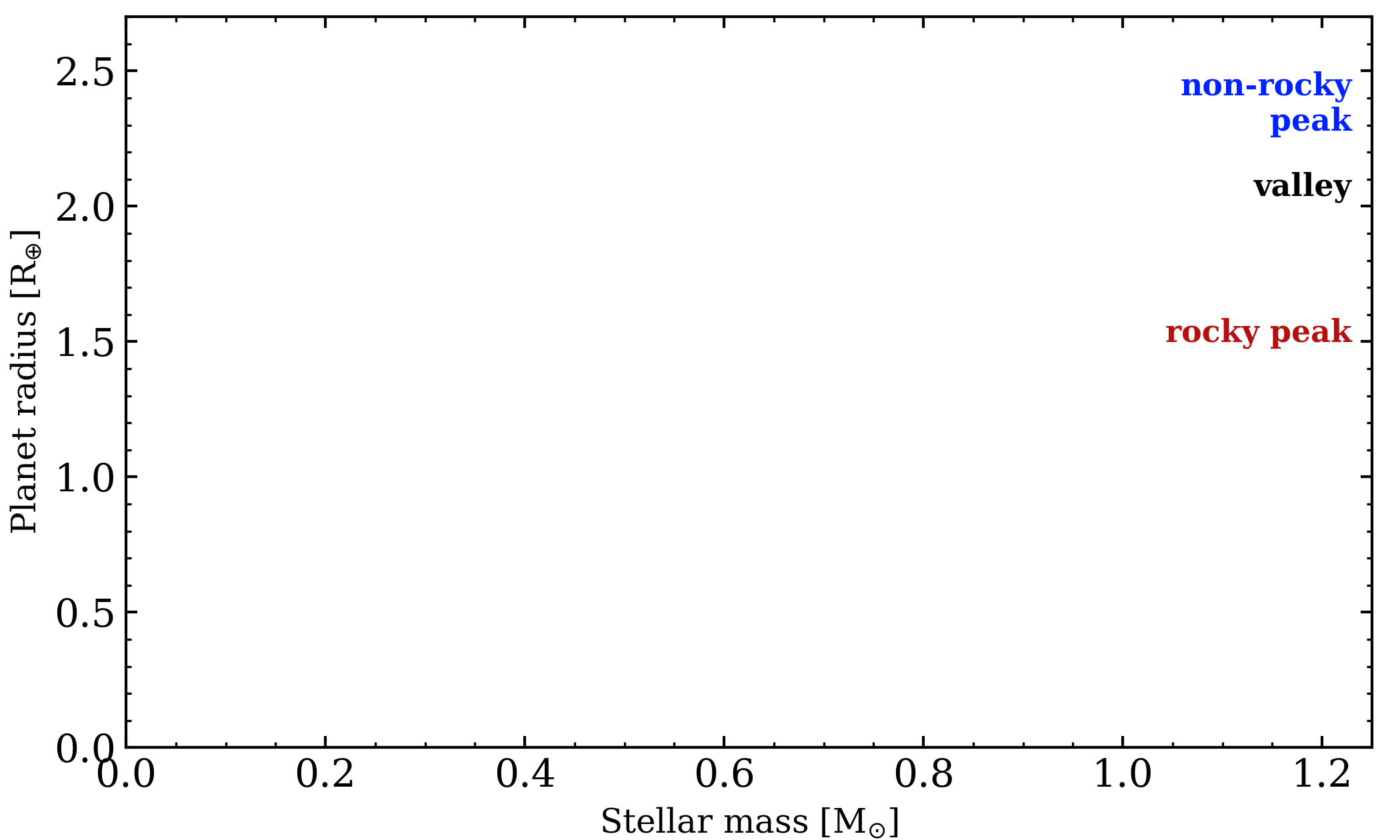}%
  \hspace{-.97\hsize}%
  \begin{ocg}{fig:off4M}{fig:off4M}{0}%
  \end{ocg}%
  \begin{ocg}{fig:on4M}{fig:on4M}{1}%
  \includegraphics[width=.97\hsize]{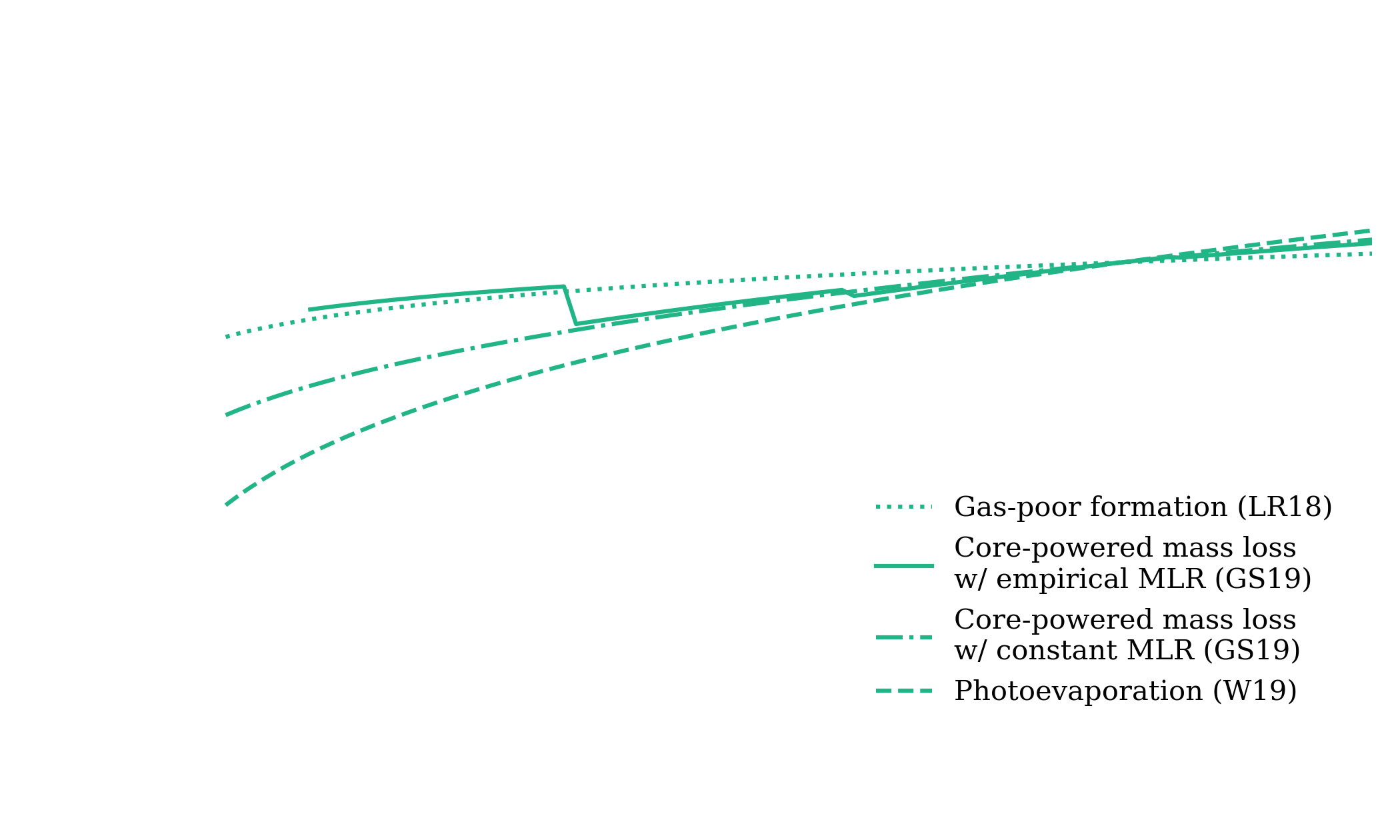}%
  \end{ocg}
  \hspace{-.97\hsize}%
  \begin{ocg}{fig:off1M}{fig:off1M}{0}%
  \end{ocg}%
  \begin{ocg}{fig:on1M}{fig:on1M}{1}%
  \includegraphics[width=.97\hsize]{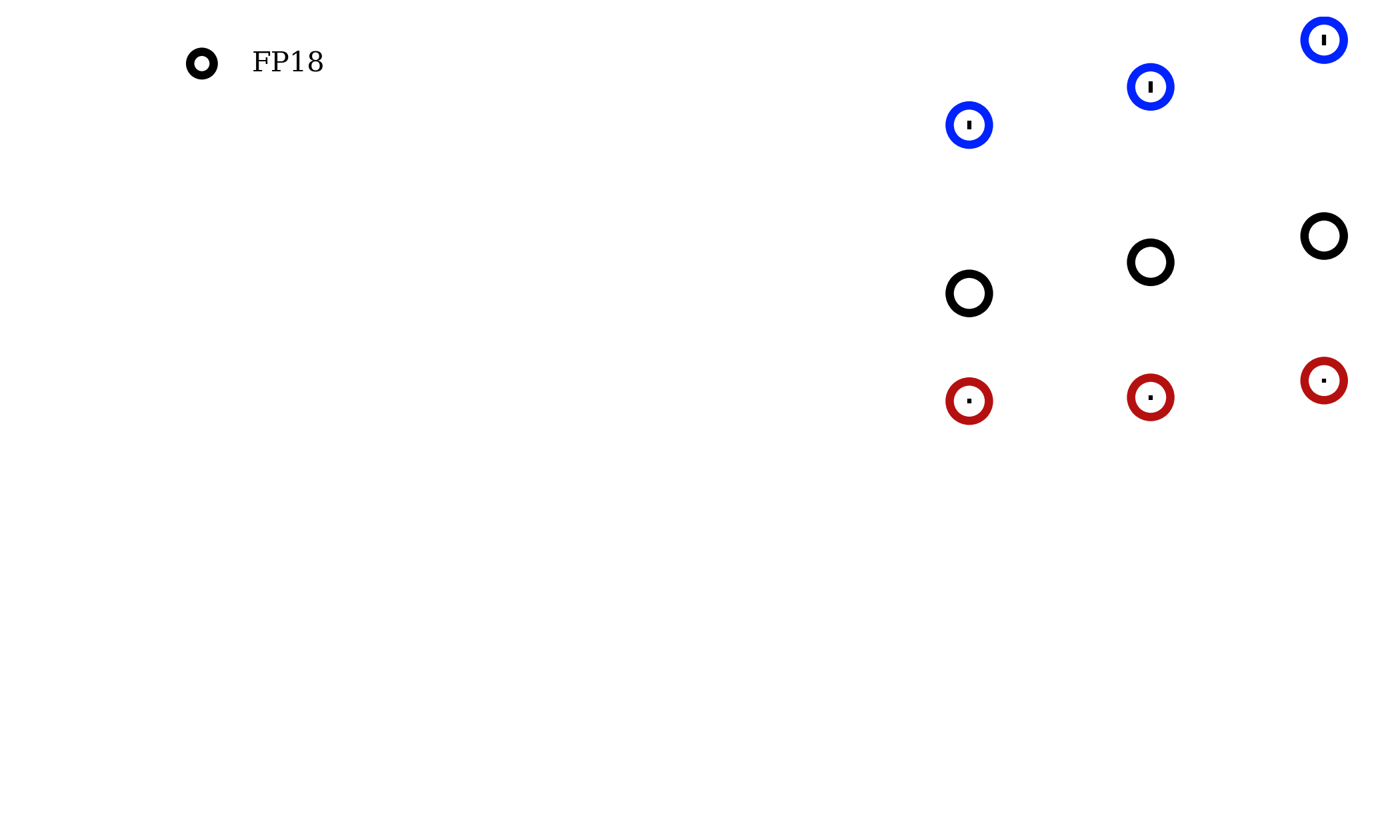}%
  \end{ocg}
  \hspace{-.97\hsize}%
  \begin{ocg}{fig:off3M}{fig:off3M}{0}%
  \end{ocg}%
  \begin{ocg}{fig:on3M}{fig:on3M}{1}%
  \includegraphics[width=.97\hsize]{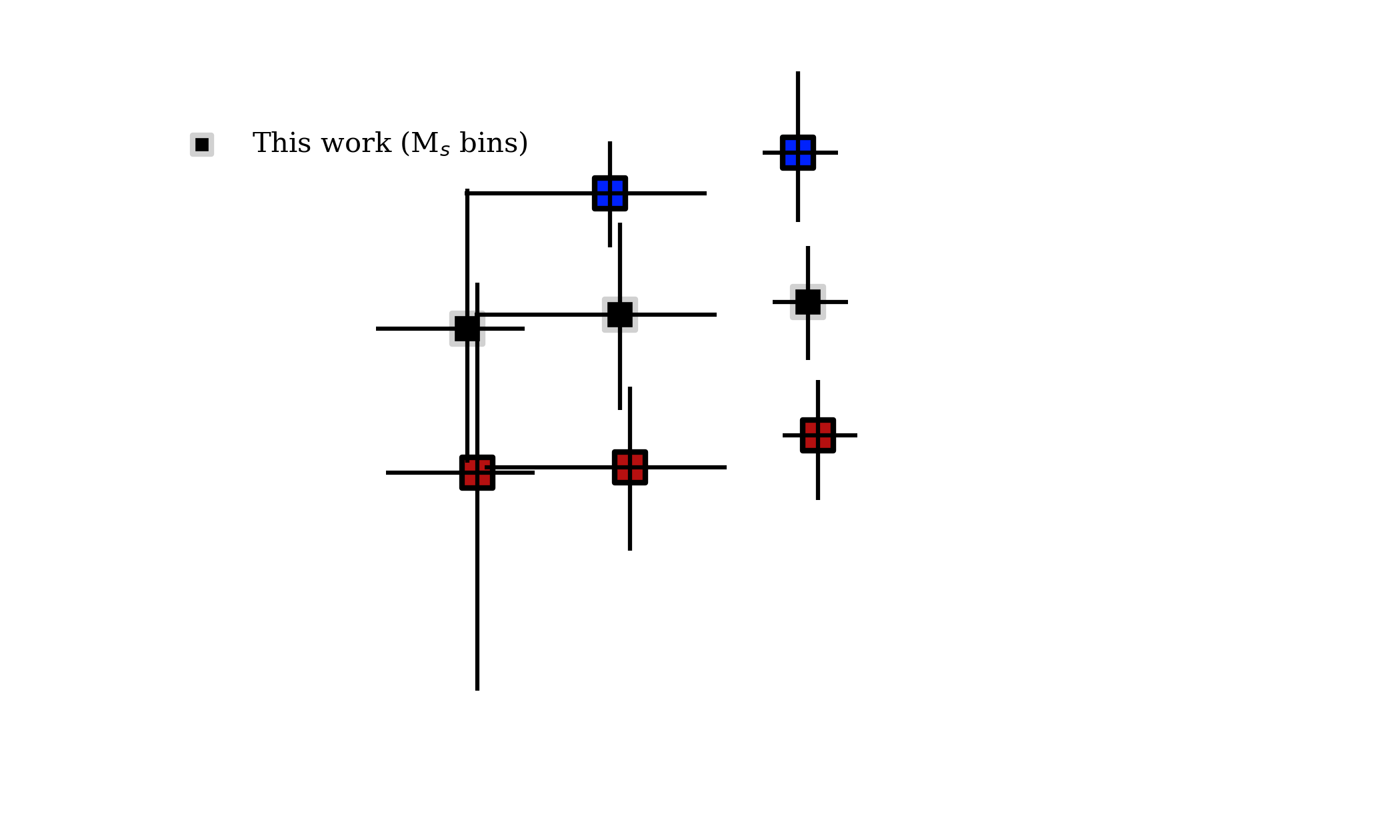}%
  \end{ocg}
  \hspace{-.97\hsize}%
  \begin{ocg}{fig:off2M}{fig:off2M}{0}%
  \end{ocg}%
  \begin{ocg}{fig:on2M}{fig:on2M}{1}%
  \includegraphics[width=.97\hsize]{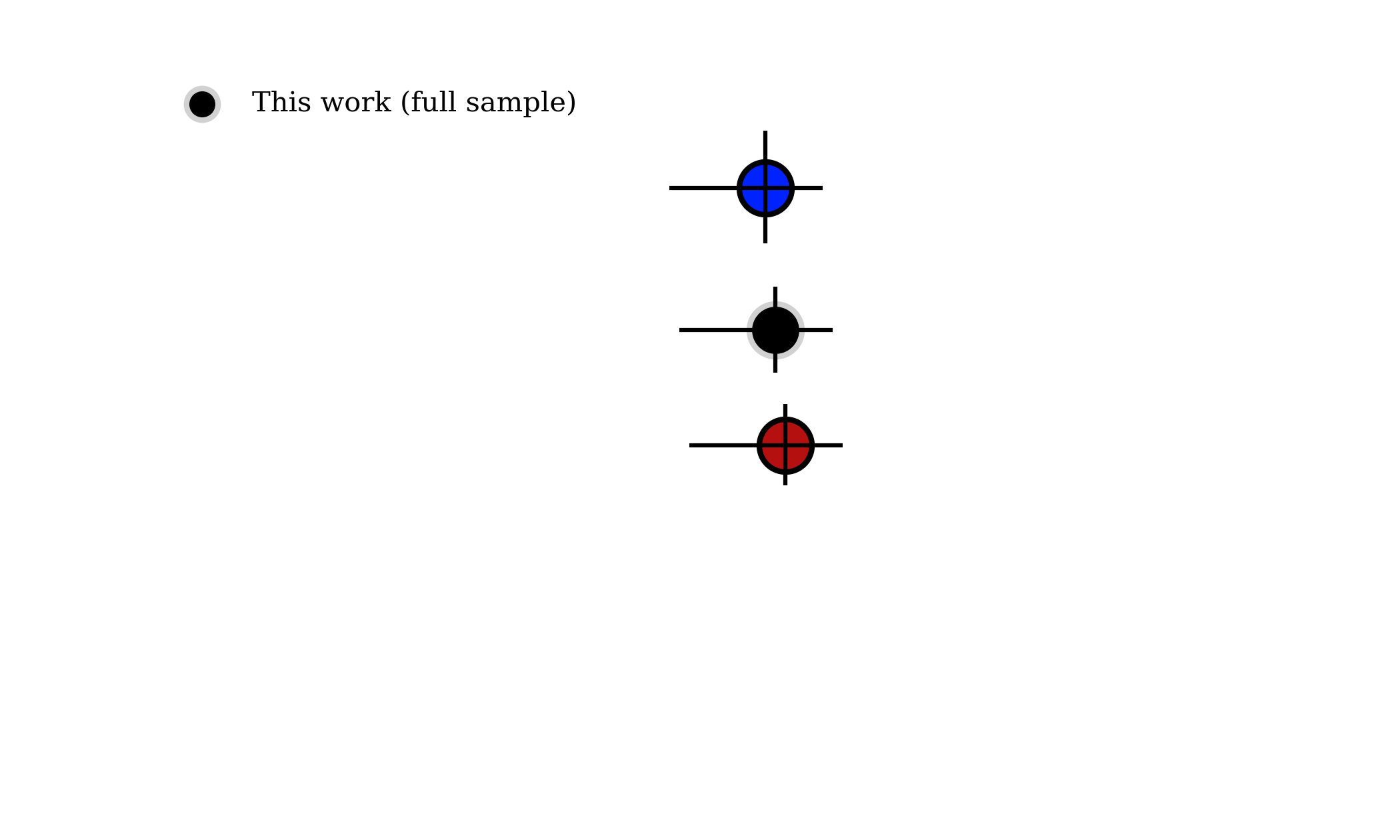}%
  \end{ocg}
  \hspace{-.97\hsize}%
  \caption{(\emph{Interactive figure}) Evolution of the radius valley features with stellar mass. \emph{Solid markers}:
    the occurrence rate-weighted locations of the non-rocky planet peak (\emph{blue markers}), the
    radius valley (\emph{black markers}) and the rocky planet peak (\emph{red markers})
    as a function of host stellar mass. Measurements around Sun-like stars with $M_s>0.8$ M$_{\odot}$ are
    retrieved from \cite{fulton18} \ToggleLayer{fig:on1M,fig:off1M}{\protect\cdbox{(\emph{open circles})}}.
    Feature radii from our full sample with a median value of
    $M_s = 0.651^{+0.058}_{-0.096}$ M$_{\odot}$ are depicted as
    \ToggleLayer{fig:on2M,fig:off2M}{\protect\cdbox{\emph{filled circles}}}.
    \ToggleLayer{fig:on3M,fig:off3M}{\protect\cdbox{\emph{Filled squares}}}
    depict the feature radii from partitioning our stellar sample into three $M_s$ bins: $M_s \in [0.65,0.93]$ M$_{\odot}$,
    $M_s \in [0.08,0.65]$ M$_{\odot}$, and $M_s \in [0.08,0.42]$ M$_{\odot}$. Markers in each stellar mass bin are slightly
    offset along the $M_s$ axis to assist in visualizing the errorbars. Uncertainties on the
    peak and valley locations are derived by sampling the measured occurrence rates and their uncertainties along
    with samples of the hyperparameters controlling map smoothing, minimum detection sensitivity, planet parameter
    binning, and the assumed feature ranges in $P$ and $r_p$. The
    \ToggleLayer{fig:on4M,fig:off4M}{\protect\cdbox{\emph{green curves}}}
    represent theoretical predictions for the evolution of the radius valley with stellar mass based on
    physical models of gas-poor terrestrial planet formation (\emph{dotted}; \citealt{lopez18}),
    core-powered atmospheric mass loss
    with an empirical mass-luminosity relation (\emph{solid}; \citealt{gupta19b}), a constant mass-luminosity
    relation (\emph{dash-dotted}; \citealt{gupta19b}), and photoevaporation (\emph{dashed}; \citealt{wu19}).
    The models only predict scaling relations with $M_s$ and as such are anchored to the measured valley location
    at $M_s \sim \text{M}_{\odot}$.}
  \label{fig:rpvMs}
\end{figure*}

The resulting locations of each radius peak and valley are depicted in \autoref{fig:rpvMs} as a function of
stellar mass. The locations and uncertainties are also given explicitly in \autoref{tab:rpvMs}.
The depicted $M_s$ values are represented by the median stellar mass within each bin and whose
uncertainties are derived from the $16^{\text{th}}$ and $84^{\text{th}}$ percentiles. In computing the
feature locations we assume that the bimodal $r_p$ distribution is present in all stellar mass bins
aside from the lowest mass bin (see \autoref{fig:rphistcomp}). In the lowest stellar mass bin we
only measure the location of the rocky planet peak and its edge which marks the transition from
rocky to non-rocky planets despite the latter being inherently rare around these types of stars.

\input{rpvMstable}

The measured feature radii are compared to those measured in \cite{fulton18} around Sun-like stars with
$M_s <0.97$ M$_{\odot}$, $M_s \in [0.97,1.11]$ M$_{\odot}$, and $M_s >1.11$ M$_{\odot}$.
Most notably, the location of each feature measured from our full stellar sample continues the trend of monotonically
decreasing towards smaller $r_p$ with decreasing $M_s$. The slopes of this decrease for the rocky and non-rocky
planet peaks measured with the three points from \cite{fulton18} and from our full stellar sample
are $\text{d}r_{p,\text{terr}} / \text{d}M_s = 0.40$ and $\text{d}r_{p,\text{gas}} / \text{d}M_s = 0.97$
respectively. The relative slopes indicate that the most common size of
non-rocky planet decreases more steeply with $M_s$ than the typical size of rocky planets. This trend
is indicative of the effective disappearance of non-rocky planets around increasingly lower mass stars
(see \autoref{fig:rphistcomp}) while terrestrial-sized planets appear to persist.
Furthermore, the reduced slope of the rocky peak may be evidence for
a characteristic planetary core size of $\approx 1$ R$_{\oplus}$ although its exact location is largely uncertain
due to the limited detection sensitivity to sub-Earth-sized planets. Furthermore, the probabilistic classification
of rocky planets as being primordially rocky or an evaporated core requires additional information
about the planet's orbit and stellar host properties \citep{neil19b}. 
For example, typical core sizes may be $M_s$-dependent
as a core size for a fixed core mass is composition dependent and the mass of heavy elements per star
is known to to be higher around M dwarfs compared to around FGK stars \citep{mulders15a,mulders15b,neil19}.

Models of the formation of the radius valley based upon photoevaporation \citep{wu19}, gas-poor formation \citep{lopez18},
and core-powered mass loss \citep{gupta19b} all make explicit predictions for the evolution of the radius valley location
with stellar mass. Predictions from the core-powered mass loss scenario are dependent on the stellar mass-luminosity
relation (MLR) $L_s \propto M_s^{\alpha}$. In \autoref{fig:rpvMs} we consider cases with a constant MLR with $\alpha=5$
\citep{gupta19b} and with the empirically-derived piecewise MLR from \cite{eker18}. All models predict a decreasing
radius valley with decreasing stellar mass but differ in their slopes. 
At the median stellar mass of our full stellar sample ($0.65$ M$_{\odot}$), the measured location of the radius valley is
$1.54\pm 0.16$ R$_{\oplus}$. This value---combined with measurements from Sun-like stars---favors
a steep $\text{d}r_{p,\text{valley}} / \text{d}M_s$ slope although we are
unable to distinguish between competing physical models given the measurement uncertainties.
Fortunately, the model predictions continue to diverge
with decreasing stellar mass such that measurements of the valley location in decreasing $M_s$ bins may be used to
rule out the operation of certain physical mechanisms in the low stellar mass regime. Although the trend of decreasing
feature radii with decreasing stellar mass appears to be upheld, the poor counting statistics in the reduced
$M_s$ bins prevent any significant inference regarding the relative strength of the competing physical mechanisms.
This problem can only be addressed by increasing the number of mid-to-late M dwarfs in transit surveys and by maximizing
the detection sensitivity to planets spanning the radius valley (see Sect.~\ref{sect:improve}).

\section{Discussion} \label{sect:discussion}
\subsection{Improving constraints on the sculpting of the radius valley using mid-M dwarfs} \label{sect:improve}
The issue of having insufficient information to distinguish between photoevaporation, core-powered mass loss, and gas-poor
formation around low mass stars can be addressed with two steps. Firstly, by expanding the low mass stellar sample in 
transiting planet searches and secondly, by quantifying the detection sensitivity in those searches. 
NASA's Transiting Exoplanet Survey Satellite \citep[\tess{;}][]{ricker15} is expected to provide hundreds of new transiting
planet discoveries in the vicinity of the radius valley \citep{barclay18}. \tess{} is particularly well-suited to the discovery of
close-in planets around low mass stars down to M5V ($M_s\sim 0.16$ M$_{\odot}$) due to its red bandpass (600-1000 nm)
and its high cadence (2-minute) observations of 200,000-400,000 stars over $\sim 94$\% of the sky by the completion of its
recently approved extended mission.

The \tess{} primary mission---lasting one year---has been ongoing since July 2018. 
Based on the photometric performance of the mission and consequently on the success of planet searches by the
Science Processing Operations Center \citep[SPOC;][]{jenkins16,twicken18,li18} at the time of writing,
we can estimate the number of low mass stars required to be observed
by \tess{} to enable robust conclusions regarding the nature of the emergence of the radius valley. These calculations
proceed by noting that the measurement uncertainty on the feature locations from binomial statistics scales as
$\sqrt{N_sP(1-P)}$ where $N_s$ is the number of observed stars and $P$ is the probability of detecting a planet close to
the radius valley 
given the detection sensitivity, the transit probability, and their inherent rate of occurrence (see Eq.~\ref{eq:prob}).
Through sectors 1-14, \tess{} has observed $N_{s,\text{TESS}} = 23,051$ stars less massive than 0.4 M$_{\odot}$
with 2-minute cadence from its Candidate Target List \citep[CTL;][]{stassun19}.
Among these stars, the SPOC has reported three objects of interest close to the radius valley
between $1.4-1.6$ R$_{\oplus}$\footnote{TOIs: 175.01, 406.01, and 667.01.}. 
Assuming a 0\% false positive rate
among these planet candidates, and the same MAP occurrence rate measured with \kepler{}
($f_{\text{valley}}\approx 0.19$ planets per star), we find the probability of \tess{} to detect a transiting planet
spanning the radius valley around a star with $M_s<0.4$ M$_{\odot}$ to be 
$P_{\text{valley,TESS}}=1.30 \times 10^{-4}$. We can compare these numbers to the \kepler{} values of $N_{s,\text{Kep}}=33$ and
$P_{\text{valley,Kep}}=8.56\times 10^{-3}$ to scale the uncertainty on $f_{\text{valley}}$---and hence on the radius
valley location---as an increasing number of mid-to-late M dwarfs are observed with 2-minute cadence with \tess{.}

\begin{figure*}
  \centering
  \includegraphics[width=0.98\hsize]{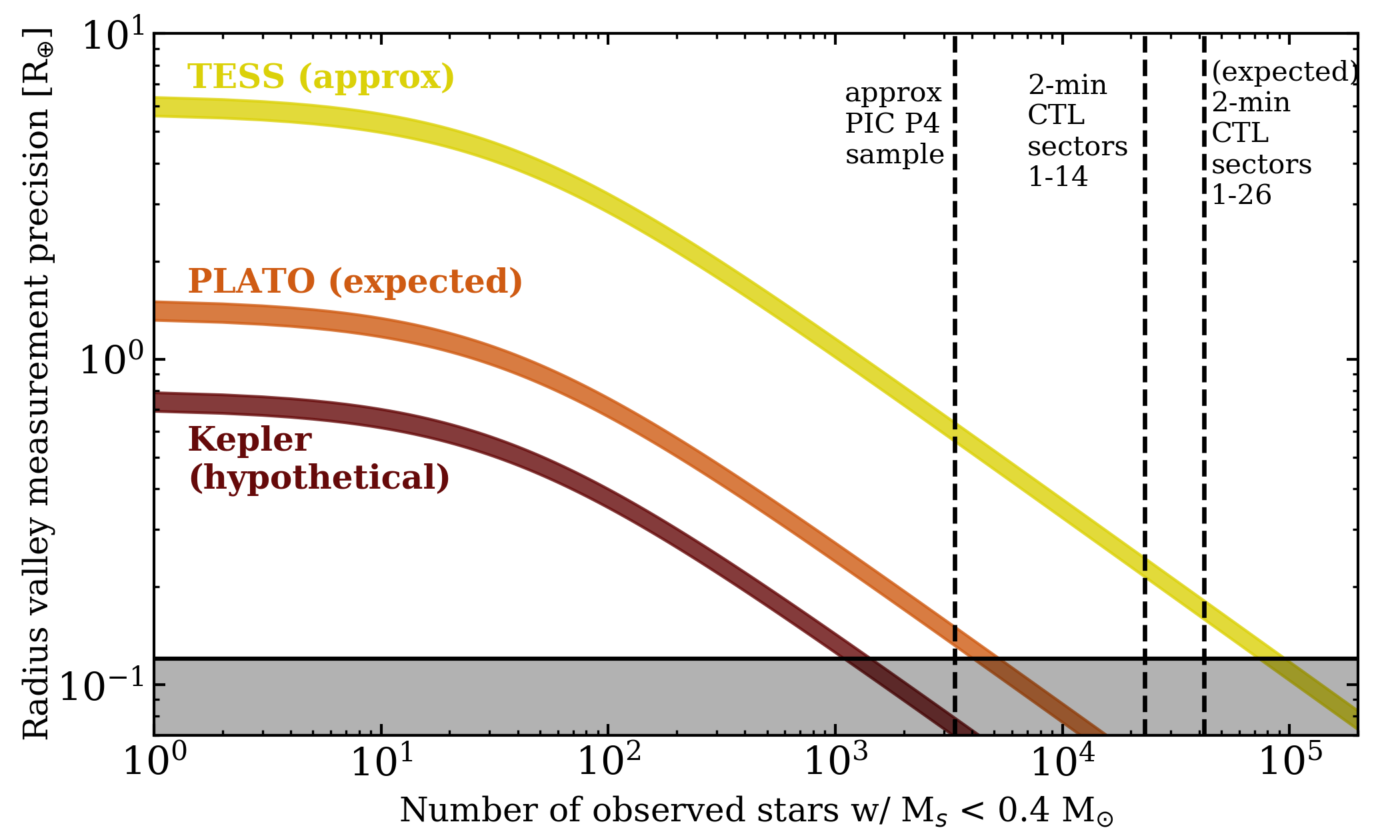}
  \caption{Expected improvement in the measurement precision of the radius valley location with the number of observed
    low mass stars. Measurement precision curves of the upper and lower uncertainties on the location of the radius valley 
    are derived from binomial statistics. These calculations use approximate and expected detection sensitivity
    values for the \tess{} (\emph{yellow}) and \plato{} (\emph{orange}) missions. The curve representing a hypothetical continuation
    of the primary \kepler{} mission is depicted in \emph{red}. The target precision of 0.12 R$_{\oplus}$ (\emph{shaded region})
    would enable models of photoevaporation to be distinguished from core-powered mass loss or gas-poor formation at $3\sigma$
    at $M_s=0.35$ M$_{\odot}$. Measurement precision curves are compared to the sample sizes of mid-to-late M dwarfs from the \tess{}
    Candidate Target List for sectors 1-14 (i.e. to the time of writing) and 1-26 (i.e. to the end of the \tess{} primary
    mission). Also included is the approximate number of mid-to-late M dwarfs included in the P4 sample of the
    PLATO Input Catalog.}
  \label{fig:improve}
\end{figure*}

The resulting improvement in the measurement precision of the radius valley with observations of additional mid-to-late
M dwarfs is shown in \autoref{fig:improve}.
The \tess{} curve reveals how precisely the location of the radius valley can be measured given \tess{'s} approximate 
detection sensitivity to planets spanning the radius valley and as the number of low mass stars observed with 2-minute
cadence is increased. Note that the improvement allotted by \tess{} should only be interpreted as an
approximation given that its detection sensitivity has not yet been adequately characterized. In our calculations, the
\tess{} detection sensitivity is estimated as a constant value as described in the preceding paragraph. 

We define a target measurement precision as that which is required to distinguish between predictions from photoevaporation
and core-powered mass loss (assuming an empirical mass-luminosity relation) at $3\sigma$ around low mass stars with a median
stellar mass of 0.35 M$_{\odot}$. Based on the model curves in \autoref{fig:rpvMs},
this required precision corresponds to a radius valley uncertainty of $\sim 0.12$ R$_{\oplus}$. A very similar level
of precision would be required to distinguish between photoevaporation and gas-poor formation as well.
The approximate \tess{} detection sensitivity implies that \tess{} will be required to observe $\sim 85,000$ mid-to-late M dwarfs
to distinguish between model predictions of photoevaporation and core-powered mass loss or gas-poor formation at $3\sigma$.
At the time of writing, only 23,051 such stars have been targeted with 2-minute cadence. Extrapolating to the end of \tess{'s}
primary mission, we expect a total of $\sim 42,000$ such stars to be observed with 2-minute cadence. If the \tess{} detection
sensitivity is well-characterized by that time and is roughly consistent with the approximate value assumed here, then \tess{}
could achieve a radius valley uncertainty of $\sim 0.17$ R$_{\oplus}$ by the end of its prime mission.
This would still be useful for constraining radius valley formation models as predictions from photoevaporation and
core-powered mass loss---or gas-poor formation---could be distinguished at $\sim 2.1\sigma$ with this level of precision.
Note that these calculations do not include non-CTL stars that may be targeted in the 30 minute \tess{} Full Frame Images
and could also contribute to the occurrence rate measurements, albeit with a reduced detection sensitivity.

Also included in \autoref{fig:improve} is the curve for a hypothetical continuation of the primary \kepler{} mission.
The calculation reveals that had \kepler{} been able to continue its prime mission and had
access to thousands of additional mid-M dwarfs than were targeted in the primary \kepler{} field, then the location
of the radius valley could have been precisely measured with $\sim 1200$ observed stars.

The stellar input catalog for the up-coming ESA \plato{} mission \citep{rauer14} has yet to be defined. The primary goal
of \plato{} is to detect and characterize transiting habitable zone planets around bright FGK stars. Despite this,
according to the mission's Definition Study
Report\footnote{\url{https://sci.esa.int/web/plato/-/59252-plato-definition-study-report-red-book}}, a subset of 
the PLATO Input Catalog (PIC) known as sample P4 will target $\geq 5000$ M dwarfs brighter than $V=16$ as part of
the mission's Long-Duration Observing Phase (LOP) lasting a minimum of two years. Furthermore, the expected random
noise in P4 is 800 ppm on one hour timescales. To compute the probability of
detecting a radius valley planet around a mid-to-late M dwarf targeted by \plato{,} we first assume that for a given
transit S/N, \plato{'s} detection sensitivity will be equivalent to that of \kepler{} (\autoref{fig:senscurves}).
We fix the transit S/N
(Eq.~\ref{eq:snr}) of a radius valley planet orbiting a mid-M dwarf using the values $r_p=1.5$ R$_{\oplus}$, $R_s=0.35$
R$_{\odot}$, $M_s=0.35$ M$_{\odot}$, CDPP$_{1\text{ hr}}=800$ ppm, and $n_{\text{transits}}=73$ for stars in the LOP observing
phase. We note the inexact nature of this calculation which
neglects the observing cadence and variations in the transit depth and photometric precision with each P4 star.
Nevertheless, assuming the \kepler{} occurrence rate we estimate that $P_{\text{valley,PLATO}}=2.35 \times 10^{-3}$. This
probability is $\sim 18$ times the estimated value for \tess{} but is about one quarter that of \kepler{.} The expected
radius valley measurement precision with \plato{} is also depicted in \autoref{fig:improve}. 

Although the exact P4 M dwarf sample is not yet defined, recent developments
at the September 2019 PIC Workshop in Italy\footnote{\url{https://indico.ict.inaf.it/event/806/}} concluded that the
properties of the sample will be consistent with M dwarf stars in the solar vicinity. Although this statement
if very rough and not binding to the final definition of the PIC, we combine this expectation with knowledge of M dwarfs
in the solar neighborhood to estimate the number of mid-to-late M dwarfs in the P4 sample. We do so by retrieving
the M dwarf sample within 25 pc from \cite{winters19}. Noting that this volume-limited sample is $\sim 33$\% complete
(J. Winters private communication), we identify $\sim 2829$ M dwarfs within 25 pc and with $V<16$.
We then scale up the M dwarf population beyond 25 pc until 5000 stars with $V<16$ are included assuming a homogeneous M
dwarf population beyond 25 pc. Of those stars representing probable targets within the P4 sample, 3358 have masses
$<0.4$ M$_{\odot}$. With this many mid-to-late M dwarfs targeted by \plato{} we expect the radius valley uncertainty
to reach $\sim 0.14$ R$_{\oplus}$ which would enable models of photoevaporation is be distinguished from models of
core-powered mass low or gas-poor formation at $\sim 2.6\sigma$.

\subsection{Implications for RV planet searches around low mass stars}
Many existing and up-coming radial velocity (RV) spectrographs
will be partially focused on characterizing the masses of planets spanning the radius valley in order to
improve our physical understanding of the nature of those planets. In particular, a subset of those spectrographs
operating in the near-IR will focus heavily on M dwarf planetary systems
(e.g. CARMENES; \citealt{quirrenbach14}, HPF; \citealt{mahadevan12}, IRD; \citealt{kotani14},
MAROON-X; \citealt{seifahrt18}, NIRPS; \citealt{bouchy17}, SPIRou; \citealt{donati18}). In defining target samples
that are equally complete on either side of the radius valley, it is critically important that the location
of the transition between rocky and non-rocky planets is known. In our full stellar sample, which includes
mid-to-late K dwarfs, the radius valley is centered at $1.54\pm 0.16$ R$_{\oplus}$. Although we remind the
reader that the exact value is dependent on the planet's separation (see \autoref{fig:fmapF}). A consistent value of
$1.55^{+0.52}_{-0.50}$ is also recovered---albeit with reduced significance---around stars later than about M2.5V.
This value is slightly lower than the valley locations measured around Sun-like stars of
$\sim 1.9$ R$_{\oplus}$ for $M_s \sim 1.2$ M$_{\odot}$ and $\sim 1.7$ R$_{\oplus}$ for $M_s \sim 0.85$ M$_{\odot}$
\citep{fulton18}.

\begin{figure}
  \centering
  \includegraphics[width=\hsize]{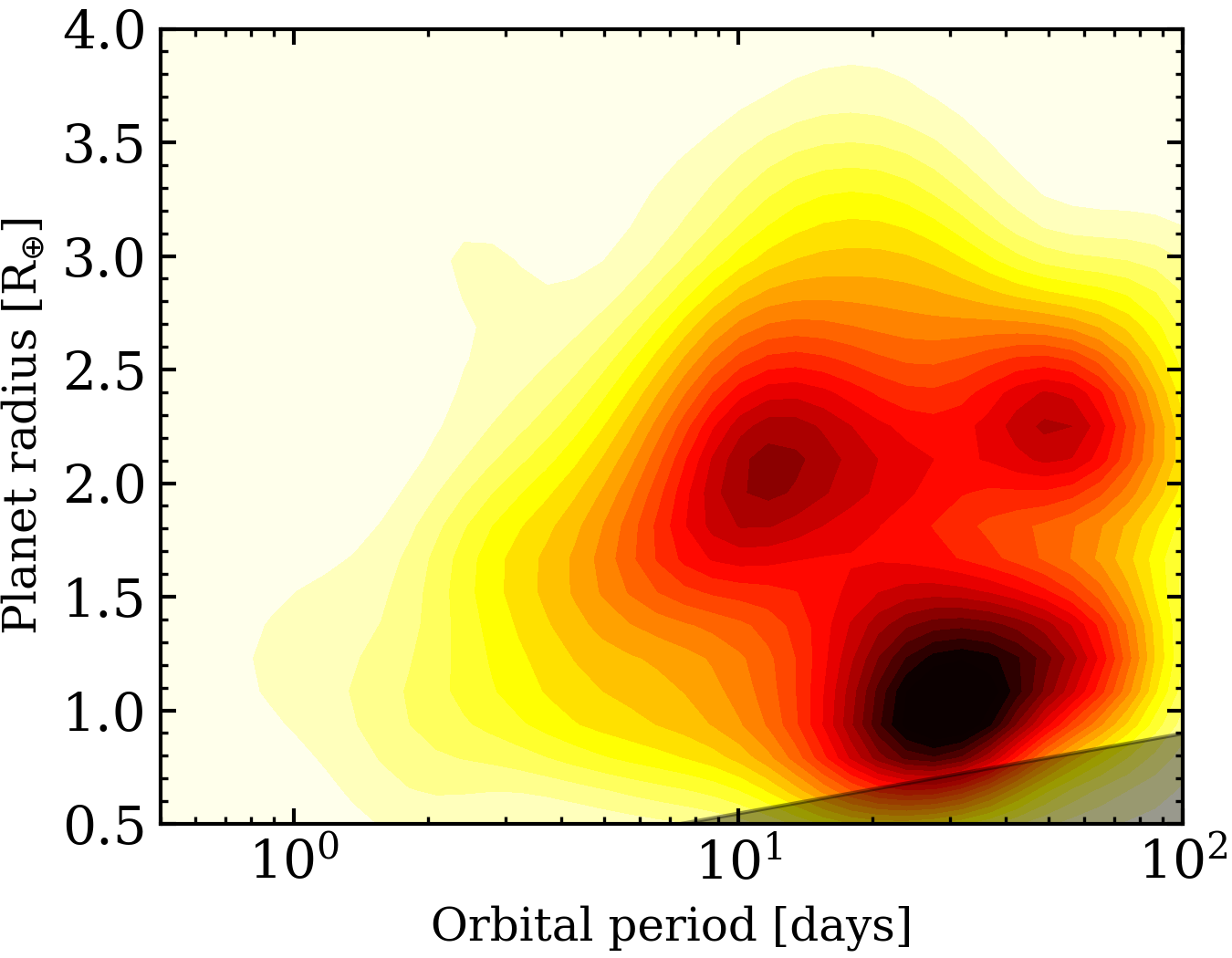}%
  \hspace{-\hsize}%
  \begin{ocg}{fig:off4pr}{fig:off4pr}{0}%
  \end{ocg}%
  \begin{ocg}{fig:on4pr}{fig:on4pr}{1}%
  \includegraphics[width=\hsize]{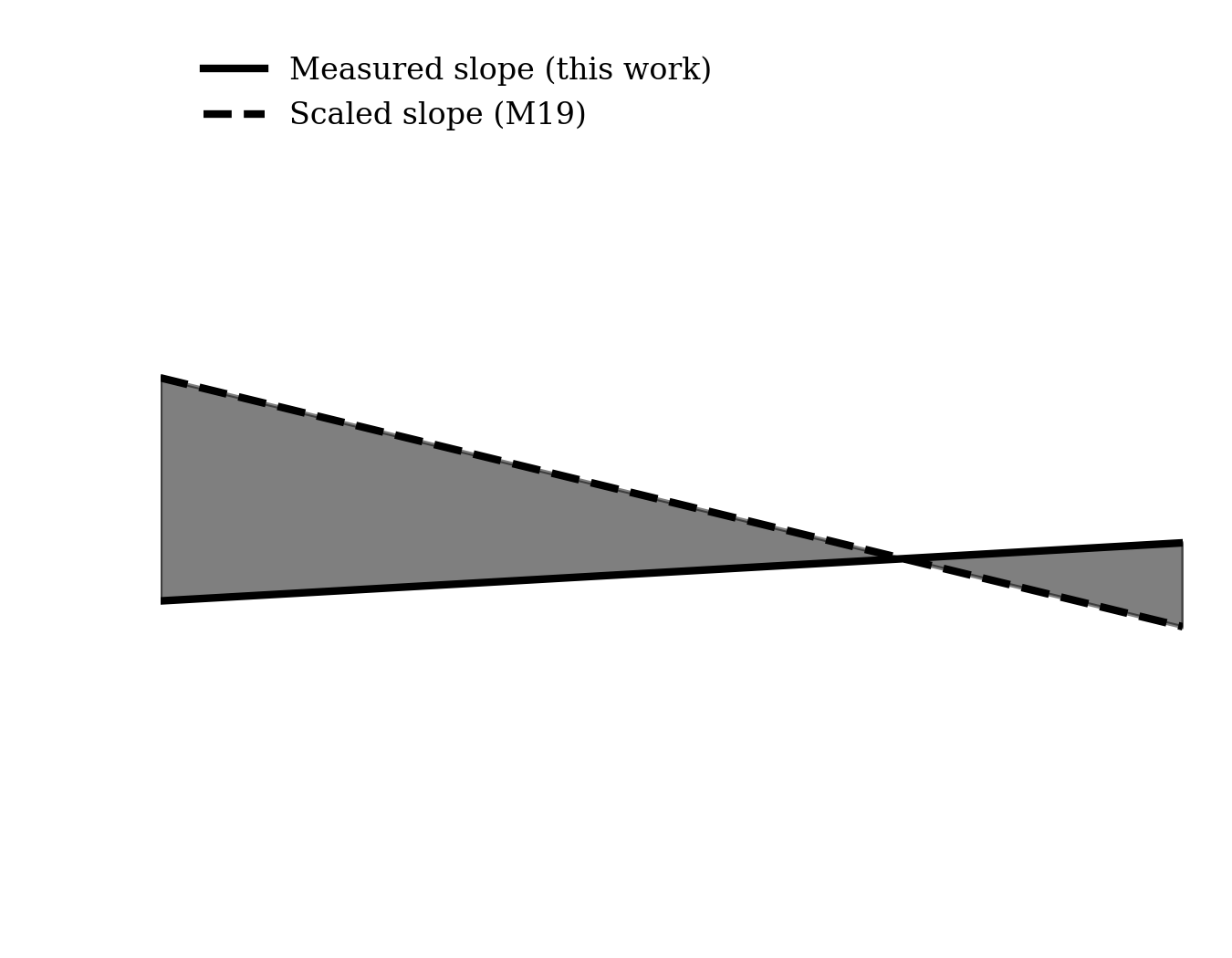}%
  \end{ocg}
  \hspace{-\hsize}%
  \begin{ocg}{fig:off1pr}{fig:off1pr}{0}%
  \end{ocg}%
  \begin{ocg}{fig:on1pr}{fig:on1pr}{1}%
  \includegraphics[width=\hsize]{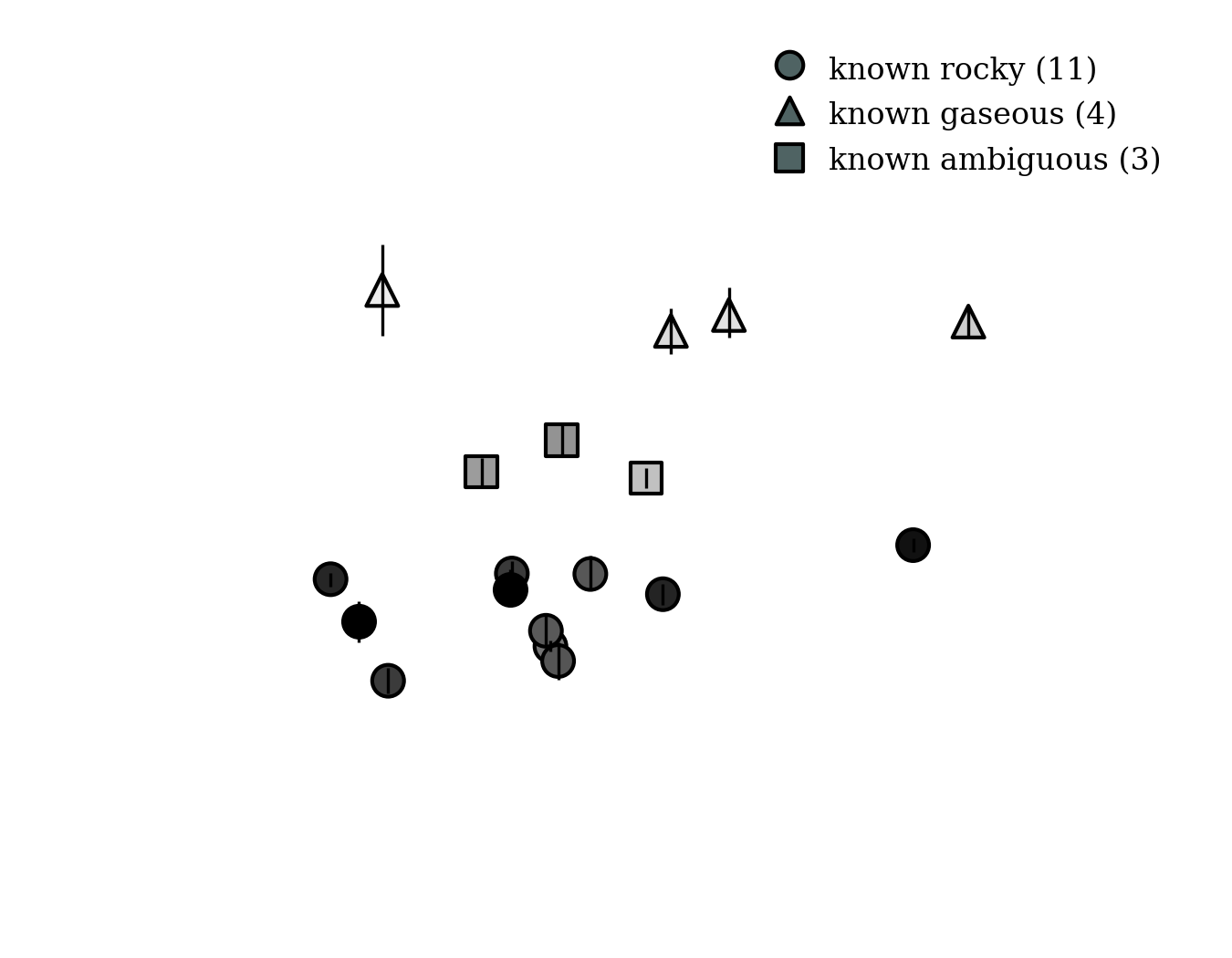}%
  \end{ocg}
  \hspace{-\hsize}%
  \begin{ocg}{fig:off3pr}{fig:off3pr}{0}%
  \end{ocg}%
  \begin{ocg}{fig:on3pr}{fig:on3pr}{1}%
  \includegraphics[width=\hsize]{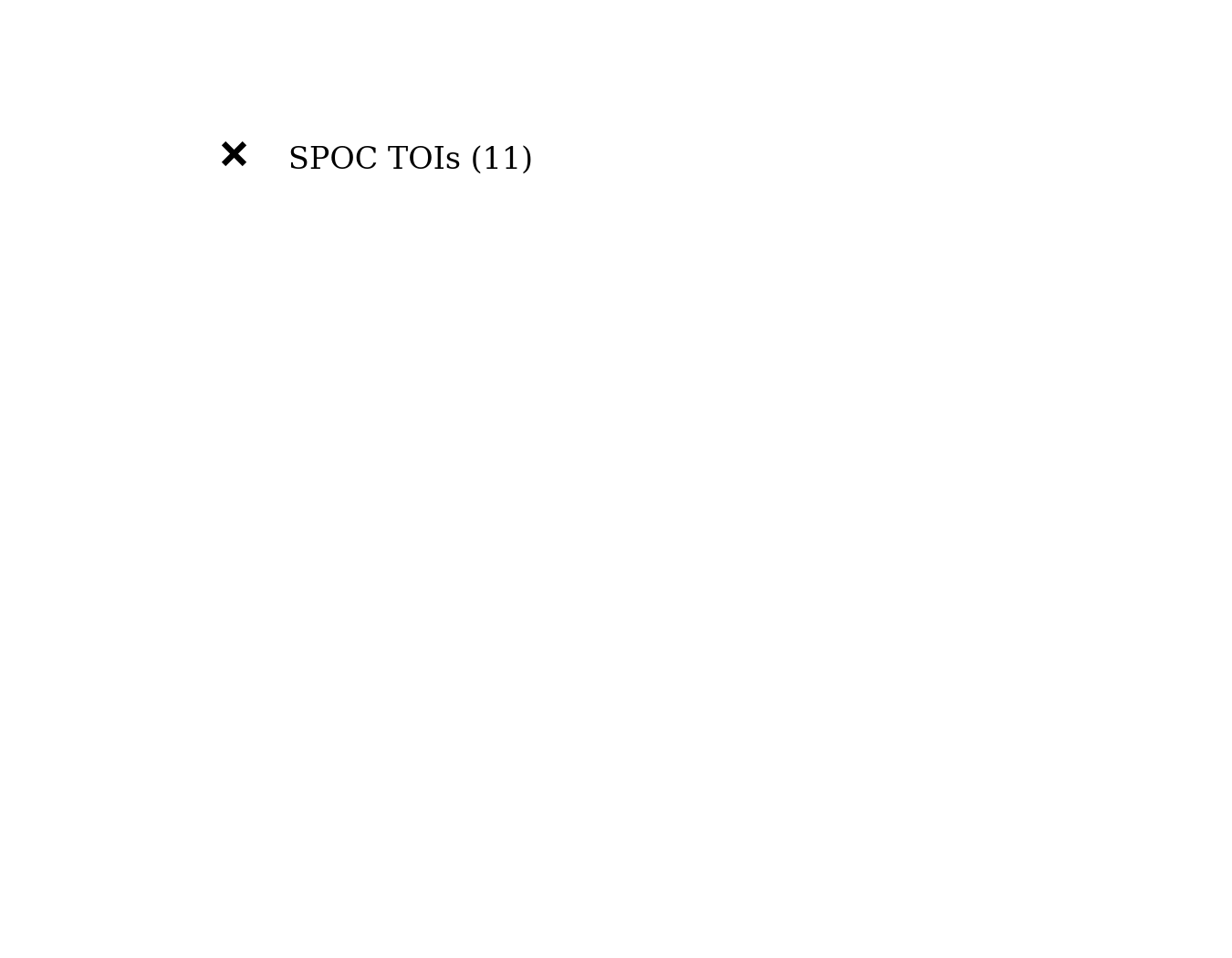}%
  \end{ocg}
  \hspace{-\hsize}%
  \caption{(\emph{Interactive figure}) Regions of interest for resolving the transition between rocky and non-rocky planets
    around low mass stars. The colormap depicts the occurrence rate of small close-in planets from
    \autoref{fig:fmap}. The rocky to non-rocky transition measured in this work is highlighted
    by the \emph{solid black line}. The transition measured around FGK stars from \cite{martinez19}
    is highlighted by the \emph{dashed black line} after being scaled to the median stellar mass of
    our sample. The \ToggleLayer{fig:on4pr,fig:off4pr}{\protect\cdbox{shaded regions}} bounded by
    these curves represent the subset of the $P-r_p$ parameter space of interest for resolving
    the transition around low mass stars with precise bulk density measurements. The eleven
    \ToggleLayer{fig:on3pr,fig:off3pr}{\protect\cdbox{TOIs}}
    that fall within this region are marked by an `x'. Also overplotted are
    \ToggleLayer{fig:on1pr,fig:off1pr}{\protect\cdbox{planets}}
    with $\geq 3\sigma$ bulk density measurements from the literature that are classified as having either
    a rocky (\emph{circles}), a gaseous (\emph{triangles}), or an ambiguous (\emph{squares}) bulk
    composition. Marker colors are indicative of the MAP planet bulk densities.}
  \label{fig:transition}
\end{figure}

Furthermore, the opposing slope signs of the radius valley around Sun-like and low mass
stars (c.f. \autoref{fig:fmapF} and \autoref{table:slopes})
has implications for where in the planetary parameter space one expects to find
predominantly rocky and non-rocky planets. \autoref{fig:transition} highlights the region of interest for resolving
the rocky to non-rocky transition in the $P-r_p$ space. This region is defined by the intermediate region between the
radius valley slope measured in this work to the slope measured around Sun-like stars from the CKS sample \citep{martinez19}
where the latter is first scaled
from its median stellar mass of 1.01 M$_{\odot}$ to the median stellar mass of our sample (0.65 M$_{\odot}$) using the
$M_s$-dependent scaling relation of the radius valley under photoevaporation \citep{wu19}.

The measured transitions from each stellar mass regime
intersect at $P\sim 23.52$ days such that two regions of interest emerge and are bounded by

\begin{equation}
  r_p \in 
  \begin{cases}
        [r_{p,\text{valley,this work}}, r_{p,\text{valley,M19}}], & P < 23.52 \text{ days,} \\
        [r_{p,\text{valley,M19}}, r_{p,\text{valley,this work}}], & P > 23.52 \text{ days.} \\
  \end{cases}  \label{eq:transition}
\end{equation}

\noindent where

\begin{align}
  r_{p,\text{valley,this work}} &= 0.11\cdot \log_{10}(P)+1.52, \\
  r_{p,\text{valley,M19}} &= -0.48\cdot \log_{10}(P)+2.32.
\end{align}

\noindent These subsets of the $P-r_p$ space define the regions of interest for resolving the
  rocky to non-rocky transition around low mass stars. For example, at $P\lesssim 23.52$ days thermally driven
  atmospheric mass loss, such as that from photoevaporation, predicts that planets in the set defined by
  Eq.~\ref{eq:transition} should be predominantly rocky. Whereas the gas-poor formation scenario, whose predicted
  radius valley slope differs in sign from that of thermally driven mass loss, predicts that those planets should be
  predominantly non-rocky. These predictions can be robustly tested by targeting planets within the
  Eq.~\ref{eq:transition} regions of interest and obtaining precise planetary bulk density measurements.
  As seen in \autoref{fig:transition}, there are only $\sim 5$ planets in the region of interest with $\geq 3\sigma$ bulk
  density measurements. These planets reveal the decrease in bulk density with increasing $r_p$ although insufficient
  information is available to resolve a possibly sharp transition.
  NASA's Transiting Exoplanet Survey Satellite \citep[\tess{;}][]{ricker15} has already identified eleven
  TOIs\footnote{\tess{} Object-of-Interest.}
  around stars with \teff{} $<4700$ K that satisfy Eq.~\ref{eq:transition} and should be targeted by RV
  follow-up campaigns.\footnote{TOIs: 134.01 \citep{astudillodefru19}, 237.01, 260.01, 544.01, 702.01, 807.01, 833.01,
    836.02, 873.01, 1075.01, and 1201.01.} Note that all eleven planet candidates have $P<23.52$ days.

\subsection{Imperfect clearing of the radius valley} \label{sect:void}
As noted in Sect.~\ref{sect:fmap} and evidenced in \autoref{fig:fmap}, the radius valley is not completely
void of confirmed planets. If the radius valley around low mass stars is produced solely from the late formation
of terrestrial planets in a gas-poor environment---as the measured slope of the radius valley suggests
(\autoref{fig:fmapF})---then the partial filling of the valley may point to a continuum of formed planet sizes
rather than two distinct populations of rocky and non-rocky planets. However if the radius valley is instead
sculpted by atmospheric post processing---as the radius valley around Sun-like stars seems to be
\citep{fulton17,fulton18,vaneylen18,martinez19}---then the partial filling of the valley would imply that post
processing from photoevaporation or core-powered mass loss is less efficient around lower mass stars.
This seems plausible in the photoevaporation scenario wherein the atmospheric mass loss rate is proportional to
the XUV luminosity of the host star. X-rays in particular are the dominant driver of atmospheric mass loss by
photoevaporation and observations have revealed that Sun-like stars have higher absolute X-ray luminosities than
their low mass star counterparts \citep{mcdonald19} despite the latter exhibiting a few to ten times larger
$L_X/L_{\text{bol}}$ ratios \citep{jackson12,shkolnik14}. 

Visual
investigation of the occurrence rates of small close-in planets in the stellar mass bins considered by
\citealt{fulton18} (i.e. $M_s<0.97$ M$_{\odot}$, $0.97 \leq M_s/\text{M}_{\odot} \leq 1.11$, and $M_s>1.11$ M$_{\odot}$)
suggests that the trend of increased filling of the valley with decreasing stellar mass may hold true   
\citep[c.f. Figure 9;][]{fulton18}. However we emphasize that this hypothesis has not been tested rigorously.   
We also note that classical models of photoevaporation do not explicitly consider the effects of
planetary magnetic fields on the efficiency of atmospheric mass loss and indeed the presence of such magnetic
fields can partially stall atmospheric escape by varying amounts depending on the core composition of the
planet \citep{owen19}.

The simplest explanation for the imperfect clearing of the radius valley instead does not invoke that either
gas-poor formation or atmospheric post processing is solely responsible for the valley's formation.
In comparing the slope of the radius valley around low mass and Sun-like stars (\autoref{fig:fmapF}), it is
clear that the dominant process affecting the slope of the valley with insolation
evolves from a positive slope to a negative slope from Sun-like to low mass stars. However,
the threshold $M_s$ dividing each regime is unresolved such that the planet population considered herein
likely features the superposition of processes such as gas-poor formation and thermally driven mass loss. The
result of competing processes on the observed planet population could naturally explain the apparent
partial filling of the gap.

\section{Summary of main findings} \label{sect:conclusions}
This study presented calculations of the occurrence rate of small close-in planets orbiting low mass stars
using data from the \kepler{} and \ktwo{} transit surveys. Our main findings are summarized below.

\begin{itemize}
\item The radius valley structure in the occurrence rate of small close-in planets---previously resolved around
  Sun-like stars---is demonstrated to persist around low mass stars (i.e. mid-K to mid-M dwarfs).
\item The occurrence rate from considering confirmed \ktwo{} planets only is likely underestimated. Although
  when \ktwo{} planet candidates are included and statistically corrected for false positive contamination, the
  measured \kepler{} and \ktwo{} cumulative occurrence rates of small close-in planets around low mass stars
  are in agreement: $2.48\pm 0.32$ and $2.26\pm 0.38$ planets per star respectively.
\item The radius valley around low mass stars exhibits a negative slope with insolation
  ($r_{p,\text{valley}} \propto F^{-0.060\pm 0.025}$) unlike 
  around Sun-like stars whose measured slope is positive
  \citep[$r_{p,\text{valley}} \propto F^{0.12\pm 0.02}$;][]{martinez19}. This result 
  supports models of gas-poor terrestrial planet formation without invoking atmospheric escape from photoevaporation,
  core-powered mass loss, or erosion by planetesimal impacts.
\item The non-rocky planet peak in the bimodal occurrence rates is centered at $\sim 2$ R$_{\oplus}$ and effectively
  vanishes around mid-M dwarfs as rocky planets ($\lesssim 1.54$ R$_{\oplus}$) increasingly dominate the close-in
  planet population towards later spectral types. The relative fraction of rocky to non-rocky planets increases
  from $\sim 0.5\pm 0.1 \to 8.5\pm 4.6$ from mid-K to mid-M dwarfs.
\item The occurrence rate-weighted location of the radius valley---and the central radius of each planet peak---shift
  to smaller
  sizes with decreasing stellar mass. The slope of the non-rocky planet peak's central radius with stellar mass is
  twice that of the rocky peak's slope indicating that the planet population in each stellar mass bin
  tends to converge towards rocky planet cores of $\sim 1$ R$_{\oplus}$ around later spectral types.
\item Robust inferences to distinguish between various proposed physical mechanisms for the formation of the
  radius valley are expected to require $\mathcal{O}(85,000)$ mid-M dwarfs observed with 2-minute cadence with
  \tess{} or $\mathcal{O}(4700)$ mid-M dwarfs observed with \plato{} based on its expected performance and
  observing strategy.
\item We advocate that transiting planets around stars with \teff{} $<4700$ K, and whose orbital periods and radii are
  situated between model predictions of the location of the rocky to non-rocky transition (see \autoref{fig:transition}),
  should be targeted by RV campaigns to resolve the location and slope of this transition with precise bulk density
  measurements. 
\end{itemize}

\acknowledgements
We thank Martin Paegert for his efforts in contributing relevant data to and for his assistance in querying the \tess{}
CTL database. We also thank Jennifer Winters and Jonathan Irwin for discussions regarding the population of nearby
M dwarfs. We also thank Sam Quinn for discussions regarding past results from the \ktwo{} mission.
We also thank David Charbonneau for his thorough review of this manuscript.
We thank the Canadian Institute for Theoretical Astrophysics for use of the Sunnyvale computing cluster
throughout the early stages of this work. RC was supported by the Natural Sciences and Engineering Research Council of
Canada and a grant from the National Aeronautics and Space Administration in support of the \tess{} science mission.  KM is supported by the Natural Sciences and Engineering Research Council of Canada.

\bibliographystyle{apj}
\bibliography{refs}

\end{document}

%% file: Keplerplanettable.tex
\capstartfalse
\begin{deluxetable*}{ccccccccc}
\tabletypesize{\small}
\tablecaption{Kepler confirmed planet parameters\label{table:planetsKep}}
\tablehead{KIC & Planet & $P$ & $F$ & $F$ upper limit & $F$ lower limit & $r_p$ & $r_p$ upper limit & $r_p$ lower limit \\
& name & [days] & [F$_{\oplus}$] & [F$_{\oplus}$] & [F$_{\oplus}$] & [R$_{\oplus}$] & [R$_{\oplus}$] & [R$_{\oplus}$]}
\startdata
2556650 & Kepler-1124 b & 2.85235 & 46.5 & 4.7 & 4.6 & 1.97 & 0.08 & 0.10 \\
2715135 & Kepler-753 b & 5.74771 & 40.2 & 4.6 & 4.5 & 1.89 & 0.30 & 0.12 \\
3234598 & Kepler-383 b & 12.90468 & 20.2 & 2.8 & 2.5 & 1.54 & 0.30 & 0.17 \\
3234598 & Kepler-383 c & 31.20122 & 6.2 & 0.8 & 0.8 & 1.49 & 0.34 & 0.22 \\
3426367 & Kepler-1308 b & 2.10434 & 55.3 & 5.6 & 5.1 & 0.89 & 0.03 & 0.14
\enddata
\tablecomments{Only the first five rows are shown here to illustrate the table's content and format. The complete table in csv format is available in the arXiv source.}
\end{deluxetable*}
\capstarttrue

%% file: K2planettable.tex
\capstartfalse
\begin{deluxetable*}{ccccccccc}
\tabletypesize{\small}
\tablecaption{K2 confirmed planet parameters\label{table:planetsK2}}
\tablehead{EPIC & Planet & $P$ & $F$ & $F$ upper limit & $F$ lower limit & $r_p$ & $r_p$ upper limit & $r_p$ lower limit \\
& name & [days] & [F$_{\oplus}$] & [F$_{\oplus}$] & [F$_{\oplus}$] & [R$_{\oplus}$] & [R$_{\oplus}$] & [R$_{\oplus}$]}
\startdata
201110617 & K2-156 b & 0.81315 & 615.4 & 51.0 & 55.4 & 1.35 & 0.12 & 0.10 \\
201155177 & K2-42 b & 6.68796 & 54.8 & 6.7 & 5.7 & 2.45 & 0.27 & 0.25 \\
201205469 & K2-43 c & 2.19888 & 81.8 & 8.5 & 7.9 & 1.43 & 0.09 & 0.08 \\
201205469 & K2-43 b & 3.47114 & 44.4 & 4.9 & 4.3 & 2.66 & 0.17 & 0.13 \\
201208431 & K2-4 b & 10.00440 & 16.5 & 1.8 & 1.6 & 2.52 & 0.34 & 0.31 
\enddata
\tablecomments{Only the first five rows are shown here to illustrate the table's content and format. The complete table in csv format is available in the arXiv source.}
\end{deluxetable*}
\capstarttrue

%% file: densitytable.tex
\capstartfalse
\begin{deluxetable*}{lccccccc}
\tabletypesize{\small}
\tablecaption{Planets with $\geq 3\sigma$ bulk density measurements around low mass stars\label{table:rhop}}
\tablehead{Planet & $P$ & $F$ & $r_p$ & $m_p$ & $\rho_p$ & Composition & Refs. \\
name & $[$days$]$ & $[$F$_{\oplus}]$ & $[$R$_{\oplus}]$ & $[$M$_{\oplus}]$ & $[$g cm$^{-3}]$ & disposition & }
\startdata
GJ 357b & 3.93072 & 13.2$\pm$ 1.4 & 1.22$\pm$ 0.08 & 1.84$\pm$ 0.31 & 5.67$^{+1.75}_{-1.35}$ & Rock & 1 \\
GJ 1132b & 1.62892 & 19.4$\pm$ 4.0 & 1.13$\pm$ 0.06 & 1.66$\pm$ 0.23 & 6.39$^{+1.36}_{-1.30}$ & Rock & 2,3 \\
GJ 1214b & 1.58040 & 22.2$\pm$ 3.0 & 2.85$\pm$ 0.20 & 6.26$\pm$ 0.86 & 1.50$^{+0.44}_{-0.34}$ & Gas & 4 \\
GJ 9827b & 1.20898 & 326.9$\pm$ 16.3 & 1.58$\pm$ 0.03 & 4.91$\pm$ 0.49 & 6.95$^{+0.81}_{-0.71}$ & Rock & 5 \\
GJ 9827c & 6.20147 & 37.0$\pm$ 1.8 & 2.02$\pm$ 0.05 & 4.04$\pm$ 0.84 & 2.71$^{+0.57}_{-0.55}$ & Am & 5 \\
HD 219134b & 3.09293 & 176.7$\pm$ 5.5 & 1.60$\pm$ 0.06 & 4.74$\pm$ 0.19 & 6.40$^{+0.75}_{-0.69}$ & Rock & 6 \\
HD 219134c & 6.76458 & 62.3$\pm$ 1.9 & 1.51$\pm$ 0.05 & 4.36$\pm$ 0.22 & 7.01$^{+0.83}_{-0.71}$ & Rock & 6 \\
K2-18b & 32.93962 & 1.2$\pm$ 0.1 & 2.71$\pm$ 0.07 & 8.63$\pm$ 1.35 & 2.40$^{+0.41}_{-0.40}$ & Gas & 7,8 \\
K2-146b & 2.64460 & 19.2$\pm$ 2.0 & 2.05$\pm$ 0.06 & 5.77$\pm$ 0.18 & 3.72$^{+0.34}_{-0.30}$ & Am & 9 \\
K2-146c & 4.00498 & 11.0$\pm$ 1.1 & 2.19$\pm$ 0.07 & 7.49$\pm$ 0.24 & 3.96$^{+0.38}_{-0.38}$ & Am & 9 \\
Kepler-80b & 7.05246 & 41.8$\pm$ 4.8 & 2.67$\pm$ 0.10 & 6.93$\pm$ 0.70 & 2.02$^{+0.43}_{-0.35}$ & Gas & 10 \\
Kepler-80c & 9.52355 & 28.0$\pm$ 3.2 & 2.74$\pm$ 0.12 & 6.74$\pm$ 0.86 & 1.82$^{+0.45}_{-0.39}$ & Gas & 10 \\
Kepler-80d & 3.07222 & 126.5$\pm$ 14.5 & 1.53$\pm$ 0.09 & 6.75$\pm$ 0.51 & 10.46$^{+2.30}_{-2.04}$ & Rock & 10 \\
Kepler-80e & 4.64489 & 72.9$\pm$ 8.3 & 1.60$\pm$ 0.08 & 4.13$\pm$ 0.95 & 5.60$^{+1.54}_{-1.32}$ & Rock & 10 \\
L 98-59c & 3.69040 & 11.9$\pm$ 1.5 & 1.35$\pm$ 0.07 & 2.46$\pm$ 0.31 & 5.55$^{+1.22}_{-1.01}$ & Rock & 11,12 \\
L 168-9b & 1.40150 & 144.2$\pm$ 12.0 & 1.39$\pm$ 0.09 & 4.60$\pm$ 0.60 & 9.51$^{+2.53}_{-2.11}$ & Rock & 13 \\
LHS 1140b & 24.73696 & 0.5$\pm$ 0.0 & 1.73$\pm$ 0.03 & 6.98$\pm$ 0.89 & 7.52$^{+1.11}_{-1.06}$ & Rock & 14 \\
LHS 1140c & 3.77793 & 6.1$\pm$ 0.5 & 1.28$\pm$ 0.02 & 1.81$\pm$ 0.39 & 4.77$^{+1.09}_{-1.01}$ & Rock & 14
\enddata
\tablecomments{References: 1) \citealt{luque19} 2) \citealt{dittmann17b} 3) \citealt{bonfils18} 4) \citealt{harpsoe13} 
5) \citealt{rice19} 6) \citealt{gillon17b}
7) \citealt{benneke17} 8) \citealt{cloutier19a} 9) \citealt{hamann19} 10) \citealt{macdonald16}
11) \citealt{kostov19} 12) \citealt{cloutier19c} 13) \citealt{astudillodefru19} 14) \citealt{ment19}.}
\end{deluxetable*}
\capstarttrue

%% file: FPtable.tex
\capstartfalse
\begin{deluxetable}{lccc}
\tabletypesize{\small}
\tablecaption{\ktwo{} false positive rates for small planets around cool stars\label{tab:FP}}
\tablehead{Reference & $N_{\text{FP}}$ & $N_{\text{VP}}$ & FP rate [\%]}
\startdata
\cite{montet15}\tablenotemark{a} & 0 & 8 & $<30.7$ \\
\cite{crossfield16b} & 2 & 39 & $4.9^{+6.0}_{-1.4}$ \\
\cite{dressing17} & 2 & 34 & $5.6^{+6.4}_{-2.0}$ \\
\cite{hirano18}\tablenotemark{a} & 0 & 16 & $<19.5$ \\
\cite{livingston18a}\tablenotemark{a}  & 0 & 14 & $<21.0$ \\
\cite{mayo18}\tablenotemark{b} & 1 & 14 & $6.7^{+12.4}_{-2.0}$
\enddata
\tablecomments{Within each study we only consider PCs with $r_p <4$ R$_{\oplus}$ and orbiting cool stars with \teff{} $<4700$ K. FP: false positive. VP: validated planet.}
\tablenotetext{a}{These studies do not detect any FPs such that the reported FP rate upper limit is represented by its 95\% confidence interval.}
\tablenotetext{b}{\cite{mayo18} did not explicitly classify their non-validated planets as FPs so we define FPs within their sample as any PC whose false positive probability exceeds 10\%.}
\end{deluxetable}
\capstarttrue

%% file: slopetable.tex
\capstartfalse
\begin{deluxetable}{lccc}
\tabletypesize{\small}
\tablecaption{Measurements and model predictions of the radius valley slope\label{table:slopes}}
\tablehead{Description & $\mathrm{d}\log{r_p} / \mathrm{d}\log{P}$ & $\mathrm{d}\log{r_p} / \mathrm{d}\log{F}$ & Refs.}
\startdata
\multicolumn{4}{c}{\emph{Empirical slope measurements}} \\
\vspace{-0.15cm} Slope around &&& \\ \vspace{-0.2cm} 
& $0.058\pm 0.022$ & $-0.060\pm 0.025$ & 1 \\ 
low mass stars &&& \\
\vspace{-0.15cm} Slope around &&& \\ \vspace{-0.2cm} 
& $-0.11\pm 0.03$ & $0.12\pm 0.02$ & 2 \\ 
Sun-like stars &&& \\
\multicolumn{4}{c}{\emph{Model-predicted slopes}} \\
\vspace{-0.15cm} Gas poor &&& \\ \vspace{-0.2cm} 
& $0.11$ & $-0.08$ & 3 \\ 
formation &&& \\
Photoevaporation & $-0.15$ & $0.11$ & 3 \\
\vspace{-0.15cm} Core-powered &&& \\ \vspace{-0.2cm} 
& $-0.13$ & $0.10$ & 4 \\ 
mass loss &&& \\
Impact erosion & $-0.33$ & $0.25$ & 5
\enddata
\tablecomments{$r_p$ in units of R$_{\oplus}$, $P$ in units of days, and $F$ in units of $F_{\oplus}$. \\
\textbf{References:} 1) this work 2) \citealt{martinez19} 3) \citealt{lopez18} 4) \citealt{gupta19b}
5) \citealt{wyatt19}.}
\end{deluxetable}
\capstarttrue

%% file: reltable.tex
\capstartfalse
\begin{deluxetable}{cccc}
\tabletypesize{\small}
\tablecaption{Relative occurrence rates of close-in rocky and non-rocky planets around low mass stars\label{tab:rel}}
\tablehead{Stellar mass & $f_{\mathrm{rocky}}$ & $f_{\mathrm{non-rocky}}$ & $f_{\mathrm{rocky}}/f_{\mathrm{non-rocky}}$ \\
range $[\mathrm{M}_{\odot}]$ & $r_p \in [1,1.6]$ & $r_p \in [1.6,2.5]$ &}
\startdata
$[0.08,0.90]$ & $0.68\pm 0.07$ & $1.02\pm 0.08$ & $0.66\pm 0.09$ \\
$[0.63,0.90]$ & $0.69\pm 0.11$ & $1.28\pm 0.16$ & $0.54\pm 0.11$ \\
$[0.08,0.63]$ & $1.10\pm 0.16$ & $1.02\pm 0.16$ & $1.08\pm 0.23$ \\
$[0.08,0.42]$ & $1.64\pm 0.43$ & $0.19\pm 0.09$ & $8.46\pm 4.62$
\enddata
\end{deluxetable}
\capstarttrue

%% file: boundstable.tex
\capstartfalse
\begin{deluxetable*}{ccccccc}
\tabletypesize{\small}
\tablecaption{Assumed boundary ranges on the locations of radius valley features\label{tab:bounds}}
\tablehead{Stellar mass & $\log{P}$ lower & $\log{P}$ upper & Rocky & Rocky  & Non-rocky  & Non-rocky \\
  range & boundary & boundary & peak lower $r_p$ & peak upper $r_p$ & peak lower $r_p$  & peak upper $r_p$  \\
  $[\mathrm{M}_{\odot}]$ & $[\text{days}]$ & $[\text{days}]$ & boundary $[\text{R}_{\oplus}]$ & boundary $[\text{R}_{\oplus}]$ & boundary $[\text{R}_{\oplus}]$ & boundary $[\text{R}_{\oplus}]$}
\startdata
$[0.08,0.90]$ & $\mathcal{U}(\log{0.5},\log{2})$ & $\mathcal{U}(\log{50},\log{100})$ & $\mathcal{U}(0.8,1)$ & $\mathcal{U}(1.2,1.5)$ & $\mathcal{U}(1.6,1.9)$ & $\mathcal{U}(2.3,2.5)$ \\
$[0.63,0.90]$ & $\mathcal{U}(\log{0.5},\log{2})$ & $\mathcal{U}(\log{50},\log{100})$ & $\mathcal{U}(0.8,1)$ & $\mathcal{U}(1.3,1.5)$ & $\mathcal{U}(1.8,2)$ & $\mathcal{U}(2.4,2.7)$ \\
$[0.08,0.63]$ & $\mathcal{U}(\log{0.5},\log{2})$ & $\mathcal{U}(\log{50},\log{100})$ & $\mathcal{U}(0.6,0.9)$ & $\mathcal{U}(1.2,1.4)$ & $\mathcal{U}(1.8,2)$ & $\mathcal{U}(2.1,2.3)$ \\
$[0.08,0.42]$ & $\mathcal{U}(\log{0.5},\log{2})$ & $\mathcal{U}(\log{50},\log{100})$ & $\mathcal{U}(0.5,0.7)$ & $\mathcal{U}(1.3,1.4)$ & $\mathcal{U}(1.7,1.8)$ & $\mathcal{U}(1.8,2)$
\enddata
\tablecomments{The $r_p$ boundaries on the radius valley are given implicitly by the upper $r_p$ limit on the rocky peak and the lower $r_p$ limit on the non-rocky peak.}
\end{deluxetable*}
\capstarttrue

%% file: rpvMstable.tex
\capstartfalse
\begin{deluxetable}{cccc}
\tabletypesize{\small}
\tablecaption{Radius valley features versus stellar mass\label{tab:rpvMs}}
\tablehead{Stellar mass & Rocky peak & Radius valley & Non-rocky peak \\
$[\mathrm{M}_{\odot}]$ & $[\text{R}_{\oplus}]$ & $[\text{R}_{\oplus}]$ & $[\text{R}_{\oplus}]$}
\startdata
$0.651^{+0.058}_{-0.096}$ & $1.118^{+0.151}_{-0.148}$ & $1.543^{+0.160}_{-0.160}$ & $2.068^{+0.211}_{-0.205}$ \\
$0.684^{+0.040}_{-0.035}$ & $1.154^{+0.205}_{-0.239}$ & $1.647^{+0.207}_{-0.215}$ & $2.197^{+0.301}_{-0.256}$ \\
$0.500^{+0.097}_{-0.146}$ & $1.036^{+0.297}_{-0.308}$ & $1.599^{+0.340}_{-0.352}$ & $2.048^{+0.191}_{-0.199}$ \\
$0.343^{+0.057}_{-0.092}$ & $1.017^{+0.700}_{-0.807}$ & $1.548^{+0.515}_{-0.496}$ & -
\enddata
\tablecomments{As depicted in Fig.~\ref{fig:rpvMs}.}
\end{deluxetable}
\capstarttrue